\documentclass[12pt]{article}
\usepackage{fancyhdr}
\usepackage{nomencl}
\usepackage{cite}
\usepackage{enumerate}
\usepackage{supertabular}
\usepackage{graphicx} 
\usepackage{rotating,psfrag,epsfig}
\usepackage{subfigure}
\usepackage{graphics}
\usepackage{subfigure}
\usepackage{latexsym}
\usepackage{setspace}
\usepackage{tweaklist}
\newcommand{\be}{\begin{equation}}
\newcommand{\ee}{\end{equation}}
\newcommand{\bea}{\begin{eqnarray}}
\newcommand{\eea}{\end{eqnarray}}
\newcommand{\bean}{\begin{eqnarray*}}
\newcommand{\eean}{\end{eqnarray*}}

\newcounter{sectno}
\newcounter{subsectno}
\newcounter{subsubsectno}

\begin{document}
\baselineskip 24pt
\begin{center}
{\Large \bf {Numerical analysis of turbulent forced convection and fluid flow past a triangular cylinder with control plate using standard $\kappa-\epsilon$ model}}
\end{center}
\begin{center}
Smruti Ranjan Jena$ \footnote[1]{E-mail:smrutiranjanjenaiter@gmail.com}$, Amit Kumar Naik$\footnote[2]{E-mail:amitnaik.babul@gmail.com}$, Amaresh Dalal$\footnote[3]{Author for Correspondence: Phone: 0091-361 2582677, E-mail: amaresh@iitg.ac.in}$, Ganesh Natarajan$\footnote[4]{E-mail: n.ganesh@iitg.ac.in}$ \\
Department of Mechanical Engineering\\
 Indian Institute of Technology Guwahati\\
 Guwahati - 781 039, India.\\ 
\end{center}
\vspace{25pt} Manuscript submitted to \emph{International Journal of Thermal Sciences}.\\ 
We declare that this manuscript has not been published elsewhere and has not been submitted for publication elsewhere.\vspace{150pt} \pagebreak
\section*{Abstract}
Turbulent flow past an equilateral triangular cylinder with splitter plate inserted downstream is numerically tested for different gap ratios (0, 0.5, 1, 1.5, 2) and plate dimensions (0, 1, 1.5) on the flow field and heat transfer characteristics. Unsteady flow simulations are carried out at Re=22,000 in a finite volume based collocated framework, on a two-dimensional unstructured mesh. Reynolds averaged momentum and energy equations are solved in conjunction with the standard $\kappa-\epsilon$ model. In this study, cylinder and control plate are subjected to constant wall temperature. It is observed that the drag force on the triangular cylinder-splitter plate system reduced with an increase in gap ratio. Vortex shedding is suppressed as Strouhal number ($St$) reduced to its least value for the maximum gap-ratio configuration studied. Heat transfer performance is also significantly improved with the inclusion of a finite gap. In addition to that, the effect of variation in length of the splitter plate has also been studied on the force coefficients, Strouhal number, local and surface averaged Nusselt number. Results show that increasing the length of the splitter plate significantly suppressed the shedding with a minimum frequency obtained for the maximum plate length of $L_{s}/h=1.5$. However, overall heat transfer reduced with the increase in plate length.\\
\section*{Nomenclature}
\begin{flushleft}
\begin{tabular}{ll}
$\overline{C_{b_{p}}} $& Time-averaged base pressure coefficient\\
$\overline{C_{d}} $& Time-averaged drag coefficient\\
$\overline{C_{d_{p}}} $& Time-averaged pressure drag coefficient\\
$C_{l_{rms}}$& Root mean square value of lift coefficient\\
$C_{p}$& Specific heat at constant pressure ($J.kg^{-1}.K^{-1}$)\\
$\overline{C_{p}} $& Time-averaged pressure coefficient\\
$\overline{C_{pl}} $& Time-averaged local pressure coefficient\\
$C_{\epsilon_{1}},C_{\epsilon_{2}},C_{\mu},f_{\mu}$& Model constants\\
$E$& Experimentally obtained constant\\
$f$& Frequency of a signal ($s^{-1}$)\\
$f_{1},f_{2}$& Damping functions\\
$F$& Drag force (N)\\
$G$ & Gap between the cylinder and the plate (m)\\
$h$ & Side-length of the triangular cylinder (m)\\
$H$ & Height of the domain (m)\\
$i,I$& Turbulent intensity\\
$k$& Thermal conductivity ($W.m^{-1}.k^{-1}$)\\
$L$ & Length of the domain (m)\\
$L_{s}$ & Length of the splitter plate\\
$L_{Vf}$ & Vortex formation length \\
$\overline{Nu}$ & Time-averaged, surface averaged Nusselt number\\
$\overline{Nu_{l}}$ & Time-averaged local Nusselt number\\
$p$& Non-dimensionalized pressure \\
$P_{k,n}$& Production term in its non-dimensional form\\
$Pr$ & Prandtl number\\
$Re$ & Reynolds number\\
$St$ & Strouhal number\\

\end{tabular}
\end{flushleft}

\newpage
\begin{flushleft}
\begin{tabular}{ll}
$t$& Non-dimensionalized temporal co-ordinate\\
$u_{x}, u_{y}$ & Non-dimensionalized stream-wise and spanwise component of velocity \\
$\overline{u'v'} $ & Time-averaged Reynolds stress\\
$\overline{u} $ & Time-averaged stream-wise velocity (m/s)\\
$U_{in}$& Inlet velocity or free-stream velocity (m/s)\\
$x,y$ & Non-dimensionalized spatial co-ordinates\\
$X_{u},X_{d}$ & Upstream distance, downstream distance (m/s)\\
$y^{+}$& Normalised wall normal co-ordinate ($\frac{yu_{\tau}}{\nu}$)\\

\vspace*{2mm}
{\bf Greek symbol}\\
$\alpha$& Thermal diffusivity ($m^2/s$)\\
$\beta$ & Blockage ratio\\
$\delta$ & Minimum edge length in a grid\\ 
$\Delta $& Maximum edge length in a grid\\
$\Delta t$& Time-step size (s)\\
$\epsilon $ & Non-dimensionalized dissipation rate of turbulent kinetic energy \\ 
$\epsilon_{inlet} $& Inlet turbulent kinetic energy dissipation rate ($m^2/s^{3}$)\\
$\kappa $ & Non-dimensionalized turbulent kinetic energy \\
$\kappa_{in}$ & Inlet turbulent kinetic energy ($m^2/s^{2}$)\\
$\mu$ & Dynamic viscosity ($kg.m^{-1}.s^{-1}$) \\
$\mu _{t,n}$& Non-dimensionalized turbulent viscosity \\ 
$\nu _{t,n}$& Non-dimensionalized turbulent kinematic Viscosity \\
$\omega_{min},\omega_{max}$& Minimum and maximum vorticity magnitude ($s^{-1}$)\\
$\rho$ & Density ($kg.m^{-3}$)\\
$\sigma_{t} $& Turbulent Prandtl number\\
$\tau$ & Time period of lift-coefficient signal\\
$\theta $ & Non-dimensional temperature\\
\end{tabular}
\end{flushleft}

\vspace*{2mm}
\newpage
\begin{flushleft}
\begin{tabular}{ll}
{\bf Abbreviations}\\
$2D$& Two-dimensional\\
$CDS$  & Central difference scheme\\
$FFT$& Fast-fourier transform\\
$PSD$& Power spectral density\\
$RANS$ & Reynolds averaged Navier-Stokes solver\\
$SIMPLE$ & Semi-implicit method for pressure linked equations\\
$TCSP$ & Triangular cylinder-splitter plate system\\
\end{tabular}
\end{flushleft}

\clearpage
\section{Introduction}
Flow involving turbulence are almost ubiquitous in all engineering applications involving fluid flow. Vortex shedding past a bluff body, otherwise called blunt body, has been a matter of intense academic and industrial research for the last few decades. When flow encounters a confined edge cylinder e.g. triangular or square cylinder flow separation occurs leading to formation of large scale vortical structures. Due to its detrimental effect on buildings and structures, suppression of shedded vortices is essential for which various near-wake and far-wake stabillization methodologies have been developed. Vortex shedding suppression using near-wake stabilisers, such as splitter plate, has been a matter of tremendous academic interest for researchers due to its widespread applications in various flow scenarios like electronic cooling, mitigation of shedding past structures, mixing enhancement etc. To shed some light on this matter, a numerical study is conducted so as to see the effects of splitter plate on vortex shedding in a turbulent unconfined flow regime.

Johansson et al. \cite{Johansson} analyzed unsteady turbulent flow past triangular-shaped flame holder at $Re=45,000$ using $\kappa -\epsilon$ model and validated their results against the experimental findings of Sjunnesson et al. \cite{Sjunnesson}. He concluded that the transport of mean momentum was more affected by the vortex shedding, than by turbulence. A numerical study on turbulent separated flow past a triangular cylinder using $\kappa-\epsilon-v^2$ model was conducted by Durbin et al. \cite{Durbin}. The onset of periodic behavior in two-dimensional laminar flow past bodies of various shapes was examined by Jackson et al. \cite{Jackson}. In their study, the critical Reynolds number was found to be 34.318 and corresponding critical Strouhal number as 0.13554 for an isosceles triangle. Valencia and Cid \cite{valencia} numerically investigated the unsteady turbulent flow past square bars using $\kappa -\epsilon$ model at a Reynolds number of 20,000. Ratnam and Vengadesan \cite{ratnam} conducted a performance assessment of various two equation models namely, standard $\kappa -\epsilon$, low-Reynolds number $\kappa -\epsilon$, non-linear $\kappa -\epsilon$, standard $\kappa -\omega$ and improved $\kappa -\omega$ model etc. They conducted  simulations on flow past a wall mounted cube at Reynolds number 1,870 and the effect of complex vortex structure on heat transfer. Three dimensional simulations on flow past a circular cylinder at Reynolds number 3,900 were performed by Ayyappan and Vengadesan \cite{ayyappan} using nonlinear turbulence model based on the $\kappa -\epsilon$ formulation and also made a comparative model assessment with standard $\kappa -\epsilon$ model of which the former was found to be better.  

Experimental investigation of flow past a circular cylinder with a splitter plate (length of 50 mm) was conducted for $Re=5,000$ by  Akilli et al. \cite{Akilli}. They analyzed the effect of gap ratio and plate thickness on vortex shedding. It was found that splitter plate has a substantial effect on vortex shedding for gap ratio varying from $0$ to $1.75D$. Similar study was conducted by Nakayama et al. \cite{Nakayama} for turbulent flow past a bluff body with a blunt trailing edge at $Re=22,000$. Yagmur et al. \cite{yagmur} investigated the flow past a triangular cylinder at different Reynolds Number both numerically (using Large Eddy Simulation) and experimentally (employing Particle Image Velocimetry). He found a good agreement between both the analyses and concluded that the drag coefficient showed a decrement with an increase in Reynolds number. Liu et al. \cite{liu} had performed an experimental study on flow over a circular cylinder in a confined channel at ${Re_D}$ = 2400 and 3000 with plate length ($L/D$) as the control parameter and further, found $L/D$ = 1 as the optimal plate length for suppression of vortex shedding. He had also observed the formation of secondary vortices in case of larger plate lengths. Dai et al. \cite{dai} studied flow over a circular cylinder in high Reynolds number regime using OpenFOAM and had indicated splitter plate to be a practical method for vortex shedding suppression. Zhang and Shi \cite{zhang} numerically studied the flow past a circular cylinder – splitter plate (CCSP) system using Lattice Boltzmann Method and thoroughly analyzed the suppression of vortex shedding near a moving wall by varying the gap ratio between the cylinder and wall. The above study confirmed that vortex shedding completely disappeared when gap ratio went below 0.2 due to the presence of a splitter plate. Zdravkovich et al. \cite{Zdravkovich} and Unal et al. \cite{Unal} have studied the effect of splitter plates on flow dynamics. 
Two dimensional wake dynamics, for a range of $Re$ number with incidence angle variation, behind triangular cylinders were studied numerically by Sheard et al. \cite{sheard}. It was found that the cylinder inclined at $60^{\circ}$ exhibits the largest fluctuations in the force profiles and further cylinder inclination has been found to have a notable effect on the separation Reynolds number. 

De and Dalal \cite{De2} numerically studied the laminar flow past a triangular cylinder in an unconfined regime and also presented an investigation on study of fluid flow and heat transfer past a triangular cylinder in a confined channel \cite{De}. 
Farhadi et al. \cite{Farhadi} numerically investigated the effect of wall proximity of a triangular cylinder on the heat transfer and flow field in a horizontal channel. Computations were carried out for Reynolds number with in a range of $100-450$ with different gap widths. It was further analyzed that the approaching triangular cylinder near the wall removes vortex shedding and with decrease in vortex shedding drag coefficient shows a noteworthy reduction. Immanuvel et al. \cite{arul} studied laminar forced convection past an elliptic cylinder in an unconfined regime and performed parametric study to see the effects on the local and surface averaged Nusselt number.
An experimental study on the effect of free stream turbulence on local mass transfer was conducted for a circular cylinder by Goldstein et al. \cite{Goldstein}. In their study, $Re$ has been varied from $3\times10^4$ to $8.3\times10^4$ with turbulence intensity varying from $0.2\%$ to $23.7\%$ and subsequently the effect of splitter plate on local mass transfer was also studied. It has also been found that splitter plate reduced the local mass transfer for low free stream turbulence values, though for higher values it seems to have little effect on local mass transfer. 
Apelt and West \cite{Apelt2} have performed an analysis on the effect of wake splitter plate length for turbulent flow past a circular cylinder for $2\leq L/D\leq 7$. Lengthening of splitter plate didn't produce any such noticeable effect. 
Bearman \cite{bearman} has experimentally verified the effects of two dimensional flow past a blunt trailing edge for Reynolds number ranging from $1.4\times10^5$ to $2.56\times10^5$ and found that the distance of fully formed vortex was inversely proportional to the base pressure coefficient. \\

From the above discussion it is observed that most of the work is focussed on flow past a triangular cylinder in a laminar and intermediate Reynolds number range. Many novel works have been done on flow modification behind a triangular bluff-body using a splitter plate but analysis in a turbulent regime has not been accomplished so far to the best of our knowledge which mainly inspired us for the present work. The purpose of this study is to understand the effect of variation in gap ratio ($G/h$) and plate length ($L_{s}/h$) on fluid flow and heat transfer.\\

\section{Governing equations and boundary conditions}
The two-dimensional continuity, momentum and energy equations for turbulent flow of an incompressible fluid having uniform density and constant thermo-physical properties are expressed below in their non-dimensional form.\\
Continuity equation:\\
\begin{equation}
\frac{\partial u_{i}}{\partial x_{i}} =0,\\
\end{equation}
Momentum Equation:\\
\begin{equation}
\frac{\partial u_{i}}{\partial t}+u_{j}\frac{\partial u_{i}}{\partial x_{j}}=-\frac{\partial p}{\partial x_{i}} + \frac{1}{Re}\frac{\partial}{\partial x_{j}}\Big[(1+\mu_{t,n})\frac{\partial u_{i}}{\partial x_{j}}\Big]\\ 
\end{equation}
Energy equation can be written as:\\
\begin{equation}
\frac{\partial \theta}{\partial t}+u_{j}\frac{\partial \theta}{\partial x_{j}}=\frac{1}{RePr}\frac{\partial}{\partial x_{j}}\Big[\Big(1+\frac{\nu _{t,n}}{\sigma _{t}}\Big)\frac{\partial  \theta}{\partial x_{j}}\Big]\\
\end{equation}
Transport equation for turbulent kinetic energy($\kappa$):\\
\begin{equation}
\frac{\partial \kappa}{\partial t}+u_{j}\frac{\partial \kappa}{\partial x_{j}}=\frac{1}{Re}\frac{\partial}{\partial x_{j}}\Big[\Big(1+\frac{\nu _{t,n}}{\sigma _{\kappa }}\Big)\frac{\partial\kappa }{\partial x_{j}}\Big] + P_{k,n}-\epsilon  \\
\end{equation}
Transport equation for dissipation rate of turbulent kinetic energy($\epsilon$):\\
\begin{equation}
\frac{\partial \epsilon}{\partial t}+u_{j}\frac{\partial \epsilon}{\partial x_{j}}=\frac{1}{Re}\frac{\partial}{\partial x_{j}}\Big[\Big(1+\frac{\nu _{t,n}}{\sigma _{\epsilon }}\Big)\frac{\partial\epsilon}{\partial x_{j}}\Big] +\frac{\epsilon }{\kappa }(C_{\epsilon _{1}}P_{k,n}-C_{\epsilon _{2}}\epsilon)\\
\end{equation}
Where,\\
\begin{equation}
  P_{k,n}= \frac{\nu _{t,n}}{Re}\Big[\Big(\frac{\partial u_{i}}{\partial x_{j}}+\frac{\partial u_{j}}{\partial x_{i}}\Big)\frac{\partial u_{i}}{\partial x_{j}}\Big]\\
\end{equation}
The model constants are employed as $C_{\mu}=0.09$, $C_{\epsilon_{1}}=1.44$, $C_{\epsilon_{2}}=1.92$, $\sigma_{\kappa}=1.0$, $\sigma_{\epsilon}=1.3$, $\sigma_{t}=0.9$. The lengths are non-dimensionalized with the side length of the cylinder ($h$), all velocity variables are non-dimensionalized with average velocity at the inlet ($U_{in}$) and pressure with $\rho {U_{in}}^2$. The turbulent kinetic energy ($\kappa$) and the dissipation rate of kinetic energy ($\epsilon$) are non-dimensionalized with ${U_{in}}^2$ and $\frac{{U_{in}}^2}{h}$ respectively. Finally time is non-dimensionalized with $\frac{h}{U_{in}}$ and temperature scale is non-dimensionalized as $\theta$=$\frac{T-T_{in}}{T_{wall}-T_{in}}$. In this study, Reynolds number is defined as $\frac{\rho U_{in} h}{\mu}$, Prandtl number is defined as $\frac{\nu}{\alpha}$ which is taken as 0.71 for air and the Strouhal number is expressed as $St = \frac{{fh}}{{{u_{in}}}}$, where $f$ refers to the vortex shedding frequency that characterizes the periodicity in the flow.\\
The boundary conditions in their non-dimensional form are given as follows.\\
$\bullet$ At the inlet boundary, \\
\begin{equation}
u_{x}=1, u_{y}=0, \theta=0, \frac{\partial p}{\partial x_{i}}=0, \kappa _{inlet}={(u_{inlet}i)^2}, \epsilon _{inlet}=\frac{C_{\mu }\kappa _{inlet}^2}{\nu r_{\mu }}.\\
\end{equation}\\
Here, constant velocity boundary condition is used at the inlet. Free stream turbulence intensity ($i$) is considered to be 2\% and viscosity ratio is taken as $r_{\mu}$=5.\\
$\bullet$ At the outlet boundary, \\
\begin{equation}
\frac{\partial u_{x}}{\partial x}=0, \frac{\partial u_{y}}{\partial x}=0, \frac{\partial \theta }{\partial x}=0, p_{outlet}=0, \frac{\partial \kappa}{\partial x}=0, \frac{\partial \epsilon  }{\partial x}=0.\\
\end{equation}
In this study, pressure is specified at the exit plane while fully developed flow boundary conditions are used for all other variables. \\
$\bullet$ At the top and bottom boundary, free-slip boundary condition is implemented i.e. \\
\begin{equation}
\frac{\partial u_{x}}{\partial y}=0, \frac{\partial u_{y}}{\partial y}=0, \frac{\partial \theta }{\partial y}=0, \frac{\partial p}{\partial y}=0, \frac{\partial \kappa}{\partial y}=0, \frac{\partial \epsilon  }{\partial y}=0. \\
\end{equation}\\
$\bullet$ At the wall boundary, isothermal condition for temperature, no-slip condition for velocity and homogeneous neumann condition for other variables are used.
\begin{equation}
u_{x}=0 ,u_{y}=0, \theta=1, \frac{\partial p}{\partial n}=0, \kappa=0, \frac{\partial \epsilon}{\partial n}=0. \\  
\end{equation}
where ``$n$'' refers to the direction normal to the surface.\\
\subsection{Non-dimensional coefficients}
The instantaneous drag coefficient is computed as ${C_d} = \frac{F}{{\frac{1}{2}\rho u_{in}^2}h}$, where $F$ refers to the drag force per unit width acting on the cylinder-splitter plate system. Further, the time-averaged drag coefficient ($\overline{C}_d$) is calculated by averaging the instantaneous drag coefficient ($C_d$) over the time period, after reaching the dynamically steady state. Similarly, the instantaneous lift coefficient is computed as ${C_l} = \frac{F_L}{{\frac{1}{2}\rho u_{in}^2}h}$, where $F_L$ refers to the lift force per unit width that acts on the cylinder-splitter plate system on the lateral direction. Further, the root mean square value of the lift coefficient ($C_{lrms}$) is obtained from the root of the mean of the squares of the instantaneous lift coefficient values, after reaching dynamically steady state.\\ 

The expressions for the various pressure coefficients that are used in the present study are detailed as follows. The instantaneous local pressure coefficient ($C_{pl}$) has been computed as ${C_{pl}} = \frac{{{p_l} - {p_0}}}{{\frac{1}{2}\rho u_{in}^2}}$, where ${p_l}$ refers to the local pressure and ${p_0}$ refers to the ambient pressure. Afterwards, the time-averaged local pressure coefficient ($\overline{C_{pl}}$) is obtained by averaging the local pressure coefficient over the time period. Furthermore, $\overline{C_{bp}}$ is referred to as time averaged base pressure coefficient which is calculated by area-averaging the time-averaged local pressure coefficient ($\overline{C_{pl}}$) over the base of the cylinder. In the present study, the backside of the triangular cylinder $i.e.$ from point “B” – through point “C” – to point “D” (as referred in Figs. \ref{fig:timeavg_local_Pressure_coefficient} and \ref{fig:timeavg_local_nusselt_number} ) is considered as the base of the cylinder.\\

The Nusselt number, being a quantitative parameter to characterize the heat transfer, is calculated as follows. The local instantaneous Nusselt number (${Nu_l}$) is calculated as the dimensionless temperature gradient. Then, the time averaged local Nusselt number ($\overline{Nu_l}$) is calculated by averaging the instantaneous local Nusselt number over a time period (after reaching dynamically steady state) as
\begin{equation}
{\overline {Nu} _l} = \frac{1}{T}\int\limits_0^T {N{u_l}dt}
\end{equation}
Further, $\overline{Nu}$ is calculated by spatially averaging the time-averaged local Nusselt number ($\overline{Nu_l}$) as
\begin{equation}
\overline {Nu}  = \frac{1}{L}\int\limits_0^L {\overline {N{u_l}} dl}
\end{equation}
\section{Numerical details}
Computation has been performed over a hybrid unstructured grid using an in-house developed solver, named AnuPravaha \cite{anu}, in a collocated framework with pressure velocity coupling through Rhie and Chow \cite{rhiechow} momentum interpolation following the SIMPLE-like algorithm of Dalal et al. \cite{Dalal}. Finite-volume method is implemented in an unsteady RANS-based solver with time step $\Delta t=10^{-3}$. Isotropic eddy-viscosity based two equation model, namely standard $\kappa-\epsilon$ model (with Launder and Spalding \cite{Launder} wall function approach) is used to provide a measure for turbulence closure. Computation of convective flux at the faces is carried out using a $CDS$-upwind hybrid scheme following the deferred correction approach of Khosla and Rubin \cite{khosla}, where as the diffusive flux is computed through green gauss reconstruction. In the current work, we make use of ILU preconditioner with BiCG solver for velocity and pressure equations and GMRES solver for $\kappa$ and $\epsilon$ equations using Krylov subspace methods. Open source LIS \cite{lis} library is used to implement these iterative methods.

\subsection{Geometrical configuration and grid generation}
\begin{figure}[htbp]
  \ \hfill \includegraphics[scale=0.4]{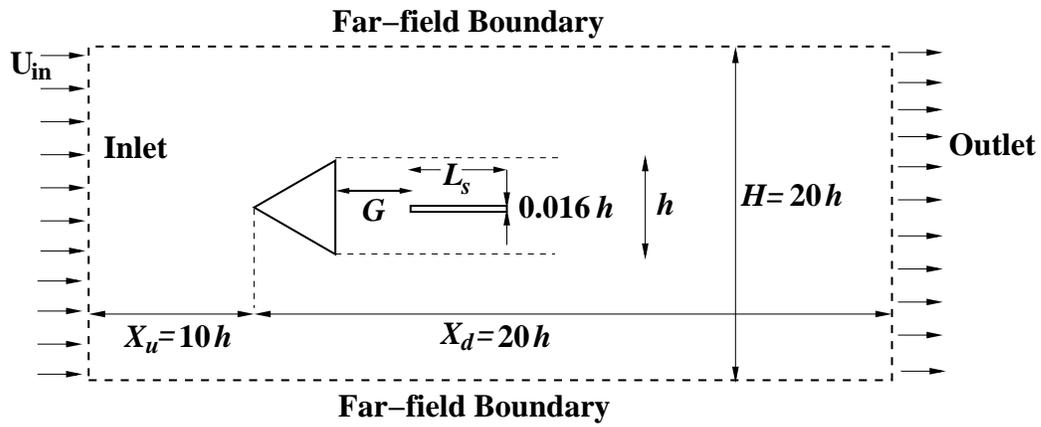} \hfill \
  \caption{Schematic of the computational domain showing unconfined flow past an equilateral triangular cylinder with a splitter plate.}
  \label{splittergeom}
\end{figure}

\begin{figure}[htbp]
  \ \hfill \includegraphics[scale=0.4]{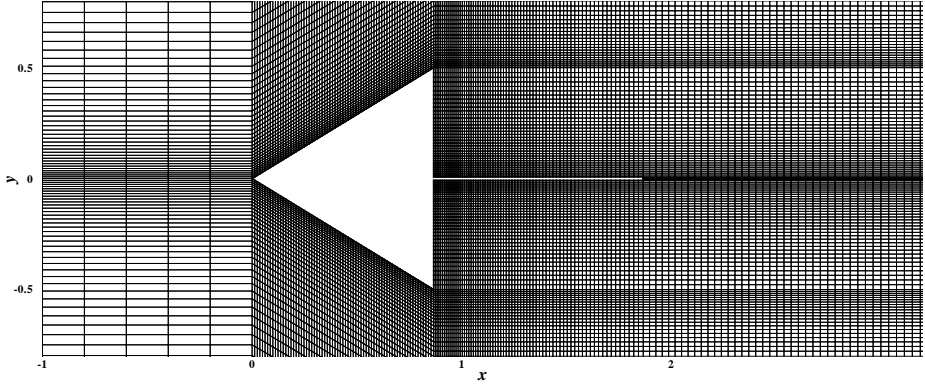} \hfill \
  \caption{Zoomed view of computational domain (a typical grid consisting of approx. 90,000 mesh elements). }
  \label{splittermesh}
\end{figure}

In this study, two-dimensional incompressible flow around an equilateral triangular cylinder with a splitter plate in an unconfined regime, as shown in Fig. \ref{splittergeom}, has been investigated. The side length of the triangular cylinder has been considered as the characteristic dimension and has been used as a basis for calculating Strouhal number and Reynolds number. The total length and height of the computational domain is chosen as $L=30h$ in the stream-wise direction and $H=20h$ in the spanwise direction. The triangular cylinder is symmetrically placed across the domain, at an upstream distance $X_{u}$=10$h$ from the inlet and downstream distance $X_{d}$=20$h$ from the outlet. The blockage ratio maintained here is $\beta$=0.05. (Data related to geometry and boundary conditions are taken from Bosch and Rodi \cite{Rodi}). Reynolds number chosen for the current study is 22,000 \cite{Rodi}. The grid structure for the aforementioned configuration is shown in Fig. \ref{splittermesh}. The total number of node points placed along the edge of the cylinder is taken to be 70. The length of the splitter plate has equal number of points along the side of the plate and the thickness of the splitter plate has been chosen to be 0.016$h$ following Akilli \cite{Akilli}.  

\subsection{Code validation}
\begin{figure}[htbp]
  \ \hfill \includegraphics[scale=0.5]{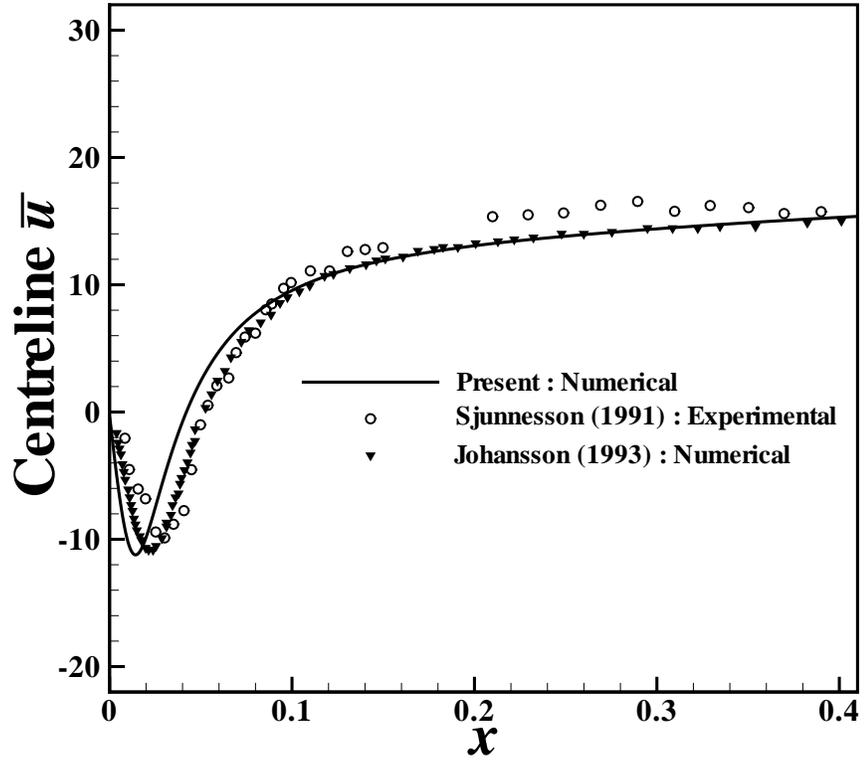} \hfill  \
  \caption{ Time-averaged stream-wise velocity ($\overline{u}$) along axial direction.}
  \label{johansson_uavg}
\end{figure}

The numerical results obtained in the present study has been validated with the experimental findings of Sjunnesson et al. \cite{Sjunnesson} on turbulent flow past a triangular cylinder in a channel for $Re=45,000$ using standard $\kappa-\epsilon$ model. Solution is found on a 2D unstructured hybrid mesh of 81,000 mesh elements. 
The time-averaged stream-wise velocity ($\overline{u}$) is plotted along geometric centreline starting from the rear end of the cylinder. It can be comprehended from the Fig. \ref{johansson_uavg} that present result has provided an accurate prediction of numerical results of Johansson \cite{Johansson} 
and it compares well with the experimental results too. 
The Strouhal number obtained from our present computation is also compared with the already established results. The Strouhal number data incurred from the experiment and from Johansson \cite{Johansson} is found to be 0.25 and 0.27 respectively, where as the results of present computation show a percentage variation of 8\% w.r.t. the experimental data, with the value being 0.267, which seems a fairly good match. To establish the validity of the solver, another validation study was also performed on flow past a free-standing square cylinder at $Re$ = 22000 using standard $\kappa - \epsilon$ model for validating results with the previously mentioned paper of Bosch and Rodi \cite{Rodi} and the comparison among the numerical and experimental results are reproduced in Table 1. The quantitative data comparison shows that the non-dimensional values ($\overline{C_{d}}$ and $St$) calculated numerically are well-matched with the reported results of Bosch and Rodi \cite{Rodi} with an agreeable percentage variation of 2.6\% and 17\% with respect to the experimental values.     

\subsection{Grid independence result}
Three different meshes with increasing number of elements 65,000, 90,000 and 1,15,000 were used to obtain a guideline on the minimum number of volumes required for a grid independent simulation. The total number of grid points distributed along the side of the cylinder are 60, 70 and 80 at these three grid levels. Here the analysis has been limited to plate length $L_{s}/h$=1 and gap ratio $G/h$=0 at $Re=22,000$ with $X_{u}/h=10$ and $X_{d}/h=20$. While performing grid refinement the minimum and maximum edge length has been maintained at $\delta$=0.008 and $\Delta$=0.5, respectively. The results for grid refinement has been summarized in Table 2. 
While the deviation in time-averaged co-efficients i.e. $C_{d}$ and $C_{d_{p}}$ on the coarse and intermediate grids was found to be 0.7\%, it was 0.3\% for $C_{Lrms}$, 0.25\% for $Nu$ and ~2\% for $St$. But the deviation deminished significantly while moving on to the finest mesh level. The effects of grid refinement seen above led us to use approximately 90,000 mesh elements in the subsequent computations. In the present study, the first grid point spacing (i.e. the first cell centre to wall distance) is so maintained that the $y^{+}$ is retained closer to 11.63 in the present computations.            

\section{Results and discussion}
In the current investigation, a detailed parametric study has been carried out to determine the effect of gap ratio ($0\leq G/h\leq 2$) and length of splitter plate ($0\leq L_{s}/h \leq 1.5$) on fluid flow past a triangular cylinder. In this regard, simulations are performed to understand the variation in flow field and heat transfer with different parametric settings.  

\begin{figure}[htbp]
\centering
\begin{tabular}{cc}
\ \hfill \includegraphics[scale=0.20]{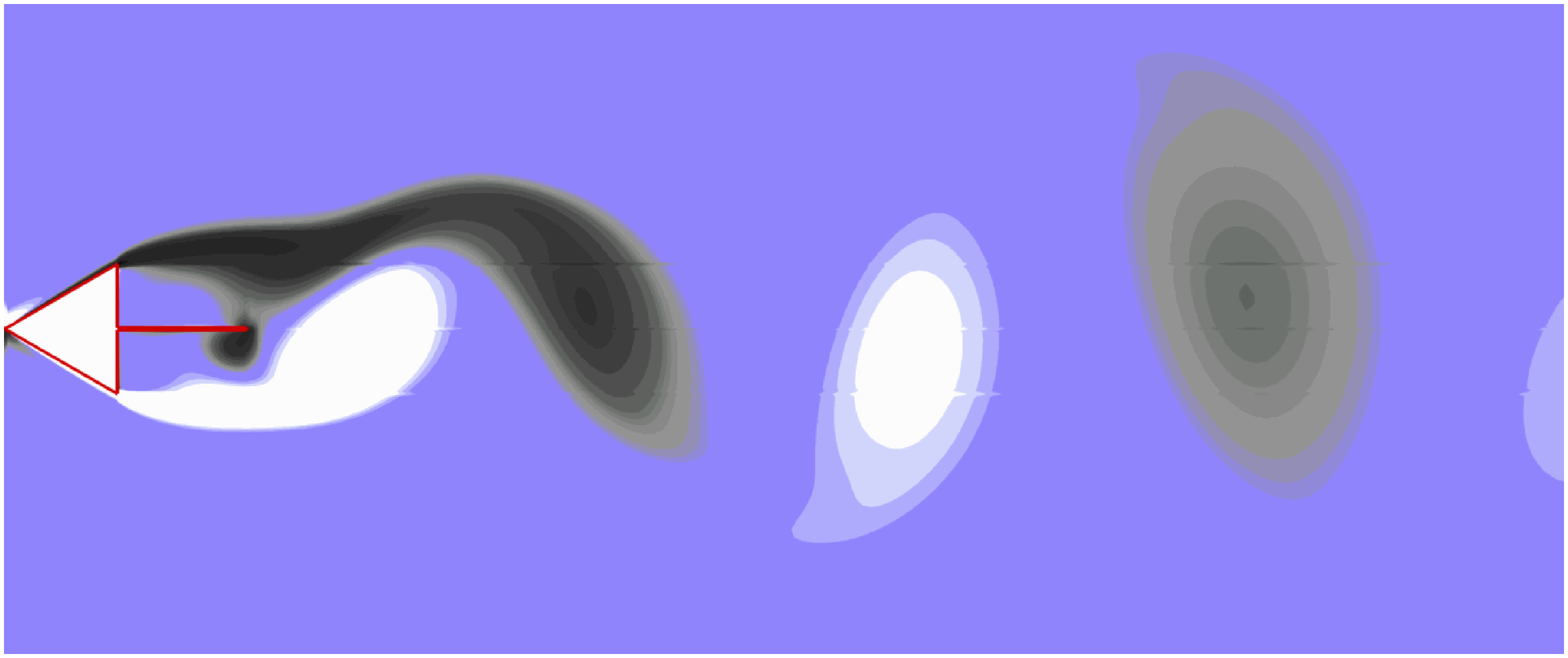}\hfill\
\ \hfill \includegraphics[scale=0.20]{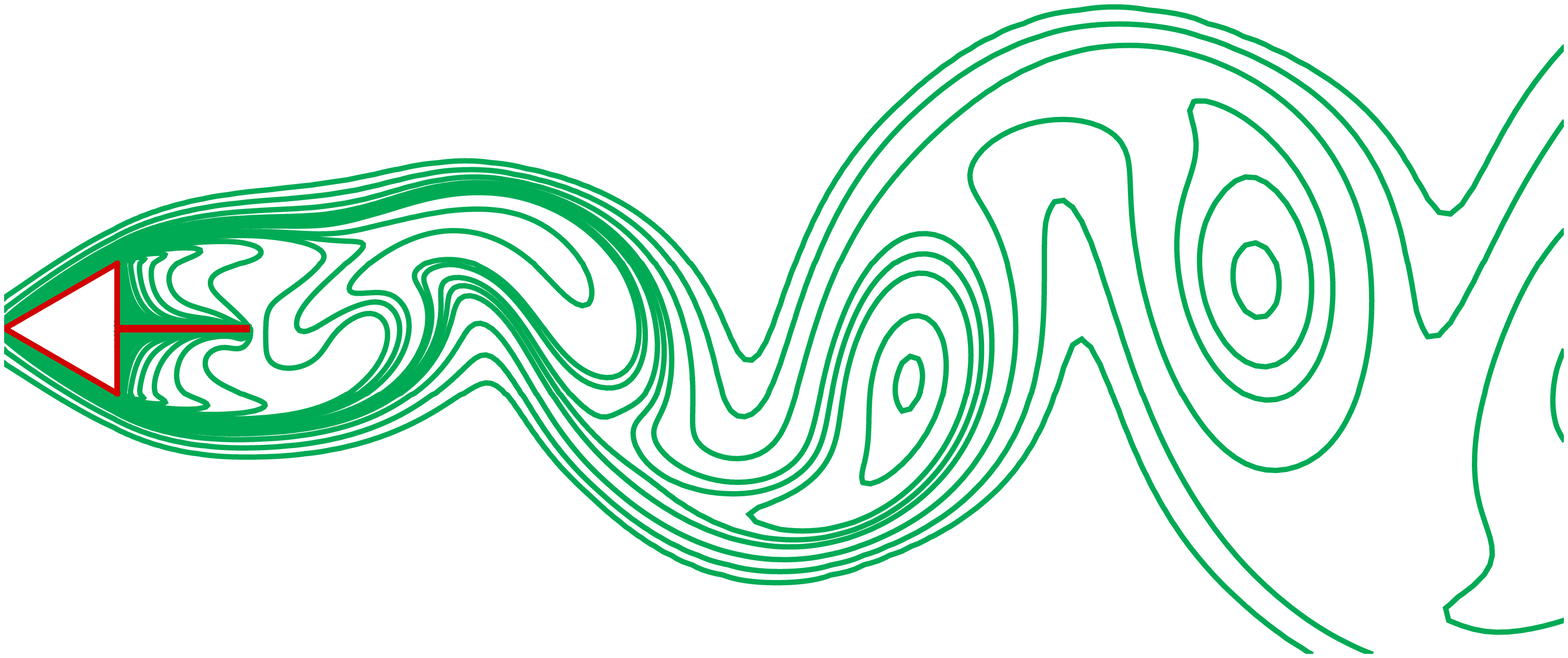} \hfill\
\end{tabular}
\begin{tabular}{cc}
\ \hfill \includegraphics[scale=0.20]{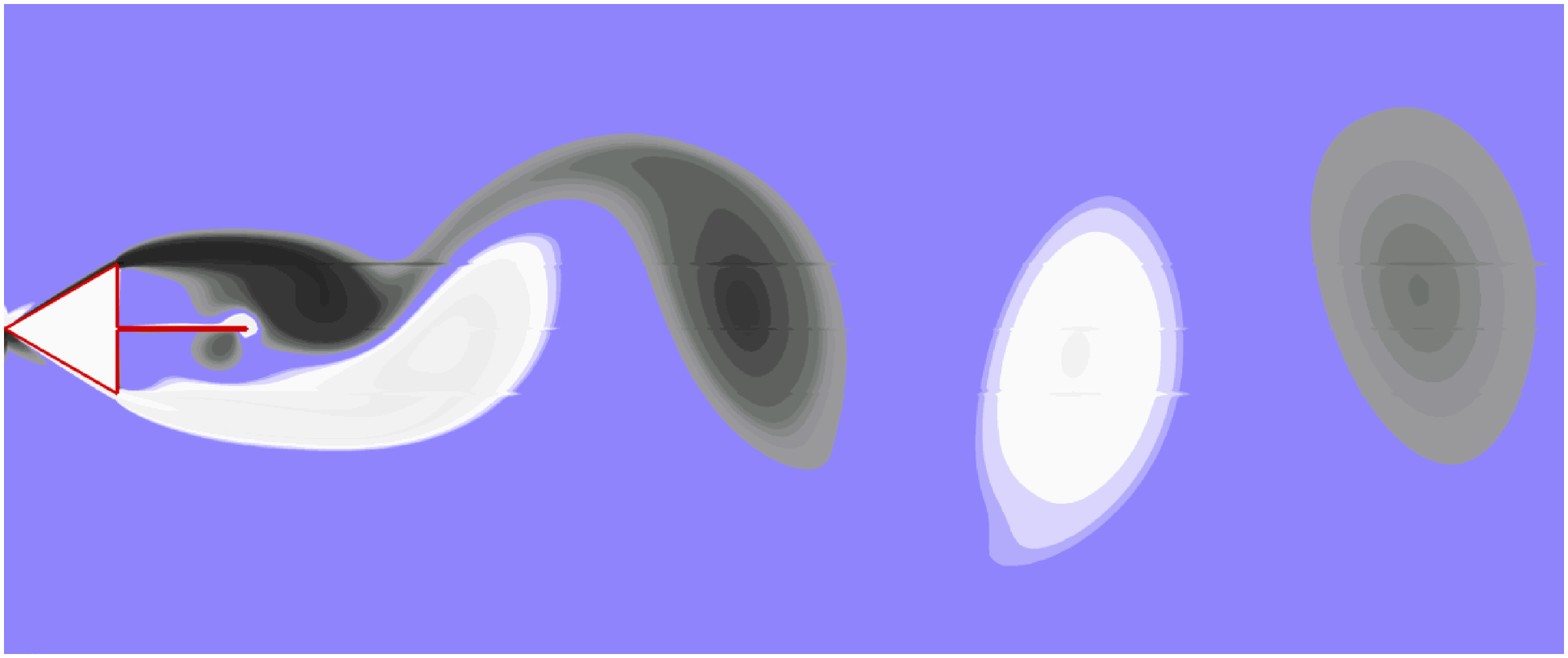}\hfill\
\ \hfill \includegraphics[scale=0.20]{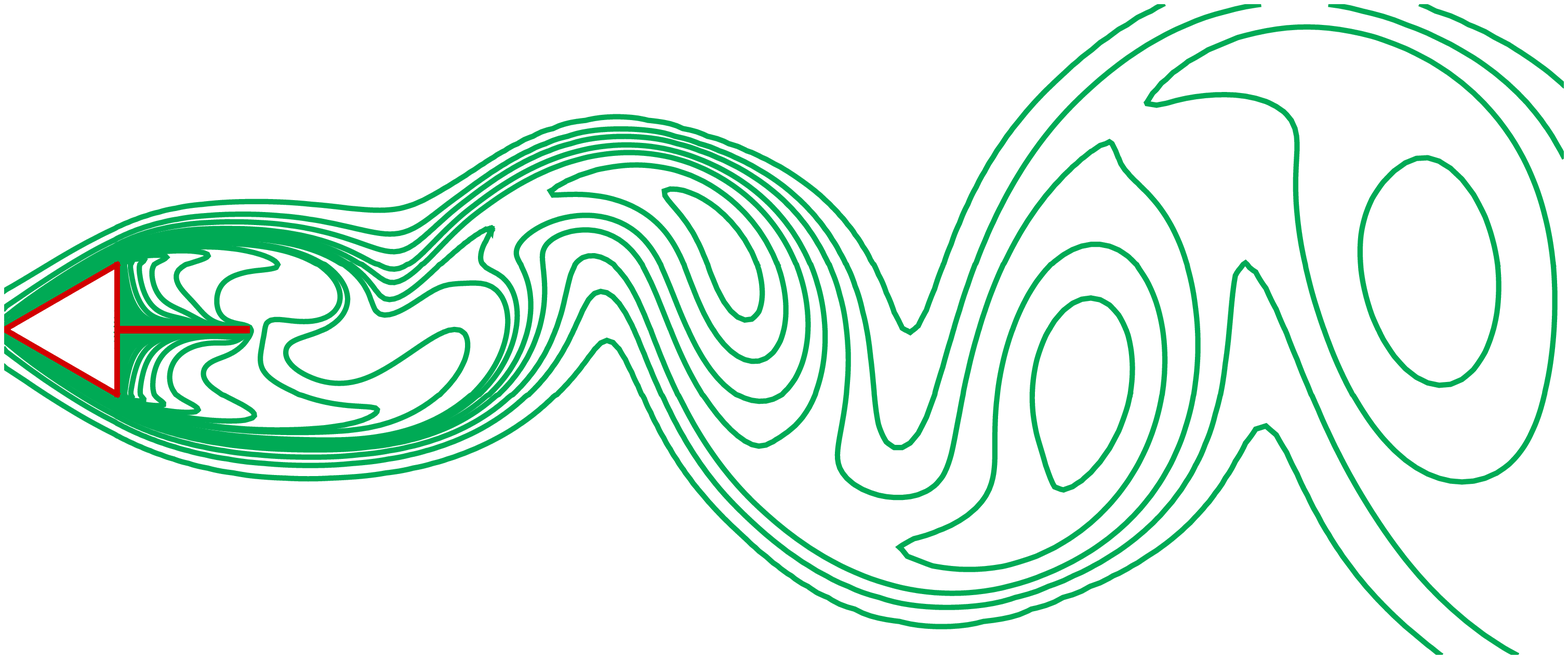} \hfill\
\end{tabular}
\begin{tabular}{cc}
\ \hfill \includegraphics[scale=0.20]{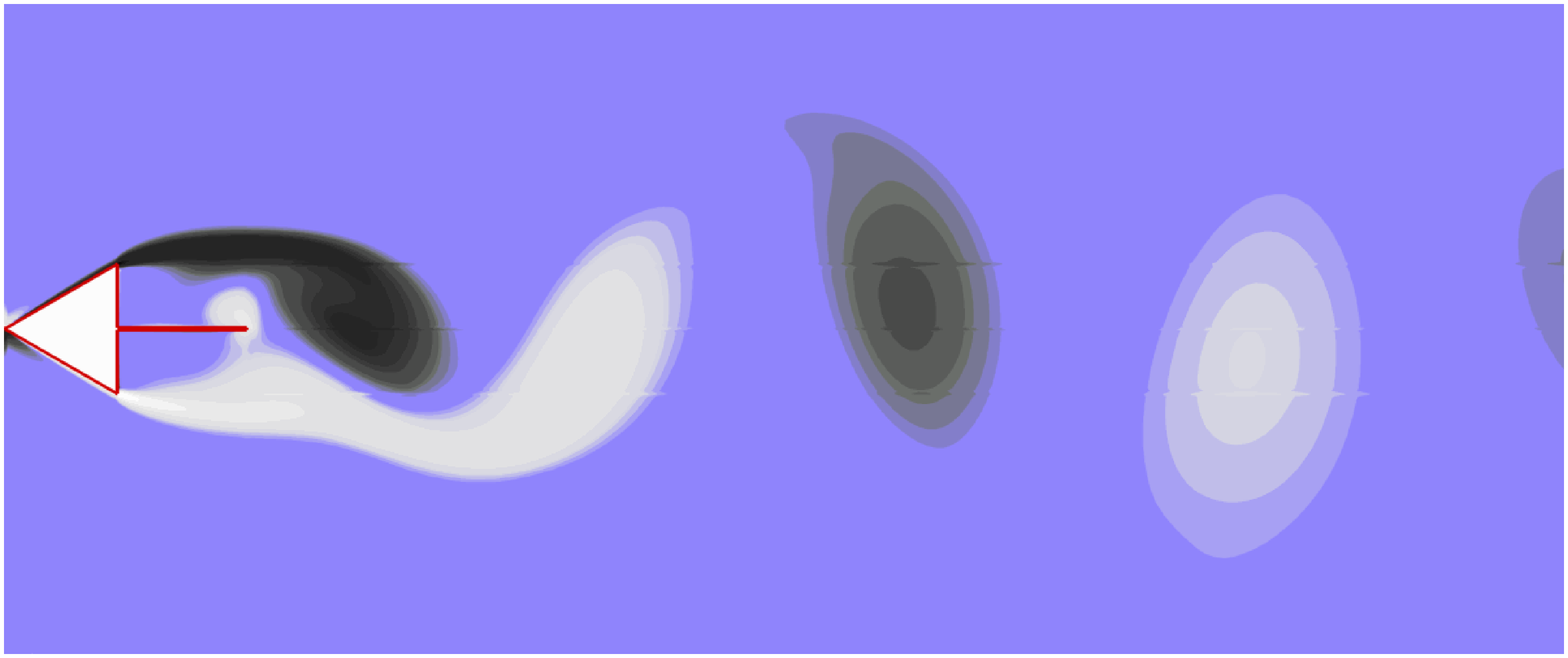}\hfill\
\ \hfill \includegraphics[scale=0.20]{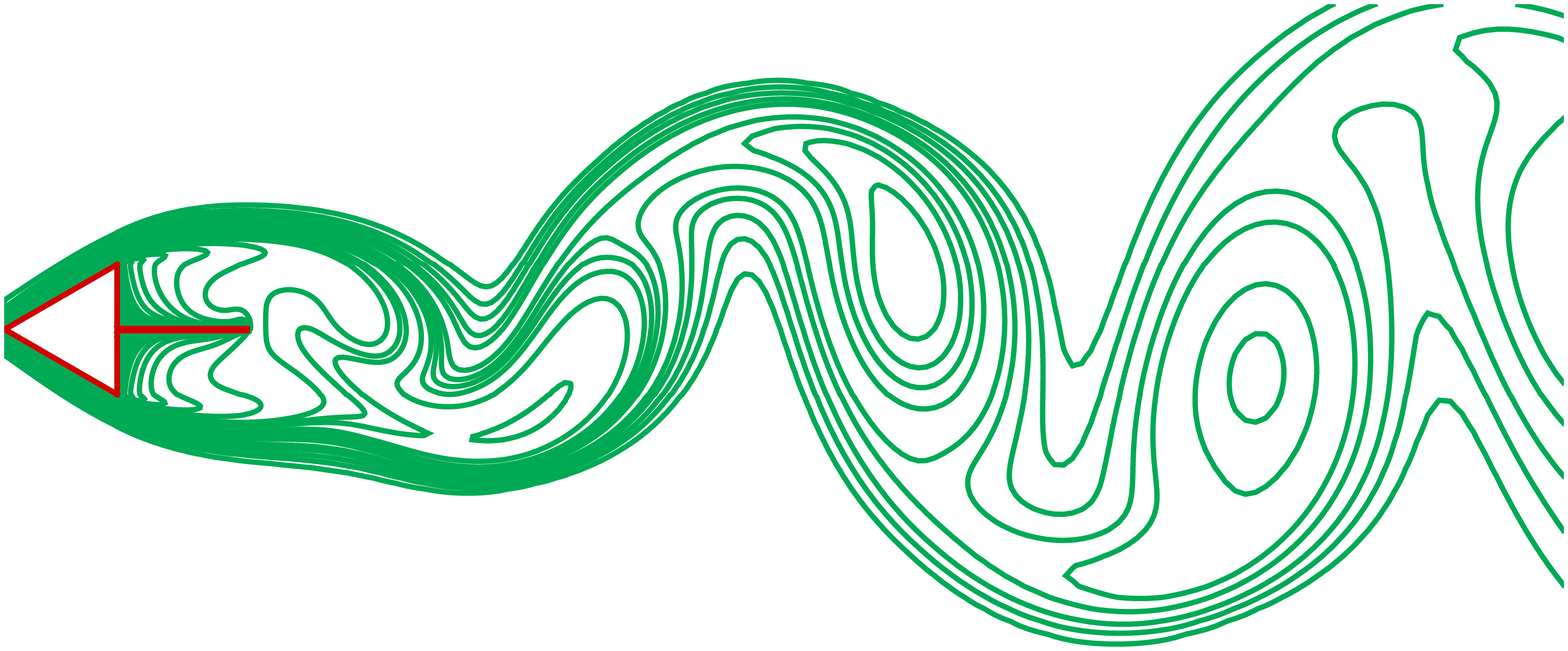} \hfill\
\end{tabular}
\begin{tabular}{cc}
\ \hfill \includegraphics[scale=0.20]{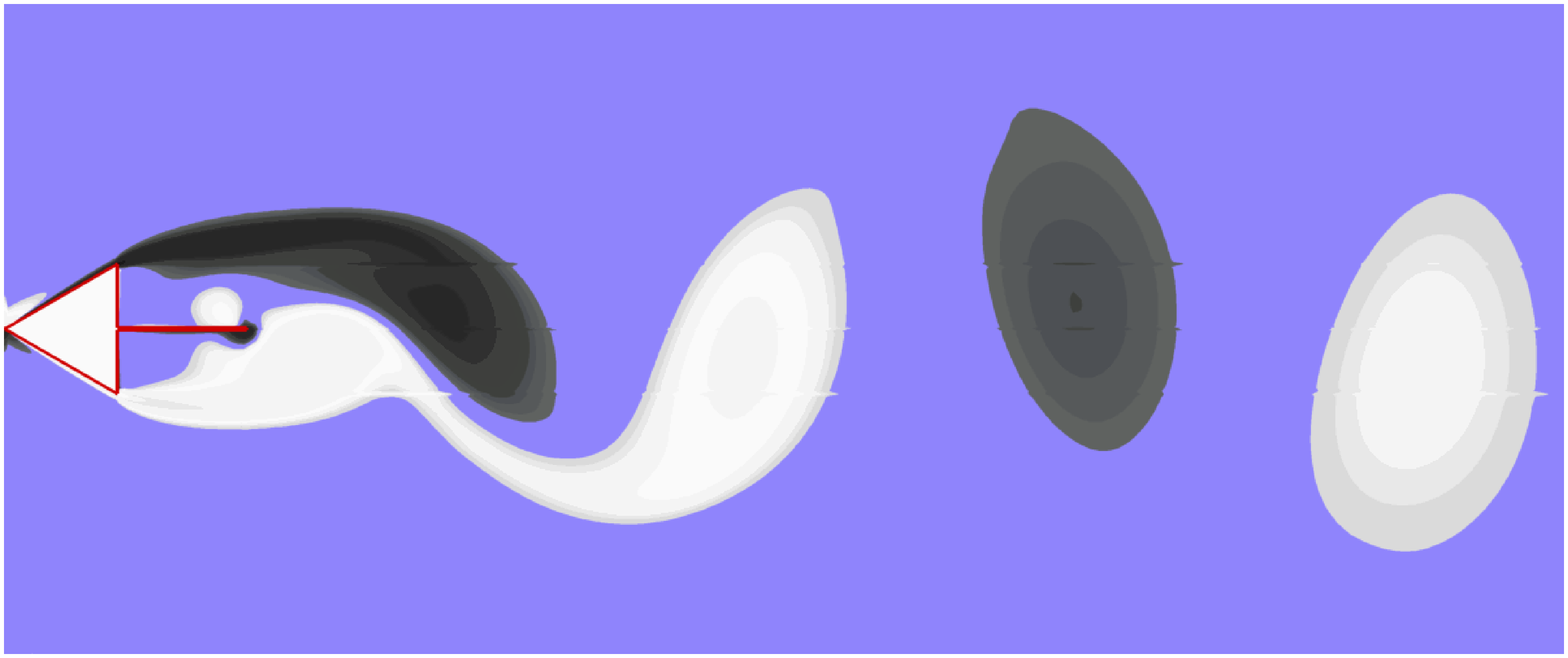}\hfill\
\ \hfill \includegraphics[scale=0.20]{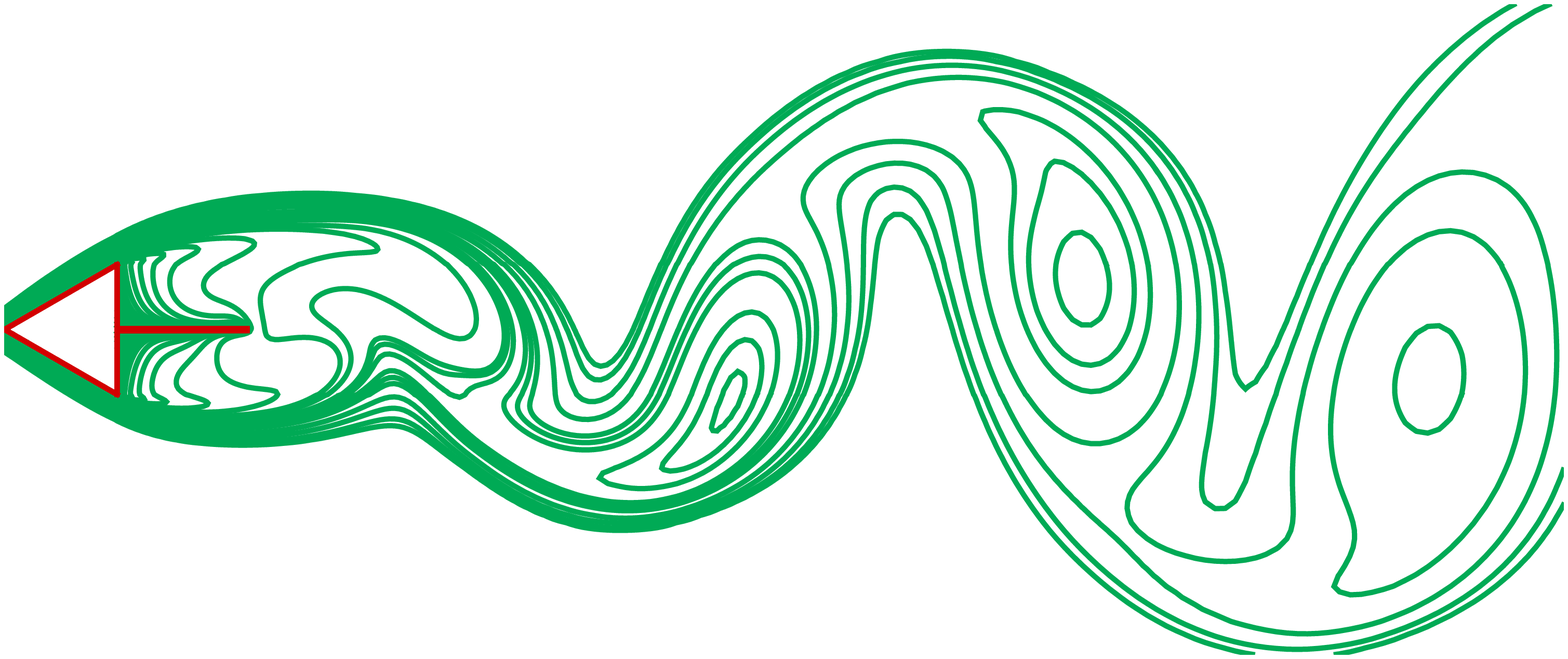} \hfill\
\end{tabular}
\begin{tabular}{cc}
\ \hfill \includegraphics[scale=0.20]{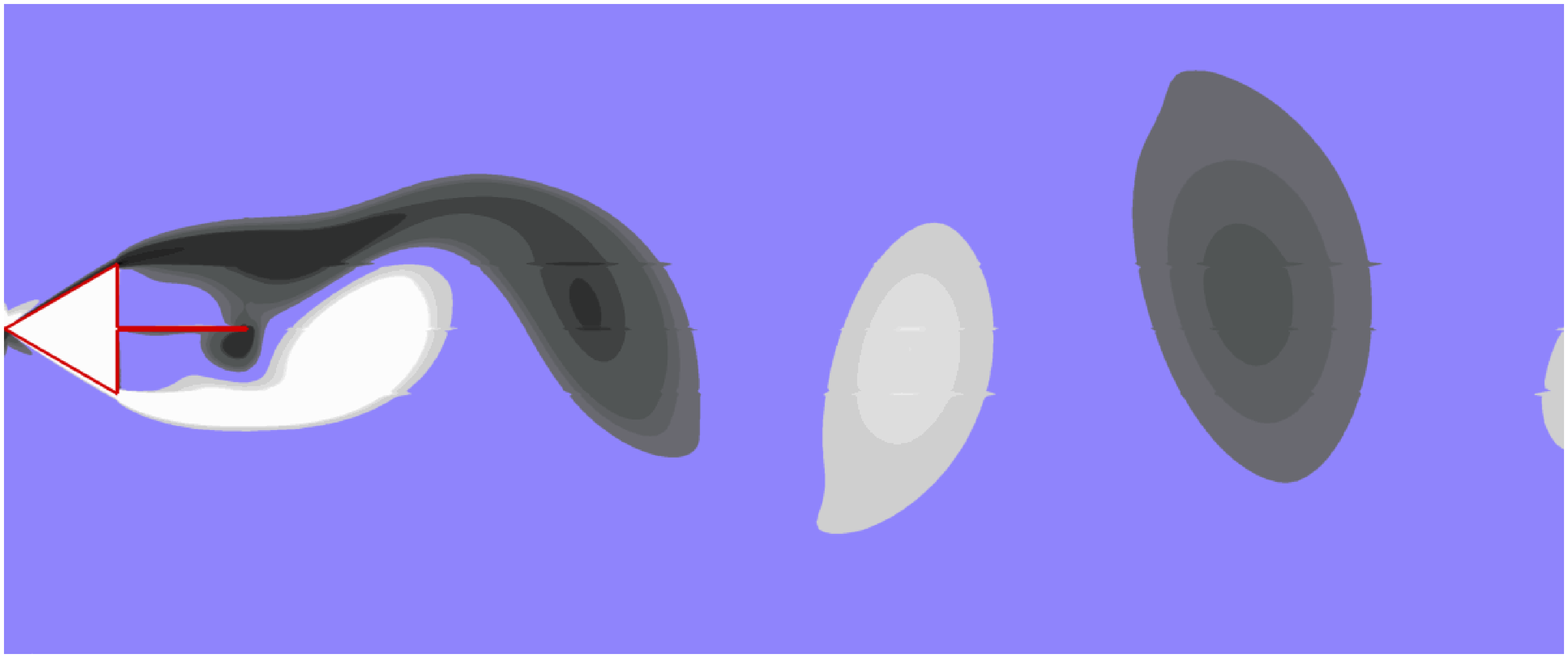}\hfill\
\ \hfill \includegraphics[scale=0.20]{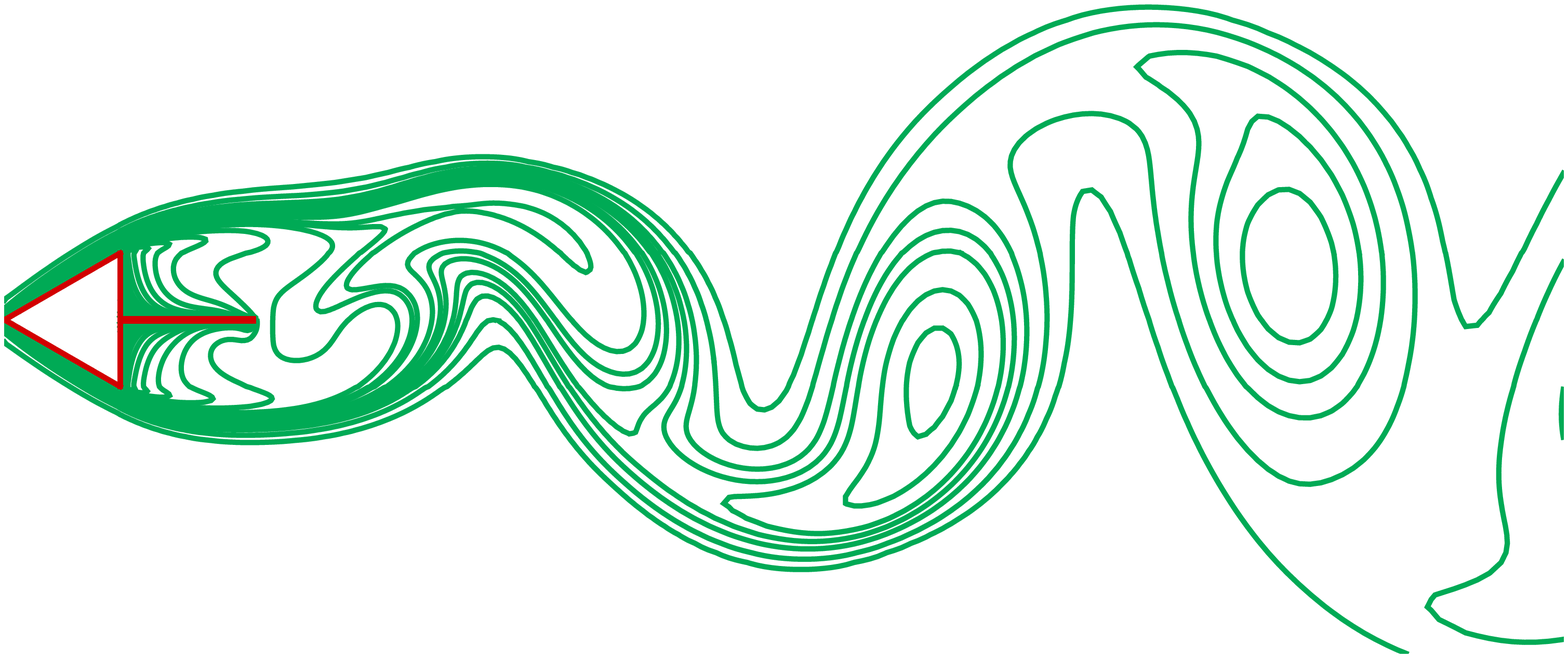} \hfill\
\end{tabular}
$~~~~~~~~~~~~~~~~$(a)$~~~~~~~~~~~~~~~~~~~~~~~~~~~~~~~~~~~~~~~~~~$(b)$~~~~~~~~~~~~~~~$
\caption{Equispaced $\tau$/4 (where, $\tau$=time period of lift-coefficient) snapshots of phase-averaged contours showing periodicity within a shedding cycle, (a) phase-averaged vorticity contours ($\omega_{min}=-60s^{-1},\omega_{max}=50s^{-1}$), and (b) corresponding phase-averaged isotherms.}
\label{fig:all8eps}
\end{figure}

The contour plot showing periodicity in Fig. \ref{fig:all8eps} suggests that boundary layers are separated from both the trailing edges of the cylinder ascribed to the presence of adverse pressure gradient and the two separated shear layers interact with each other in the flow downstream to form the well known Von-K$\acute{a}$rm$\acute{a}$n vortex street. The series of figure provided feature the details of the development of vortex and vortex shedding process.\\
\subsection{Effect of change in $G/h$ on force coefficients and Strouhal number ($St$):}
\begin{figure}[htbp]
\centering
\begin{tabular}{cc}
\ \ \centering \hfill \psfig{figure=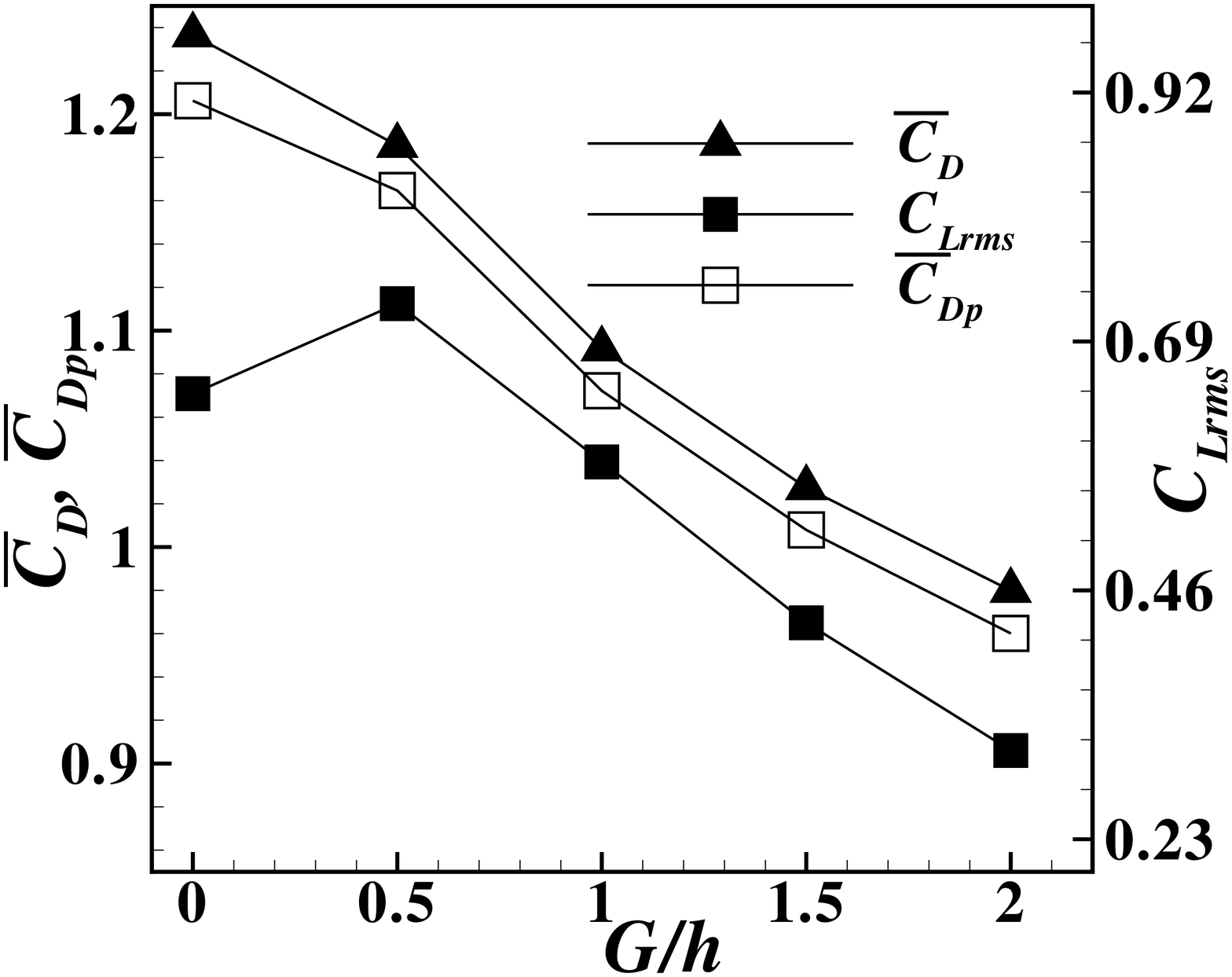,width=2.6in,height=2.0in} \hfill \ \
\ \ \centering \hfill \psfig{figure=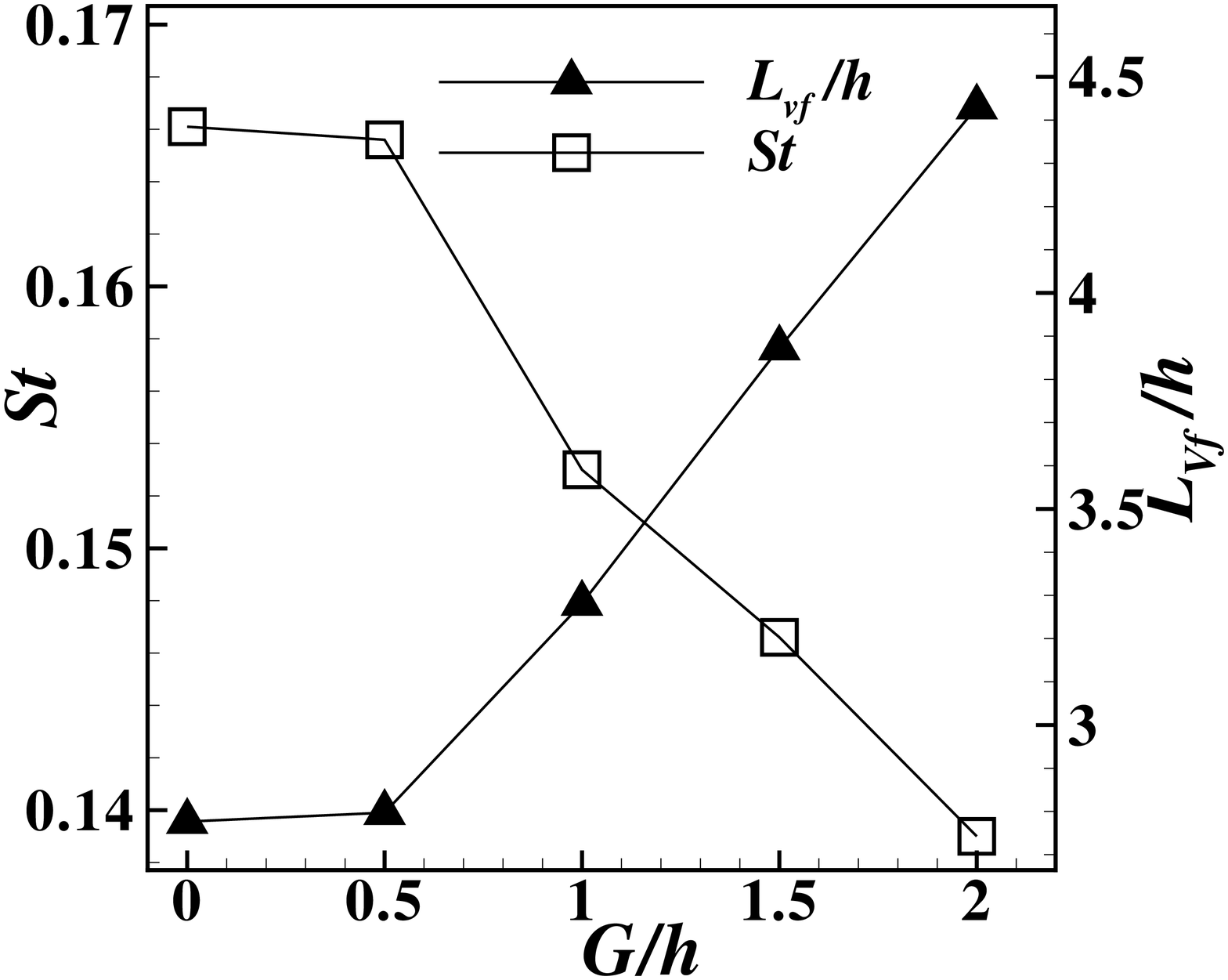,width=2.6in,height=2.0in} \hfill \ \
\end{tabular}
$~~~~~~~~~~~~~~~~~~~~~~~$(a)$~~~~~~~~~~~~~~~~~~~~~~~~~~~~~~~~~~~~~~~~~~~~~~~~~~~~~~$(b)$~~~~~~~~~~~~~~~$
\caption{Variation of (a) time-averaged integral parameters: $\overline{C_{d}}$, $\overline{C_{d_{p}}}$ \& $C_{Lrms}$; and (b) $St$ \& $L_{Vf}$ for 0$\leq G/h \leq$ 2.}  
\label{fig:integralparameters}
\end{figure}

Figure \ref{fig:integralparameters}(a) illustrates the variation of various integral parameters, namely $\overline{C_{d}}, \overline{C_{d_{p}}}$ and $C_{Lrms}$ with respect to change in gap-ratio. The time-averaged drag coefficient, $\overline{C_{d}}= 2F/(\rho U_{in}^2h)$, decreases as gap ratio increases in Fig. \ref{fig:integralparameters}(a), where $F$ refers to the net drag force per unit width on the triangular cylinder-splitter plate ($TCSP$) system. Similar trend is observed for time-averaged pressure drag coefficient too while increasing the $G/h$ from 0 to 2. However, the change in $C_{Lrms}$ with respect to gap-ratio follows a different trend altogether. Initially it increases till it attains a peak at $G/h$=0.5 and then decreases further. \\
Figure \ref{fig:integralparameters}(b) depicts that, as gap-ratio ($G/h$) increases, Strouhal number decreases continuously and attains a minimum value of 0.139 for $G/h$=2. Simulations show that $St$ has peaked to its maximum of 0.171 for $G/h$=0, the possible mechanism for explaining the flow scenario could be the placement of the plate is not adequate enough to resist the interaction of shear layers. However, after introducing a gap between the cylinder and the splitter plate, it is observed that the flow behavior has changed remarkably. It is essential to note that the presence of gap between cylinder and splitter plate plays a major role in modifying the behavior of vortex shedding. Figure \ref{fig:integralparameters}(b) shows that with increase in gap-ratio ($G/h$) vortex formation length increases, which indicates the adherence of the recirculating bubble 
to the plate surface for a longer period of time with increase in gap-ratio. This delays the vortex shedding which leads to increase in vortex formation length and decrease in shedding frequency.  In the present study, vortex formation length denotes the distance from rear stagnation point to the point at which back flow stops or the wake stagnation point in the time averaged profile.
Figure \ref{fig:integralparameters}(b) reveals that the change in $St$ for gap ratio ranging from 0 to 0.5 is not very steep and lies well below 0.3\%, which means the change in shedding frequency is not that prominent. But, with further increase in gap-ratio from 0.5 to 1, a maximum decrease of 7.6\% in $St$ is observed. The change in shedding frequency continues to decrease at the same rate till it reaches $G/h$=2. In the present study, vortex formation length denotes the distance from rear stagnation point to the point at which back flow stops or the wake stagnation point in the time averaged profile.  \\
\subsection{Effect of change in $G/h$ on other flow characteristics:}
\begin{figure}[h!]
\centering
\begin{tabular} {c}
 \ \ \hfill \includegraphics[scale=0.4]{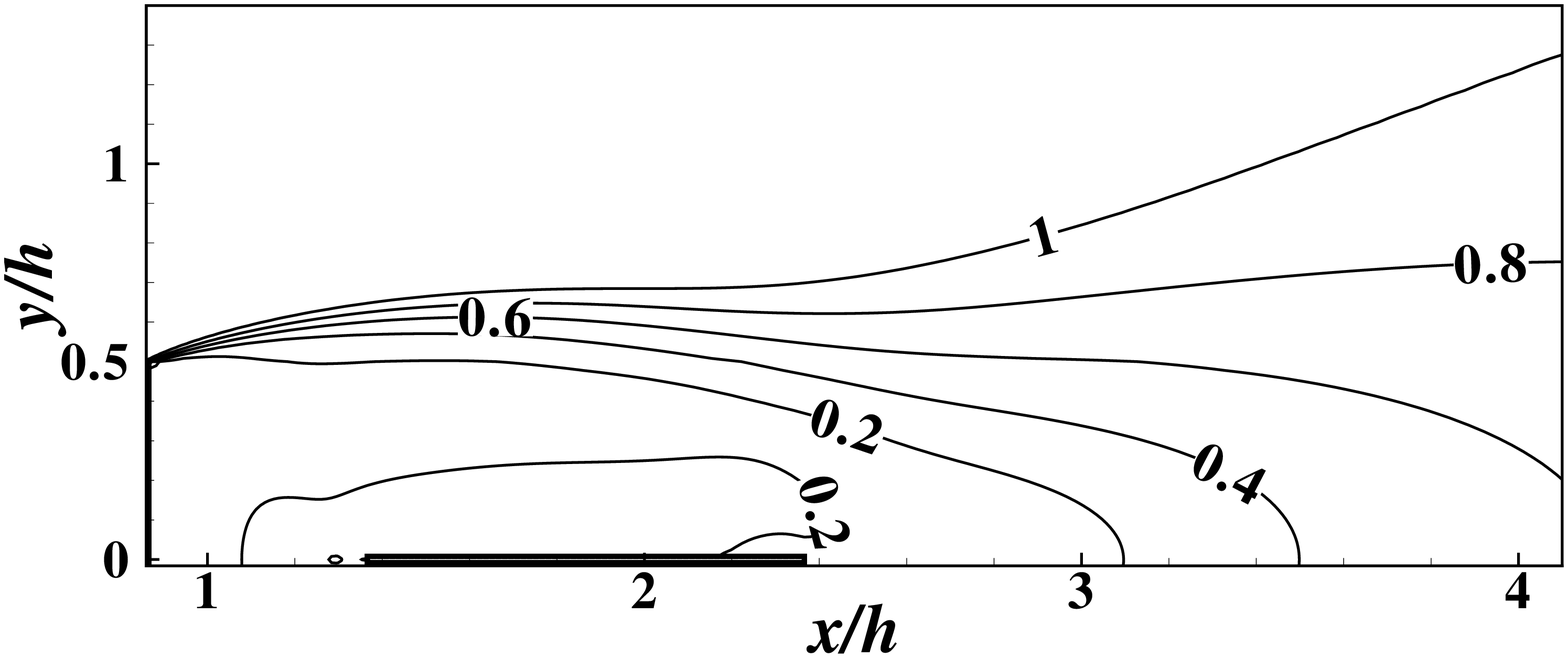} \hfill  \ \
\end{tabular}
$~~~~~~~~~~~~~~~~~~~~~~~~~~~~$(a)$~~~~~~~~~~~~~~~$
\begin{tabular} {c}
  \ \hfill \includegraphics[scale=0.4]{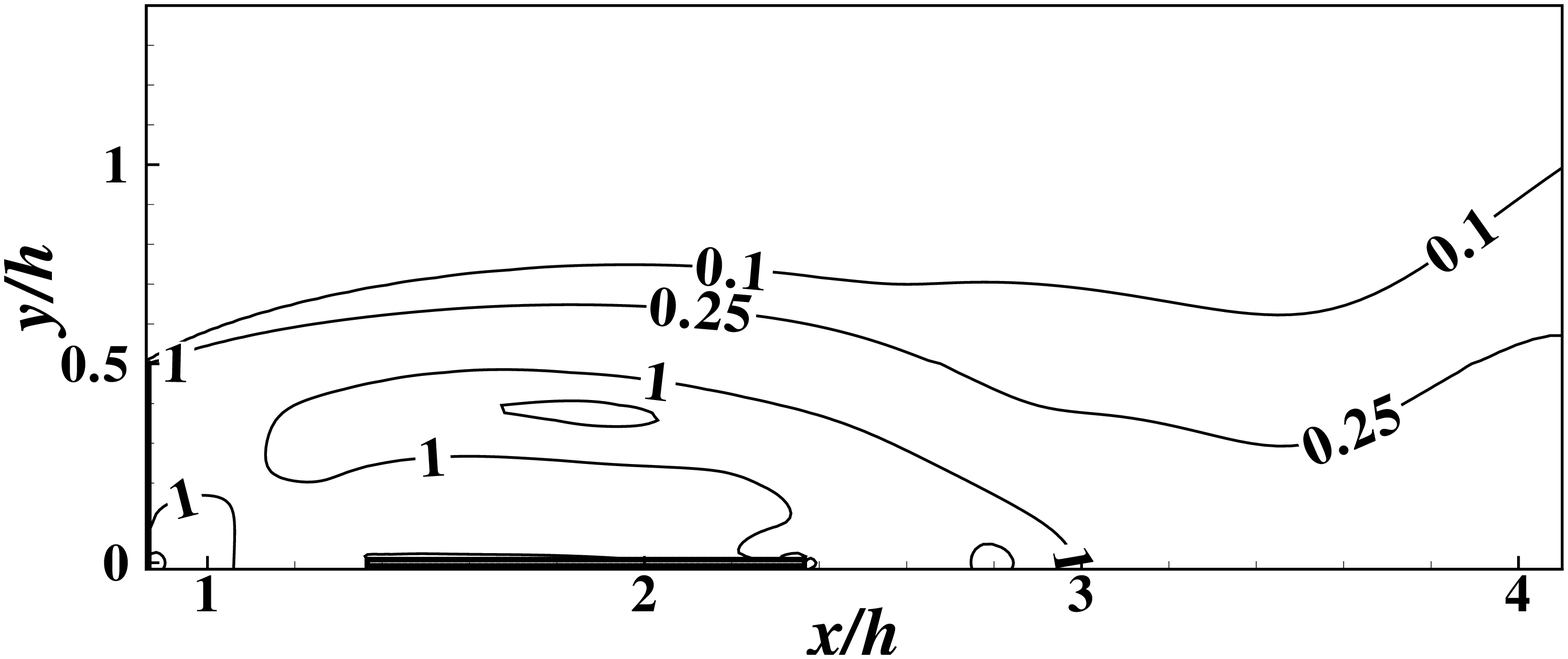} \hfill  \
\end{tabular}
$~~~~~~~~~~~~~~~~~~~~~~~~~~~~$(b)$~~~~~~~~~~~~~~~$
\caption{Zoomed view showing (a) typical mean velocity ($\overline{u}/U_{in}$) isocontours and (b) typical isocontours of turbulence intensity ($i$) for $L_{s}/h=1, G/h=0.5$. In the figure above, contours are originating from the trailing edge of the cylinder.}  
\label{fig:meanvel_intensity}
\end{figure}

\begin{figure}[htbp]
\centering
\begin{tabular} {c}
  \ \hfill \includegraphics[scale=0.4]{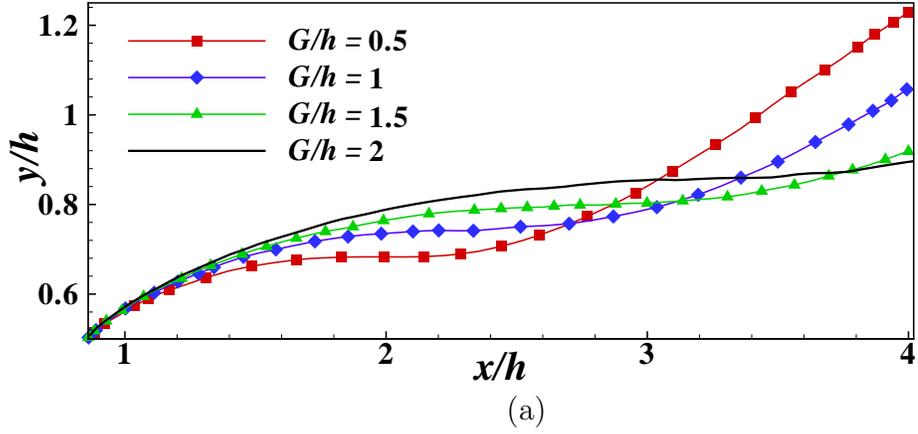} \hfill  \  
\end{tabular}
$~~~~~~~~~~~~~~~~~~~~~~~~~~~~$(a)$~~~~~~~~~~~~~~~$
\\ \vspace{0.2cm}
\begin{tabular} {c}
  \ \hfill \includegraphics[scale=0.4]{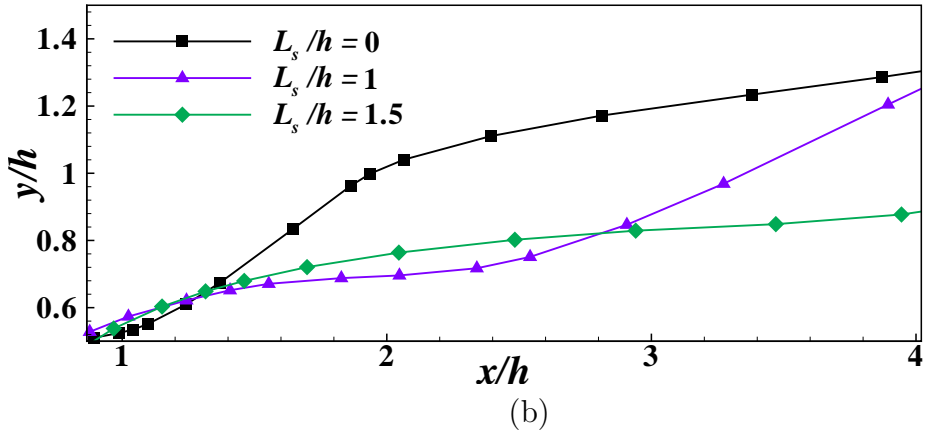} \hfill  \
\end{tabular}
$~~~~~~~~~~~~~~~~~~~~~~~~~~~~$(b)$~~~~~~~~~~~~~~~$
\caption{Wake envelope showing mean velocity ($\overline{u}/U_{in}=1$) profiles (a) for $L_{s}/h=1$ with different $G/h$; and (b) for $G/h=0$ with different $L_{s}/h$.}  
\label{fig:mean_velocity_with_gapratio}
\end{figure}

\begin{figure}[h!]
\centering
\begin{tabular}{ccc}
\ \hfill \includegraphics[scale=0.20]{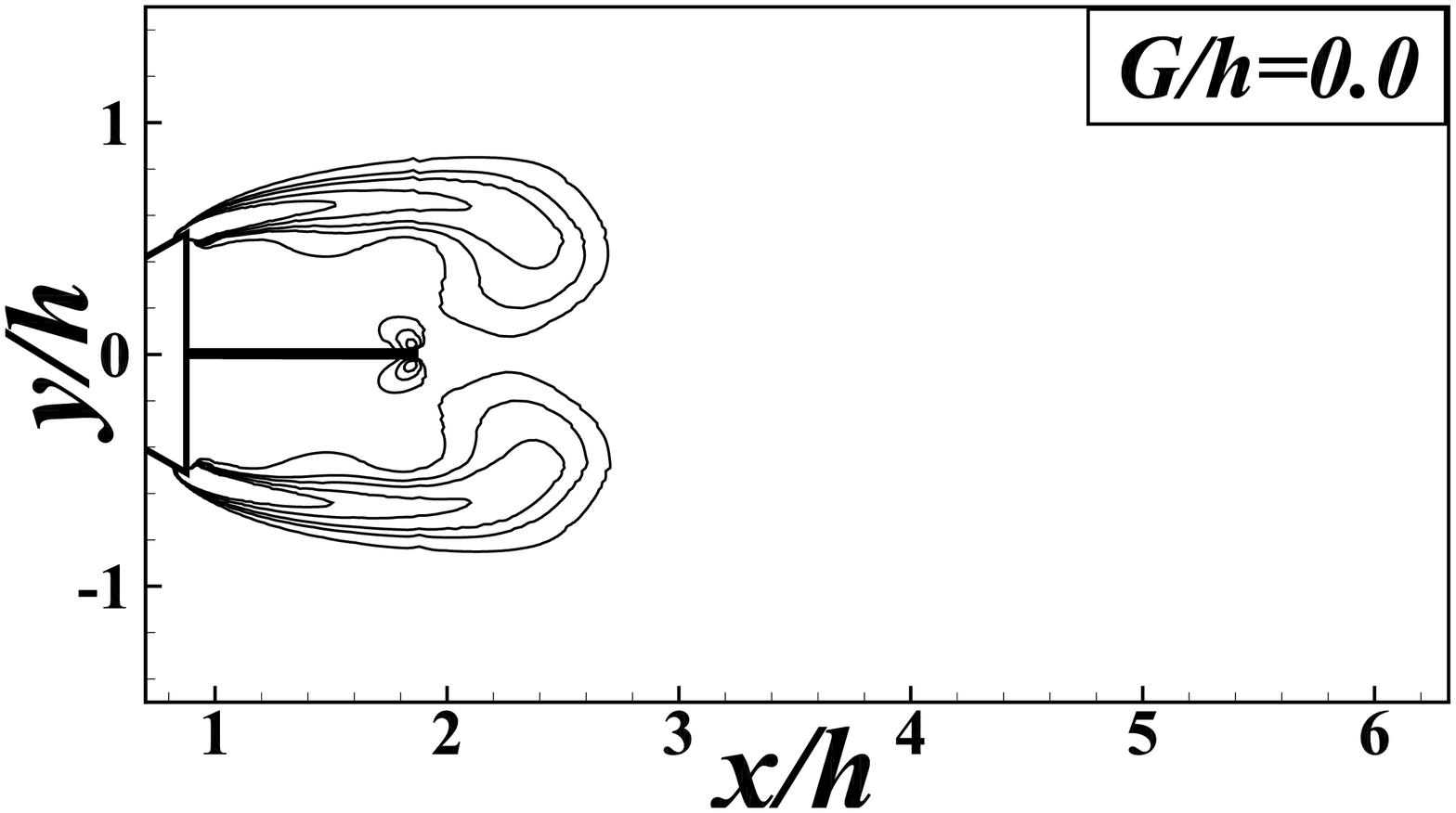} \hfill\
\ \hfill \includegraphics[scale=0.20]{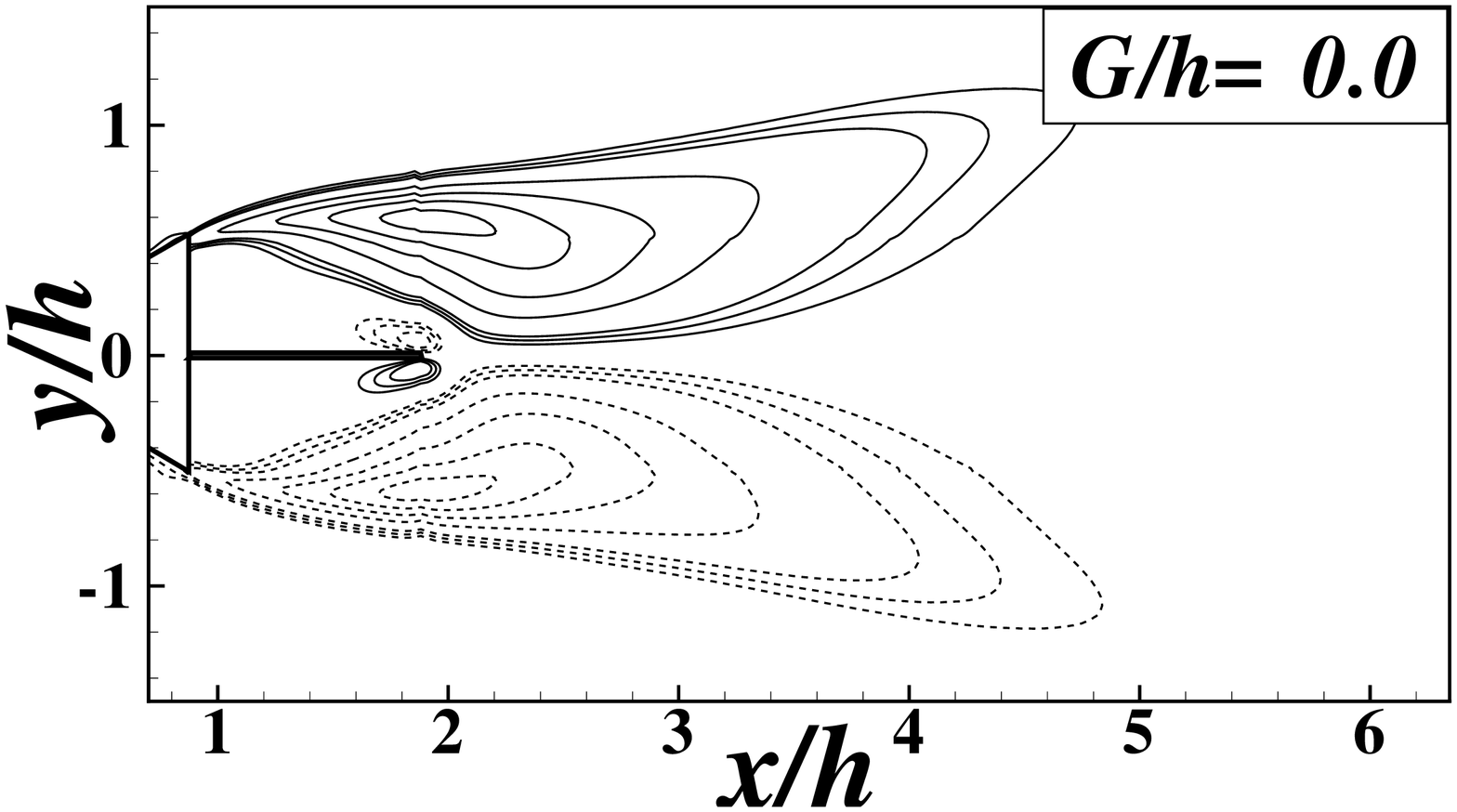}\hfill\
\ \hfill \includegraphics[scale=0.20]{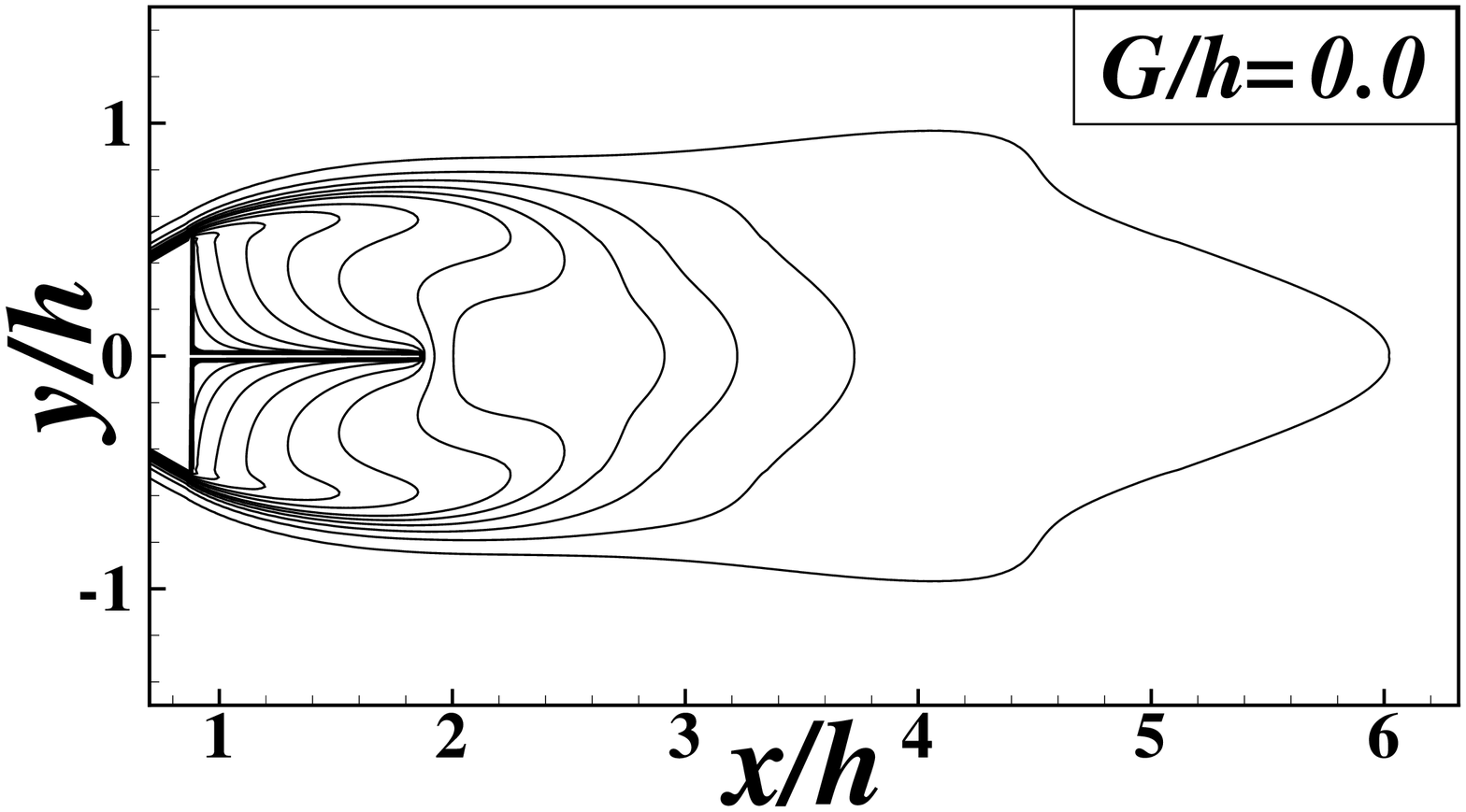} \hfill\
\end{tabular}
\begin{tabular}{ccc}
\ \hfill \includegraphics[scale=0.20]{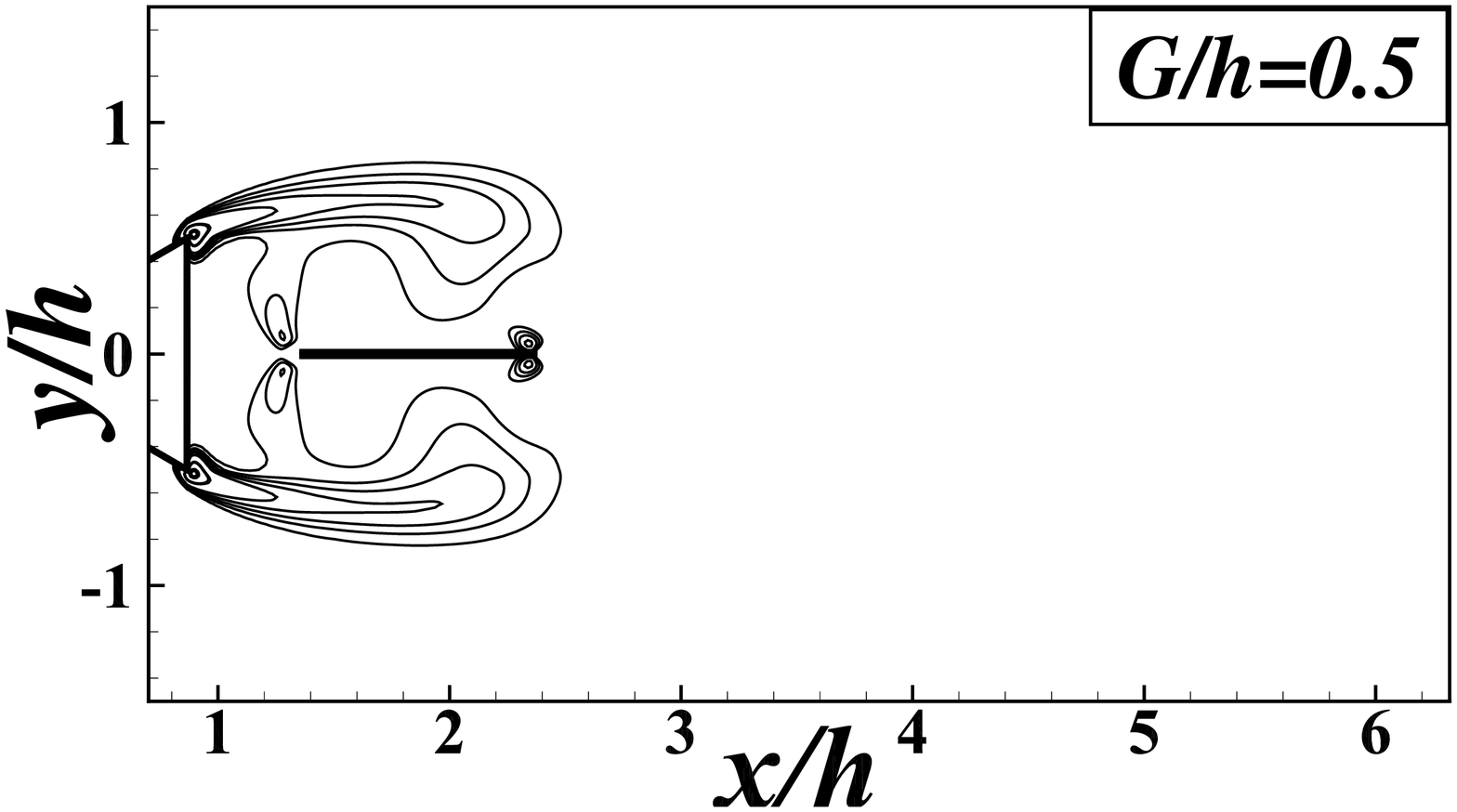} \hfill\
\ \hfill \includegraphics[scale=0.20]{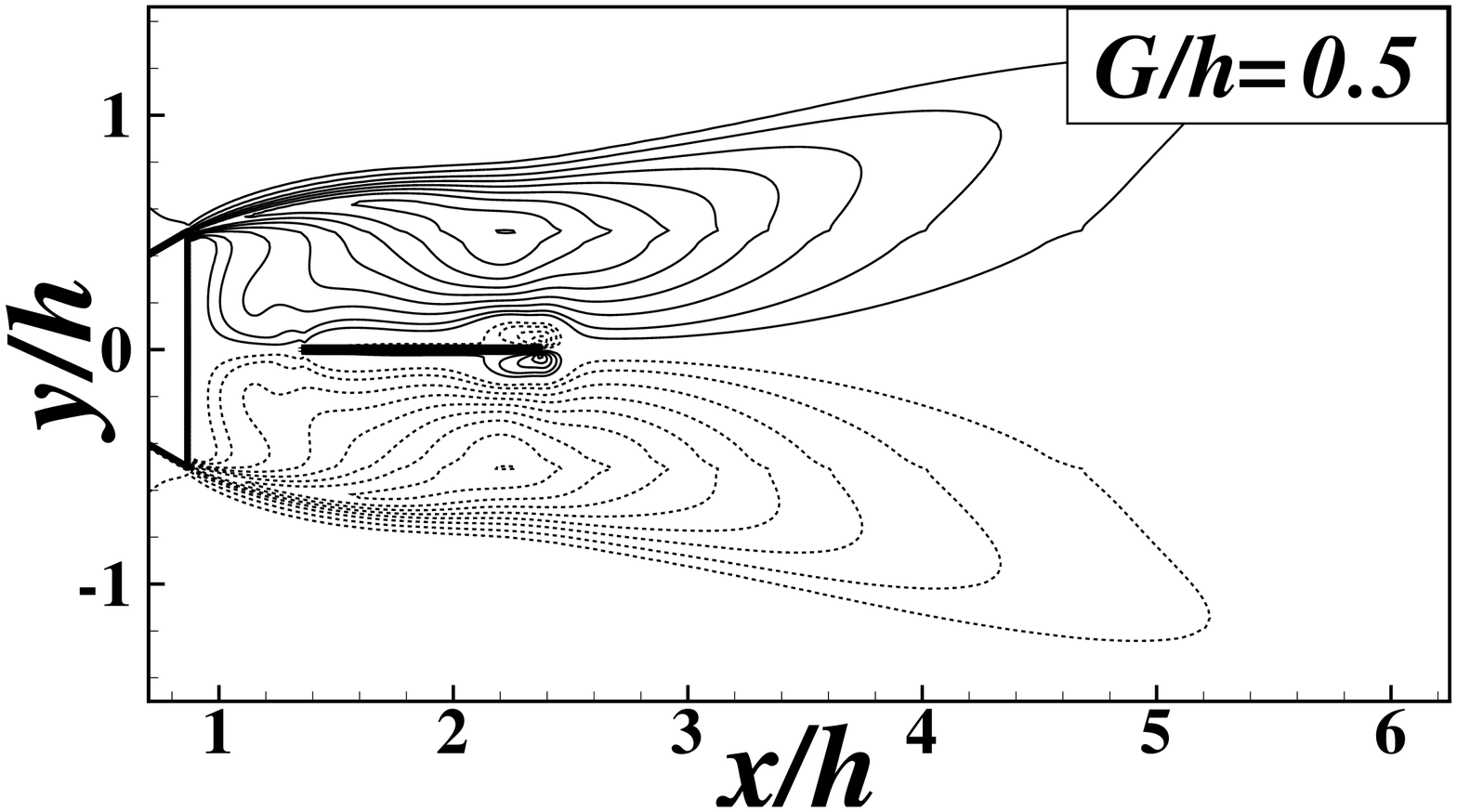}\hfill\
\ \hfill \includegraphics[scale=0.20]{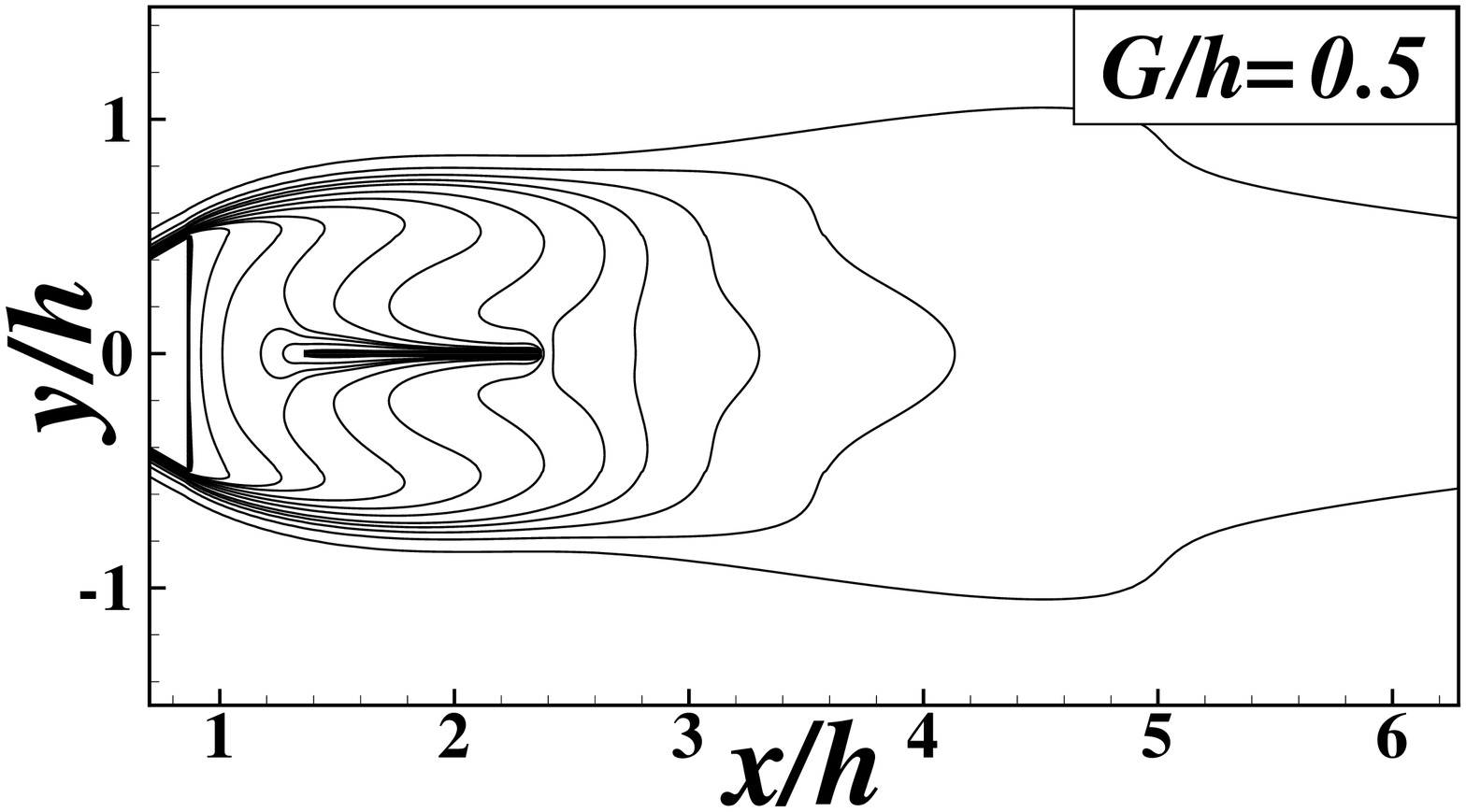} \hfill\
\end{tabular}
\begin{tabular}{ccc}
\ \hfill \includegraphics[scale=0.20]{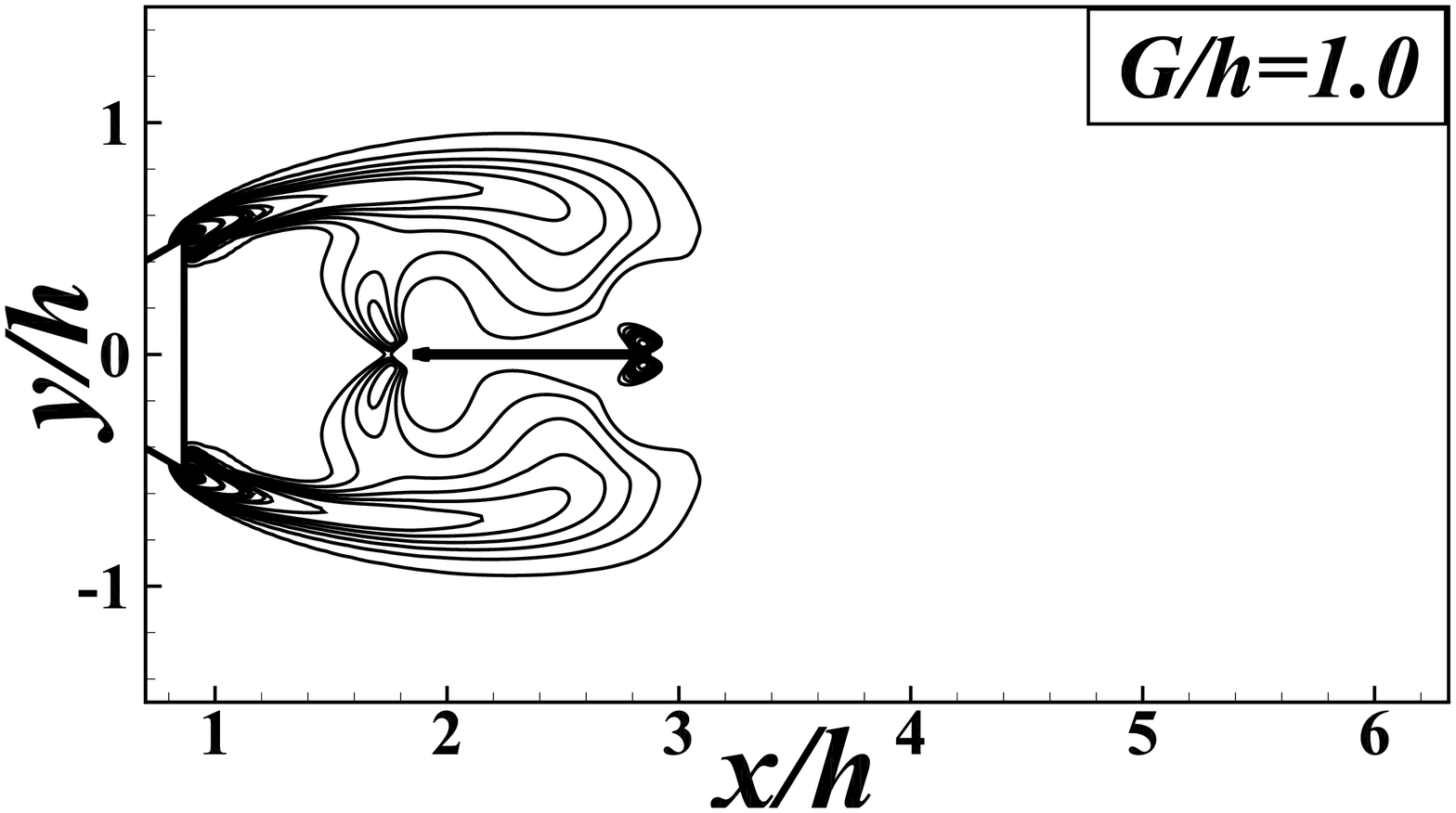} \hfill\
\ \hfill \includegraphics[scale=0.20]{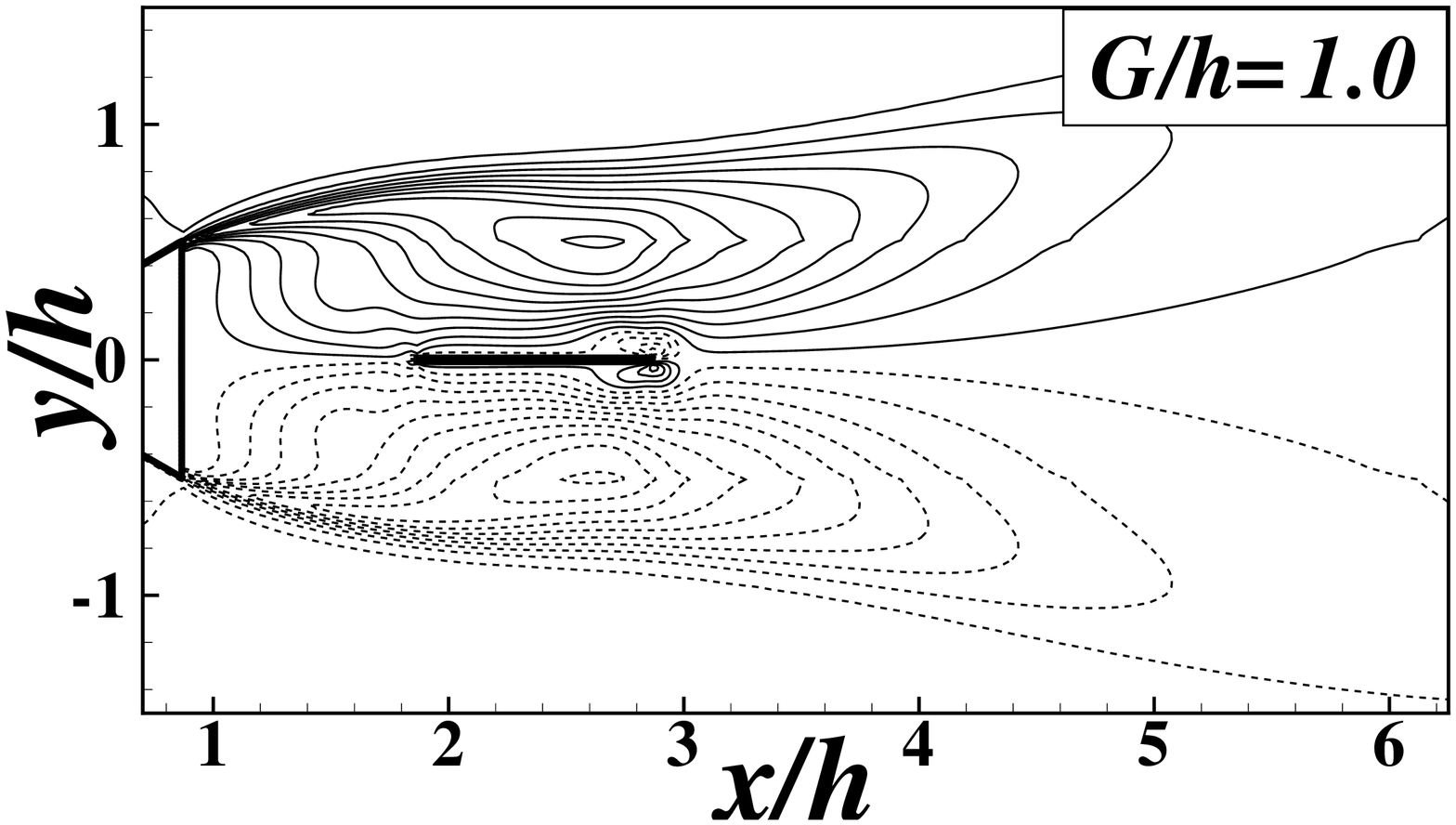}\hfill\
\ \hfill \includegraphics[scale=0.20]{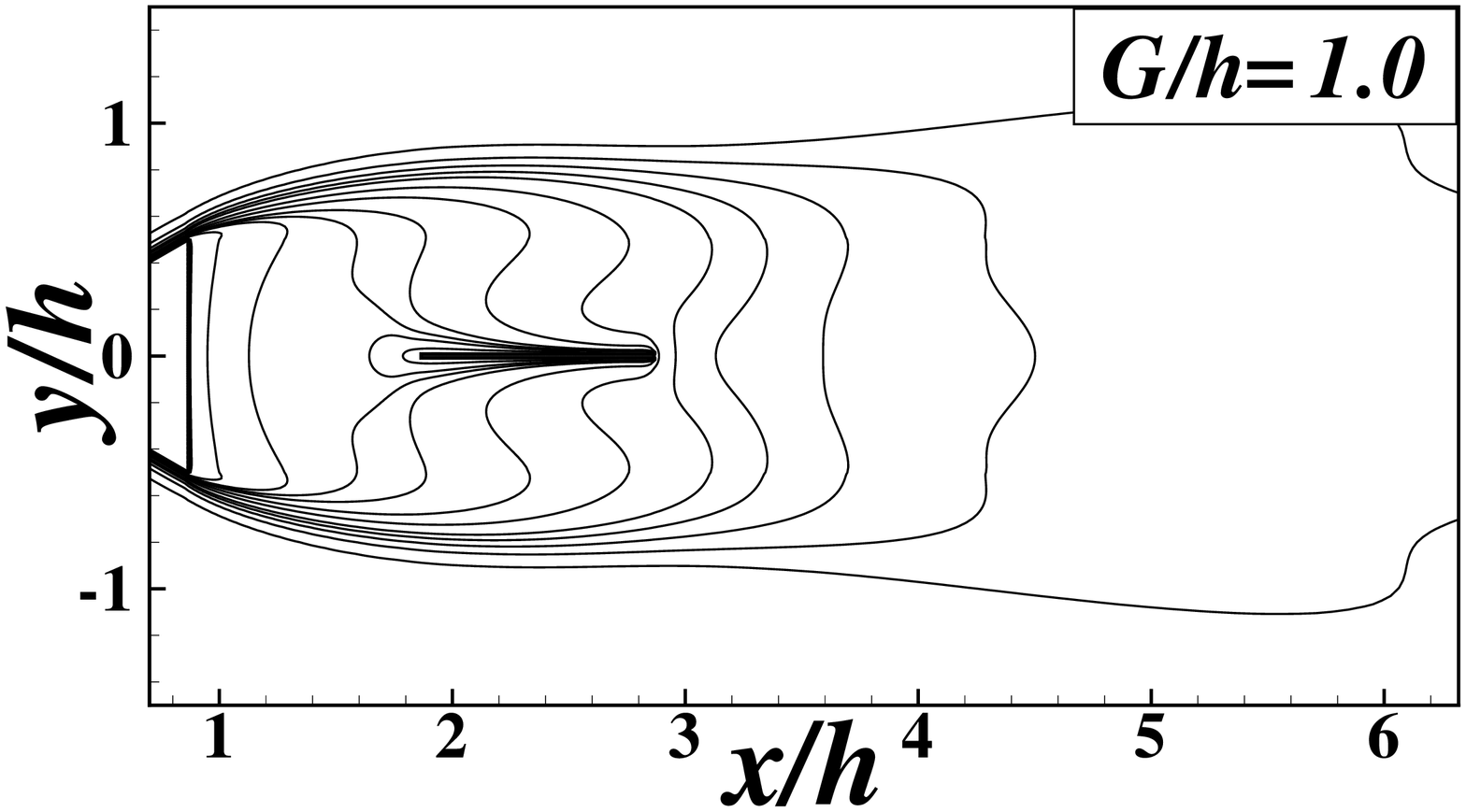} \hfill\
\end{tabular}
\begin{tabular}{ccc}
\ \hfill \includegraphics[scale=0.20]{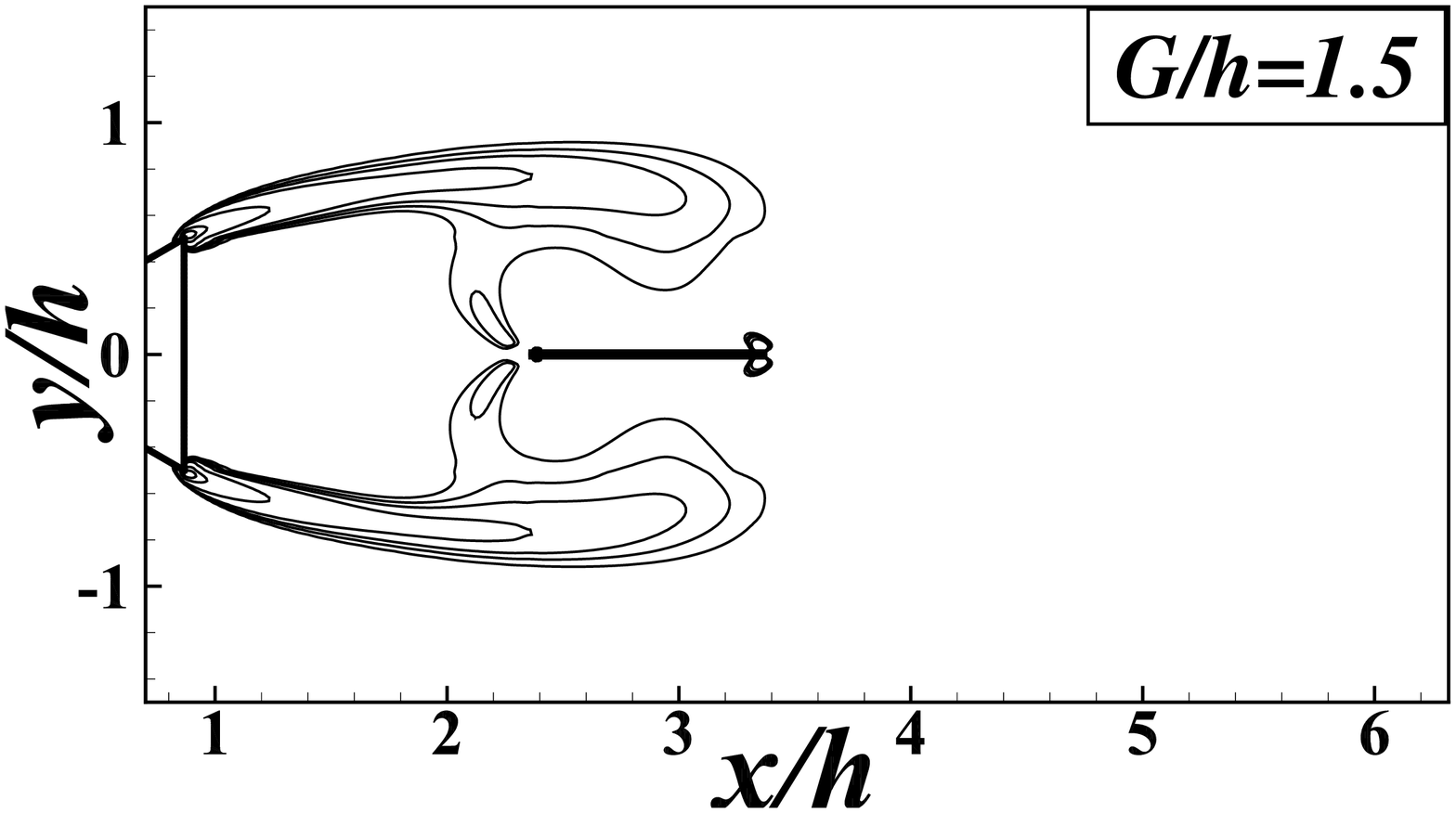} \hfill\
\ \hfill \includegraphics[scale=0.20]{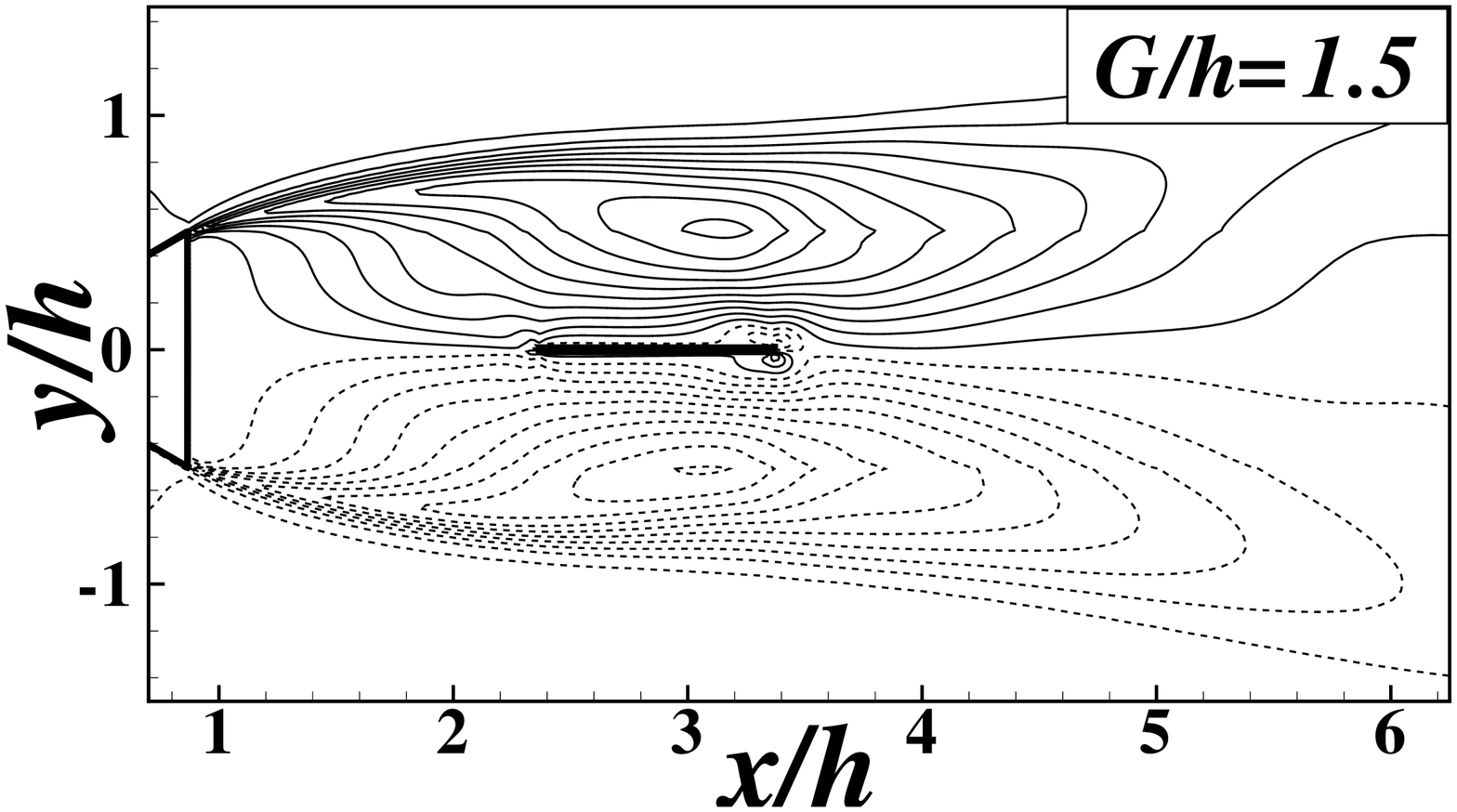}\hfill\
\ \hfill \includegraphics[scale=0.20]{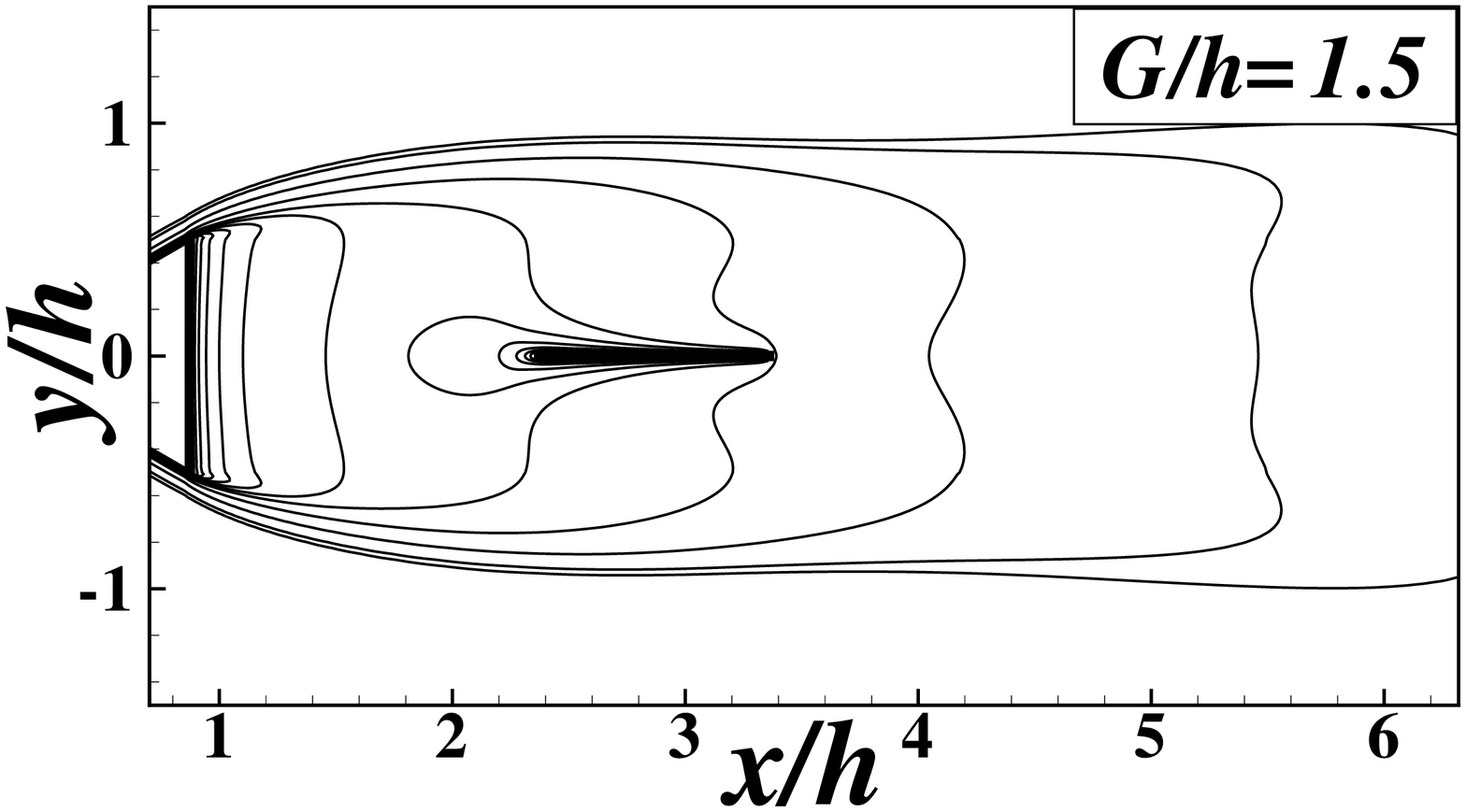} \hfill\
\end{tabular}
\begin{tabular}{ccc}
\ \hfill \includegraphics[scale=0.20]{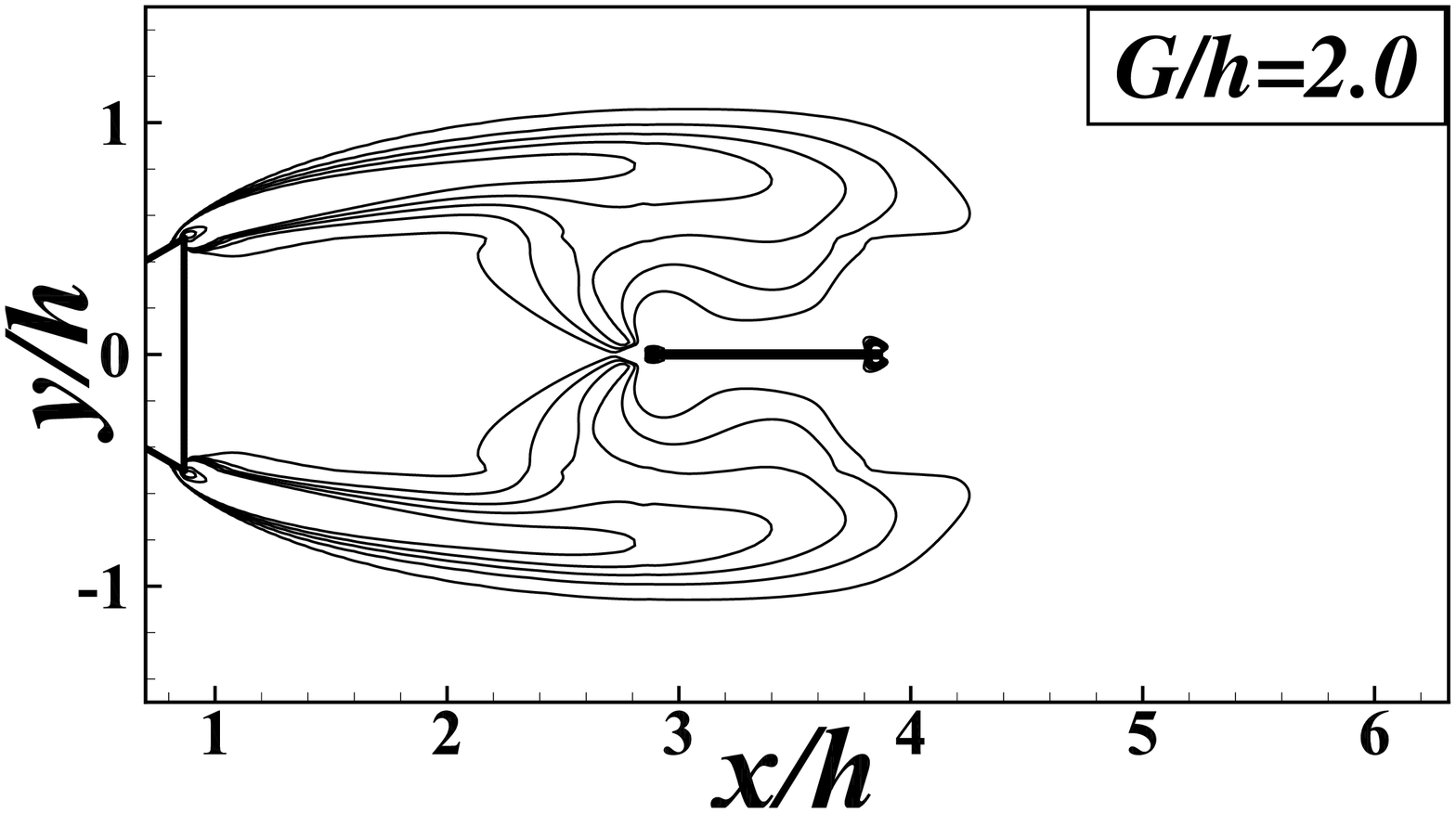} \hfill\
\ \hfill \includegraphics[scale=0.20]{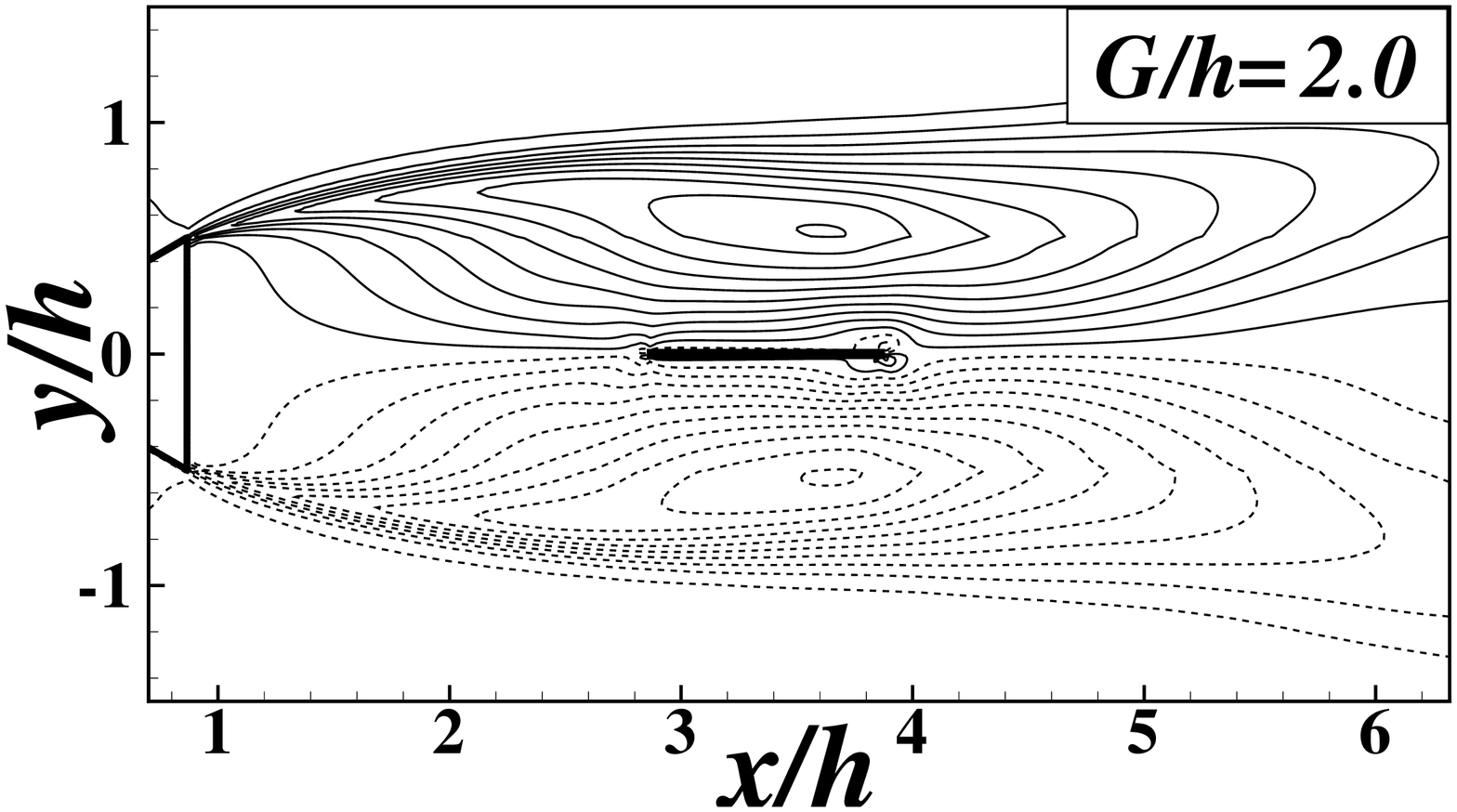}\hfill\
\ \hfill \includegraphics[scale=0.20]{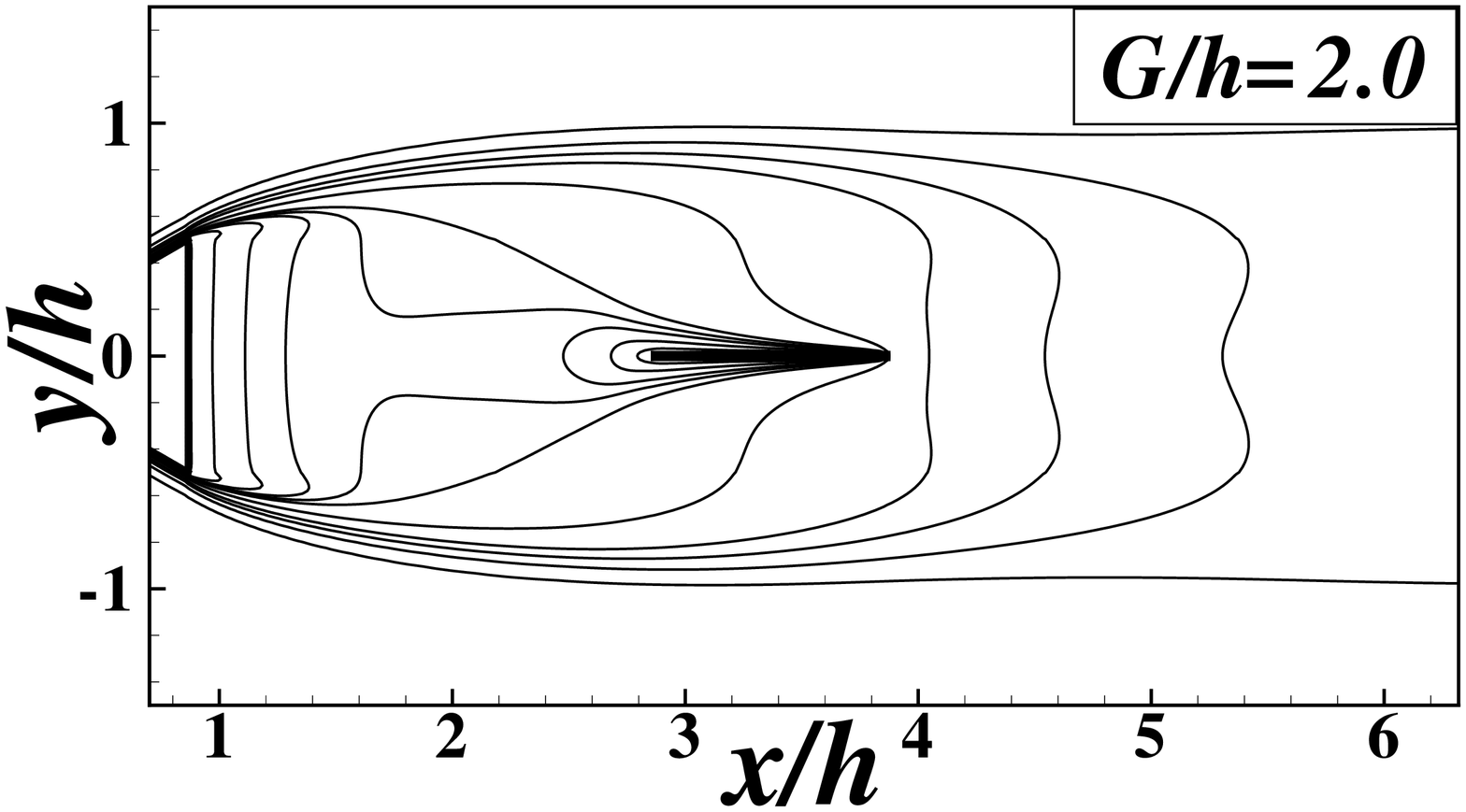} \hfill\
\end{tabular}
\caption{Time-averaged $\lambda_{2}$ isocontours, corresponding isocontours of normalised Reynolds Stress. Maximum and incremental values of Reynolds stress are defined as $[\overline{u'v'}/{U_{in}}^2]_{max}$=0.030 and $\Delta [ \overline{u'v'} /{U_{in}}^2]$=0.001, corresponding Isotherms (with $\theta _{min}=0$, $\Delta\theta=0.02$ and $\theta _{max}=1$ ).}
\label{fig:lambda_2_restress}
\end{figure}

Isocontours of mean velocity ($\overline{u}/U_{in}$) has been plotted for various values of 0.1, 0.25, 1 \& 4 and a typical set is shown in Fig. \ref{fig:meanvel_intensity} (a). Similarly, the distribution of turbulence intensity ($i$) are portrayed in Fig. \ref{fig:meanvel_intensity} (b). As suggested by Fig. 6 (a) the turbulence intensity, being a function of the fluctuating velocity components, has possessed the maximum value near to the solid surface i.e. the plate wall and the backside of the cylinder and the possible mechanism might be due to the presence of rapidly changing velocity components due to the formation and shedding of vortices. The normalized velocity contours, as portrayed in Fig. 6 (b), have given correct interpretation of physical insights as the velocity contour approaches the free stream value in the far-field region and the value diminishes as it approaches the wall following no-slip condition.  
\\The conglomerated diagram in Fig. \ref{fig:mean_velocity_with_gapratio} (a) shows the profiles of $\overline{u}/U_{in}$=1 for all the configurations analyzed. With increase in $G/h$, the attenuation in wake width seems apparent with a very narrow wake at $G/h$=2, which indicate that successive increase in gap between cylinder and splitter plate prevents the breaking of shear layers by stretching it along the stream-wise direction. Thereby it averts the free-stream momentum transfer into the wake region. Furthermore, time-averaged isocontours of $\lambda_{2}$ (as in Fig. \ref{fig:lambda_2_restress}), as a measure of visualising vortices, suggest the successive elongation in the wake region with increase in gap-ratio. 
\\The isocontours showing normalised time-averaged Reynolds stress that characterize the velocity fluctuations arising due to momentum transfer between adjacent fluid layers, are presented in Fig. \ref{fig:lambda_2_restress}. The stress contours shown with oppositely signed peaks are symmetrically located downstream for all the cases tested. As reported by Akilli \cite{Akilli} for a comparatively low Reynolds number flow past a circular cylinder, peak Reynolds Stress reduces with increase in gap ratio from 0 to 2. In the present case study, with successive increase in gap-ratio the maximum stress is reduced from 0.030 (for a bare cylinder) to 0.028 and 0.024 for $G/h$=0 and 0.5 respectively. 
However, it is observed that further increment in $G/h$ for subsequent settings alters the flow behavior with maximum stress rising to 0.027, 0.030 and 0.029 for $G/h$= 1, 1.5 and 2, respectively. As seen in Fig. \ref{fig:lambda_2_restress}, The location of maximum Reynolds stress moves a considerable distance downstream of with increase in gap ratio due to the velocity fluctuations in this region. \\
Time-averaged local pressure coefficient variation along the cylinder is shown in Fig. \ref{fig:timeavg_local_Pressure_coefficient} (a) and it posseses a local maximum value at the front stagnation point or at the leading edge (i.e. Point A). Reduction in wake width in turn reduces the pressure drop characteristics in the wake region. $\overline{C_{pl}} $ shows improvement with increase in gap-ratio along the trailing edge of the cylinder with optimal performance for $G/h=2$. Diminution in the pressure drop in the wake region in turn improves the profile of $\overline{C_{pl}} $ along the plate significantly, as seen in Fig. \ref{fig:timeavg_local_Pressure_coefficient} (b).  
\subsection{Effect of variation in plate length on non-dimensional parameters:}
\begin{figure}[h!]
\centering
\begin{tabular}{cc}
\ \hfill \includegraphics[scale=0.30]{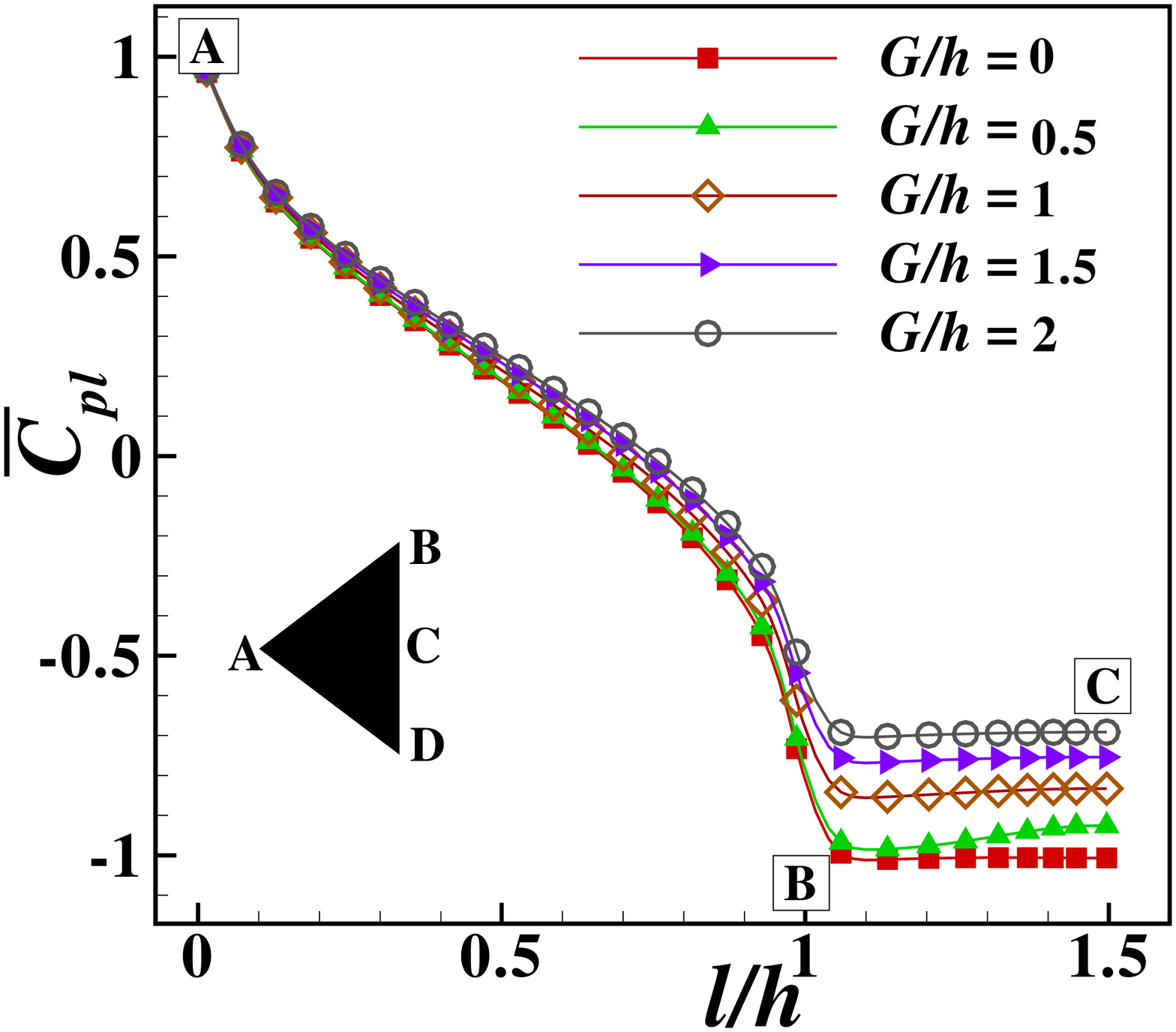} \hfill\ 
\ \hfill \includegraphics[scale=0.30]{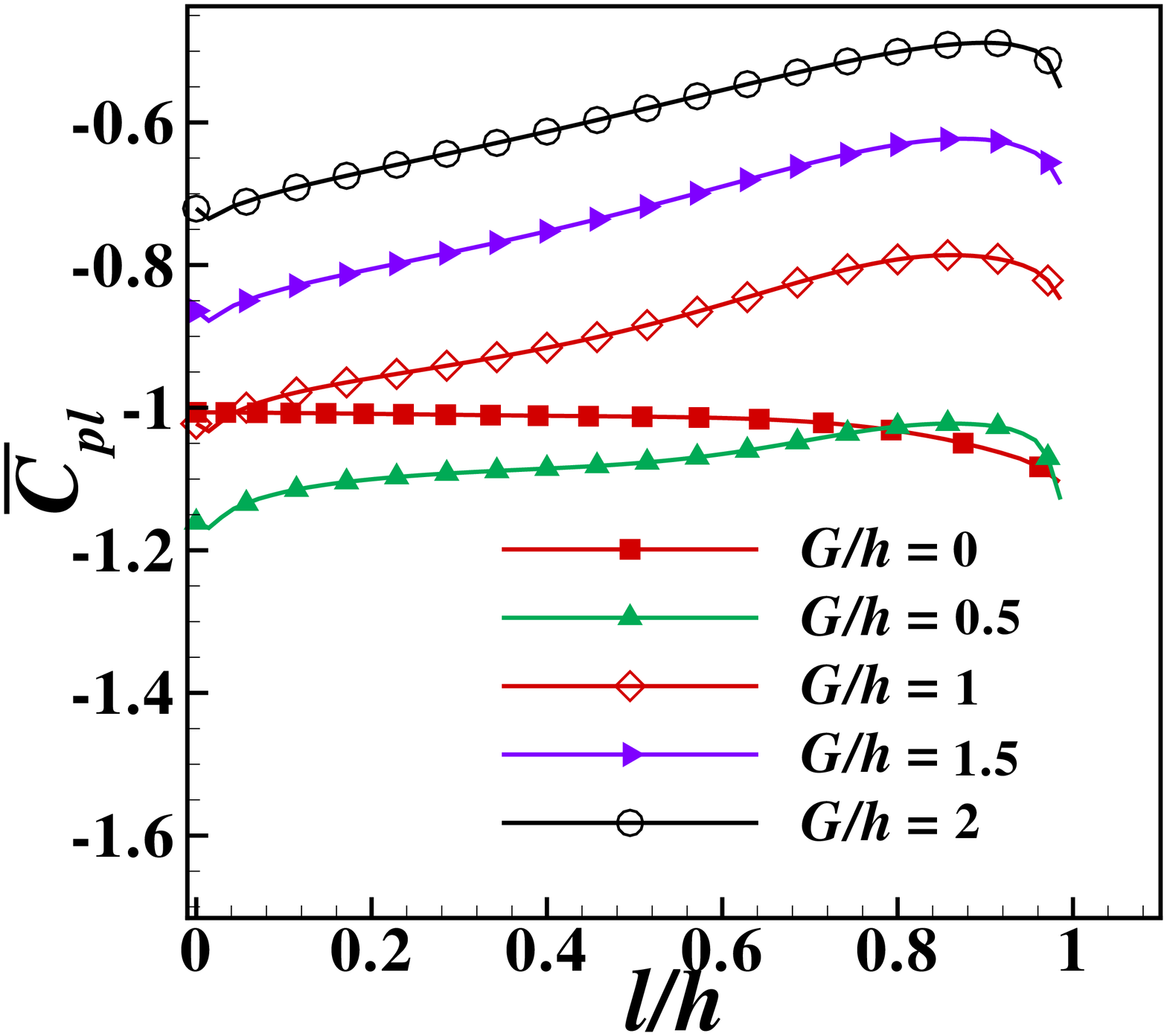}\hfill\ \\ 
\end{tabular}
$~~~~~~~~~~~~~~~~~~~~~~~$(a)$~~~~~~~~~~~~~~~~~~~~~~~~~~~~~~~~~~~~~~~~~~~~~~~$(b)$~~~~~~~~~~~~~~~$\vspace{0.2cm}
\begin{tabular} {c}
  \ \centering \hfill \psfig{figure=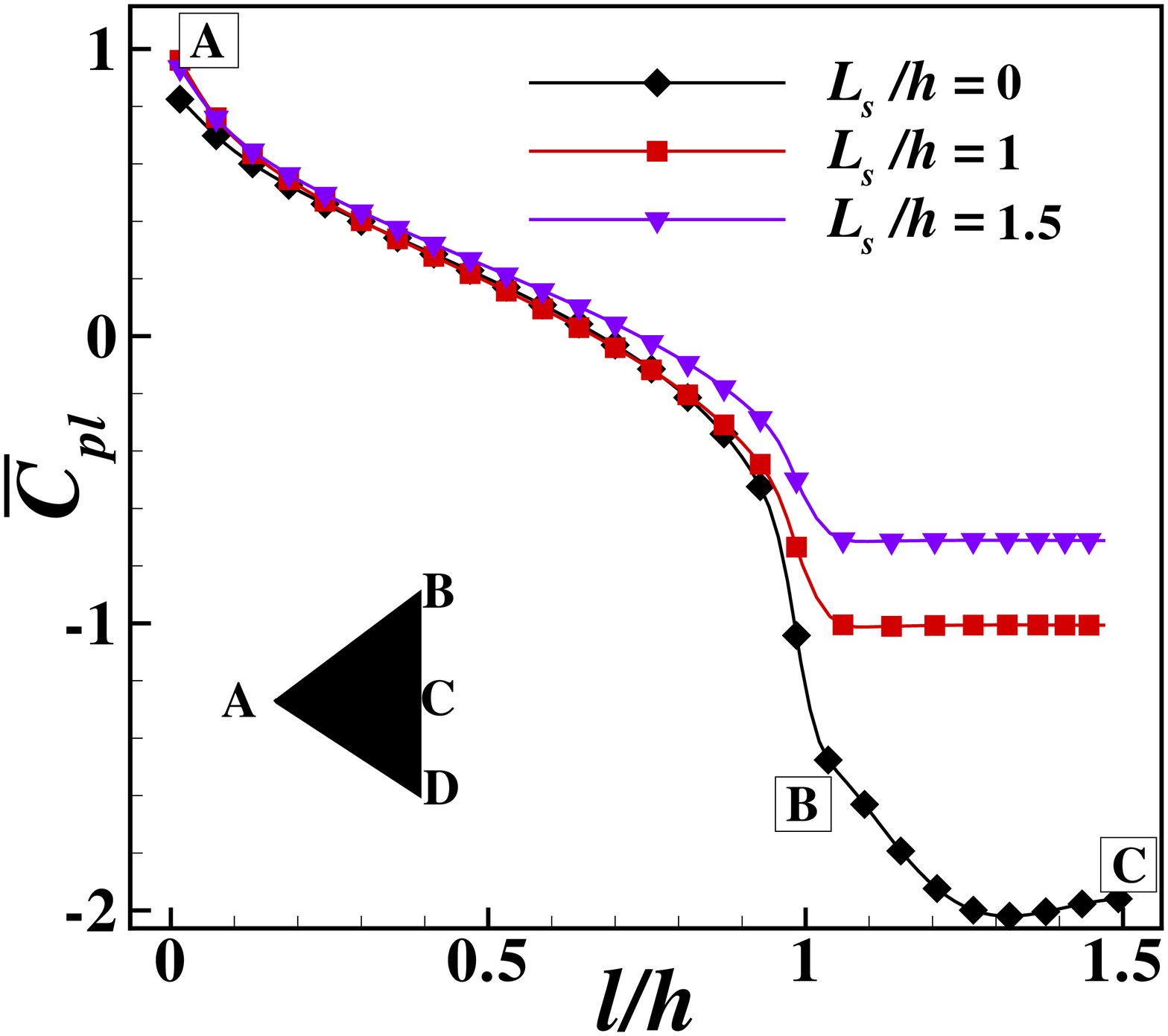,width=2.8in,height=2.3in} \hfill \ \
\end{tabular}
$~~~~~~~~~~~~~~~~~~~~~~~~~~~~~~~~~~$(c)$~~~~~~~~~~~~~~~~~~~~~~~~~~$\vspace{0.2cm}
\centering
\begin{tabular}{cc}
\ \hfill \includegraphics[scale=0.30]{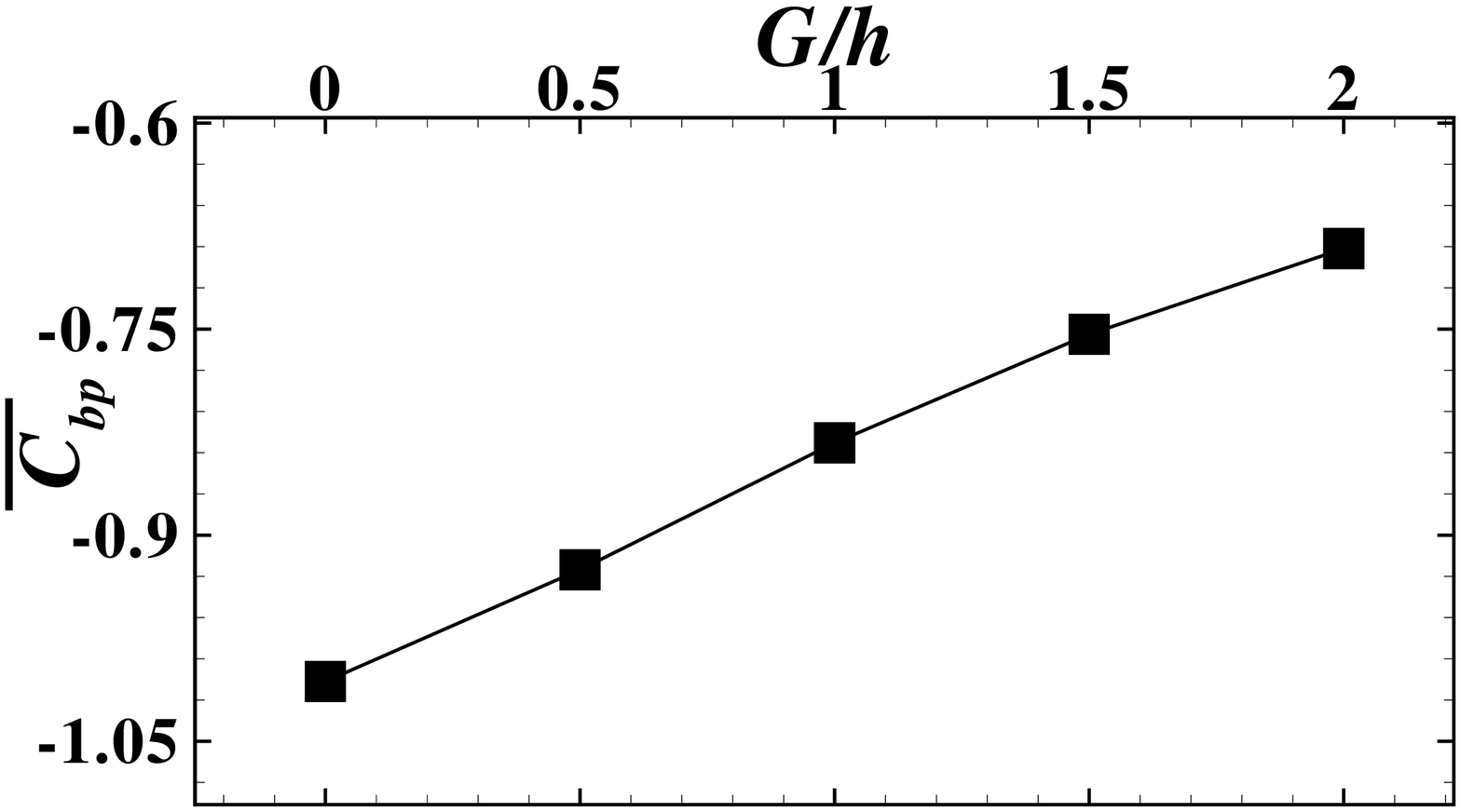}\hfill\
\ \hfill \includegraphics[scale=0.30]{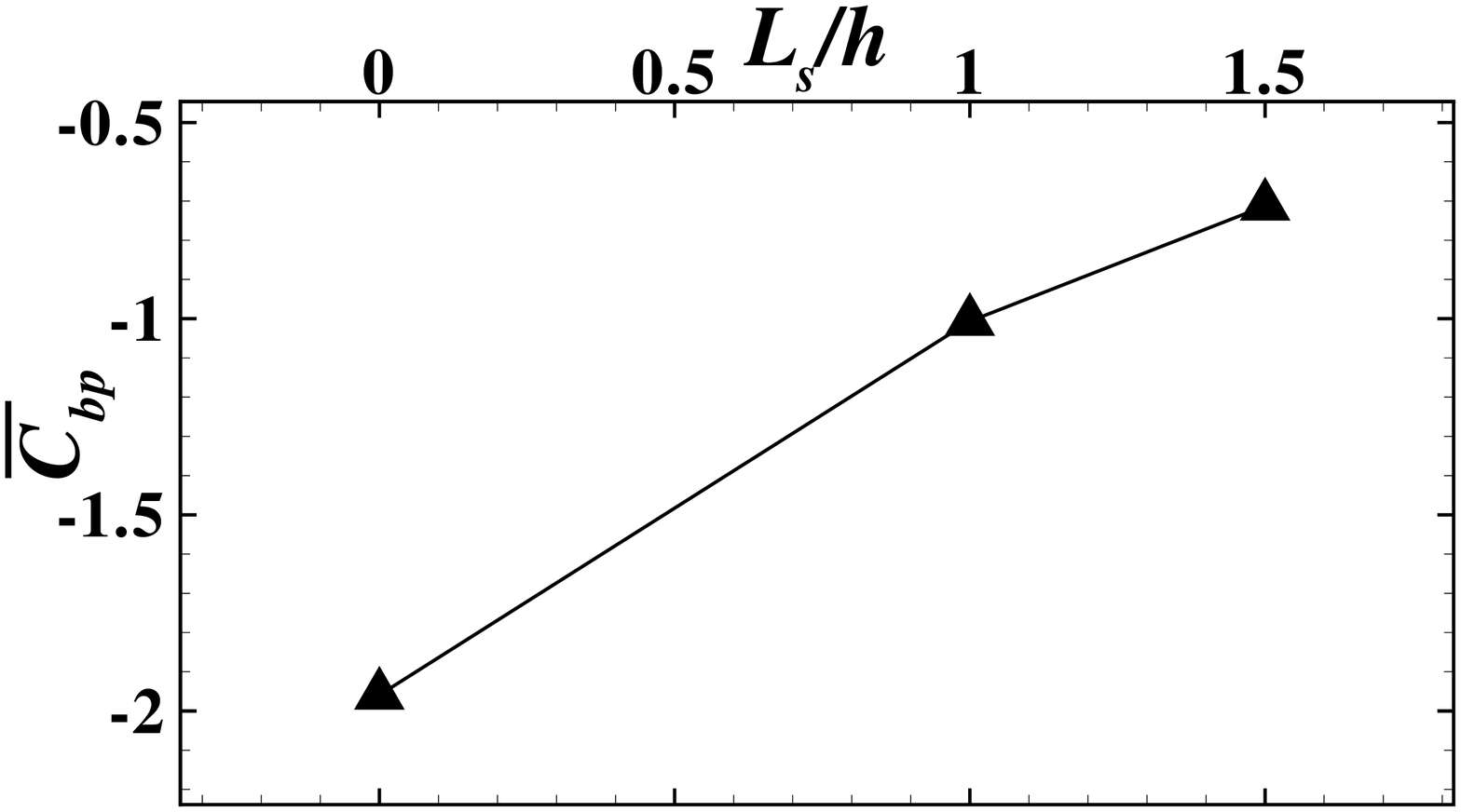} \hfill\
\end{tabular}
$~~~~~~~~~~~~~~~~~~~~~~~$(d)$~~~~~~~~~~~~~~~~~~~~~~~~~~~~~~~~~~~~~~~~~~~~~~~$(e)$~~~~~~~~~~~~~~~$
\caption{The profile of time-averaged local pressure coefficient ($\overline{C_{pl}}$) along the edge of the triangular cylinder (a) for $L_{s}/h=1$ and $0\leq G/h\leq 2$, (c) for $G/h=0$ and $0\leq L_{s}/h\leq 1.5$; (b) The profile showing variation of $\overline{C_{pl}}$ along the upper surface of the splitter plate for various gap ratio configurations; and Variation of time-averaged base pressure coefficient ($\overline{C_{bp}}$) (d) with varying $G/h$ for $0\leq G/h\leq 2$; and (e) with varying $L_{s}/h$ for $0\leq L_{s}/h\leq 1.5$.}   
\label{fig:timeavg_local_Pressure_coefficient}
\end{figure}

Non-dimensional entities namely $\overline{C_{d}}$, $\overline{C_{d_{p}}}$, $C_{lrms}$, $St$, $\overline{C_{p}}$, $\overline{Nu}$ are presented in Table 3 for different plate-length cases analyzed. Complying with the conventionally established results, increase in plate length resulted in a significant reduction in drag force on the TCSP system. The $\overline{C_{d}}$ is reduced from 2.021 (for a bare cylinder) to 1.237 (for $L_{s}/h=1$) with a percentage reduction of $\approx38.8\%$ and with subsequent increment in plate length, it reduces even further. Pressure drag, contributing the maximum proportion towards the resulting drag, follows a similar trend. While the $St$ in the absence of control has a maximum value of 0.171, with the insertion of splitter plate, shedding frequency shows a slight reduction of 2.9\% for $L_{s}/h=1$ and among shedding cases the least frequency of 0.139 was obtained for $L_{s}/h=1.5$. Introducing a splitter plate results in a decrement of $\overline{C_{p}}$ to 89\% with increment in plate length from 0 to 1 and further reduction is observed for increase in plate length which suggests the massive diminution in pressure drop in the wake region. Detailed variation of time-averaged local pressure coefficient ($\overline{C_{pl}} $) is shown in Fig. \ref{fig:timeavg_local_Pressure_coefficient} (c). Data provided in Table 3 indicates remarkable increase in vortex formation length too. While it has its minimum value for bare cylinder, it shows a monotonic increase with increase in plate length. Results for base pressure coefficient variation ($\overline{C_{bp}} $) with gap ratio and plate length are plotted in Figs. \ref{fig:timeavg_local_Pressure_coefficient} (d) and (e). Here base region refers to the midpoint (Point C as in Fig. \ref{fig:timeavg_local_Pressure_coefficient}) on the rear edge of the triangular cylinder. 
\\Apelt and West \cite{Apelt1} have studied the effect of plate length on wake width for flow past a circular cylinder with $Re$ varying over a range of $1\times10^4$ to $5\times10^4$. They observed that the wake width reduces substantially for shorter plates and for longer plates with range varying from $L_{s}/h=1 - 1.5$, the wake width reduces further, but increases thereafter for a plate length of $L_{s}/h=2$. Similar study has been performed here to examine the effect of plate length on wake width for various settings. As seen in Fig. \ref{fig:mean_velocity_with_gapratio} (b) wake envelope at the near wake suggests that wake gradually narrows with increase in plate length which is an indication of suppression in vortex shedding.  
\subsection{Effect of change in $G/h$ on heat transfer:}
\begin{figure}[h!]
\centering
\begin{tabular} {c}
\ \centering \hfill \psfig{figure=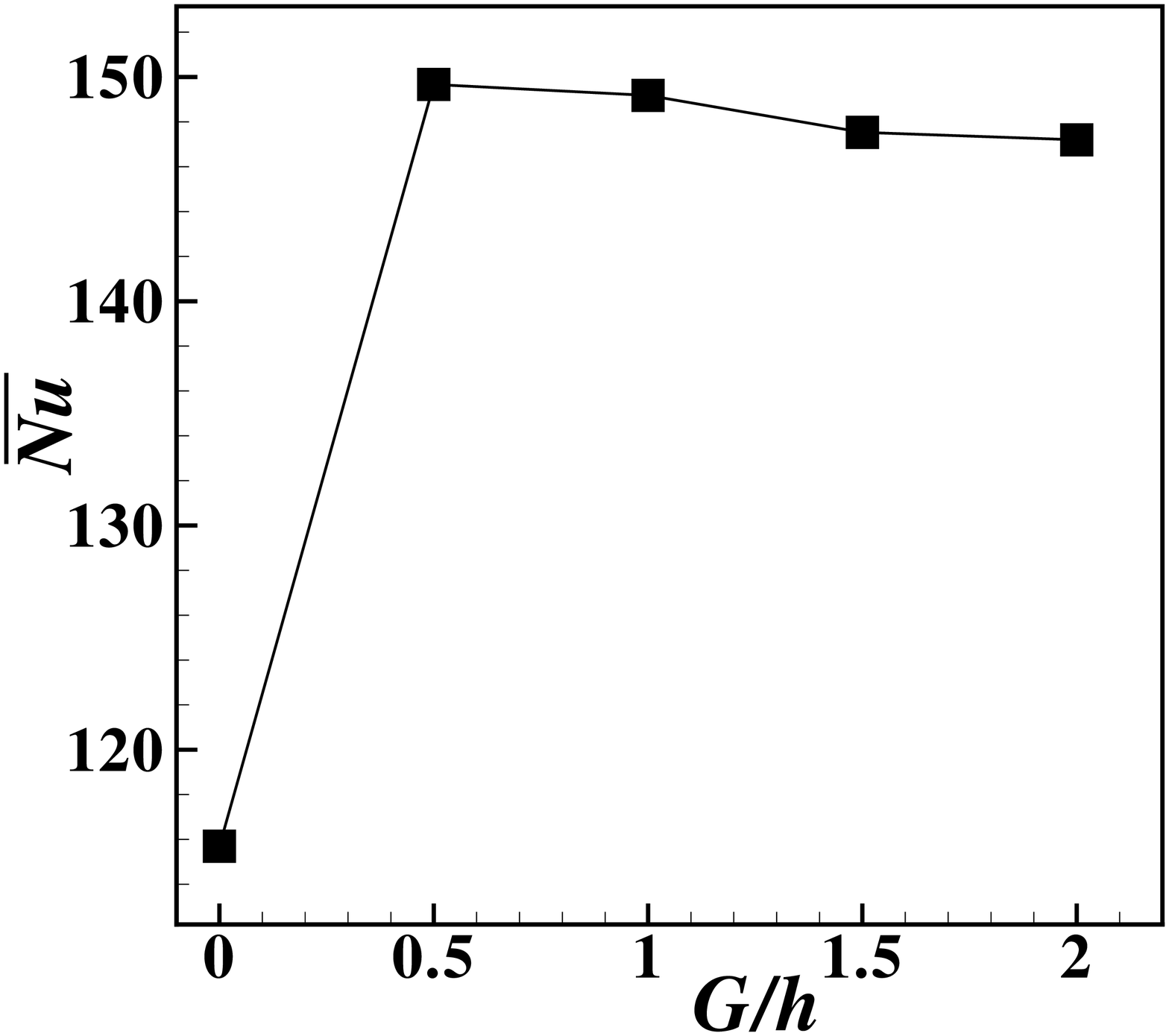,width=2.8in,height=2.3in} \hfill \ \
\end{tabular}
$~~~~~~~~~~~~~~~~~~~~~~~~~~~~~~~~~~~~~~~~~~~~~~~~~~~~~$(a)$~~~~~~~~~~~~~~~~~~~~~~~~~~~~~~~~~~~~~~~~~~~~~~~$
\begin{tabular}{cc}
  \ \hfill \includegraphics[scale=0.30]{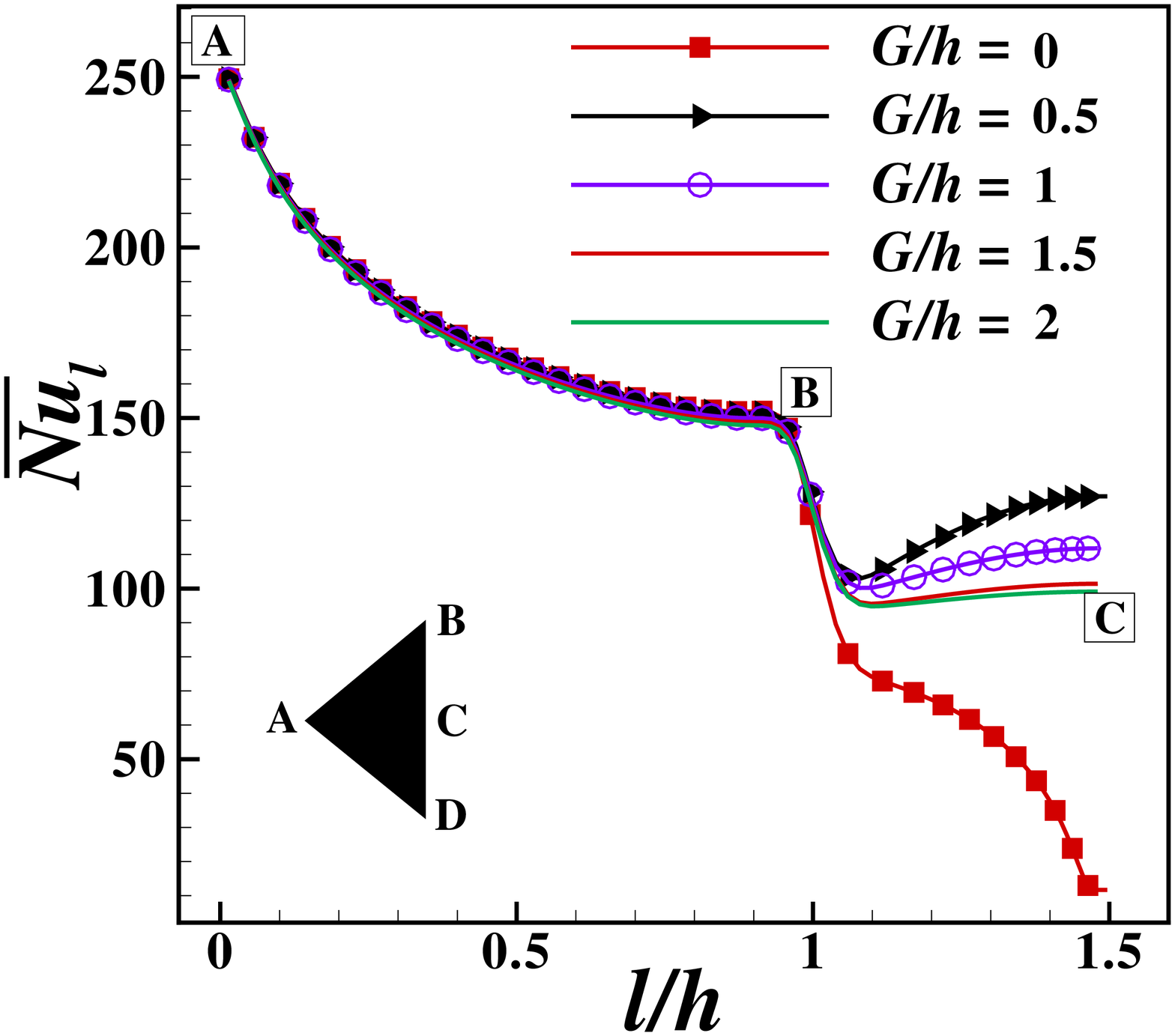} \hfill  \
  \ \hfill \includegraphics[scale=0.30]{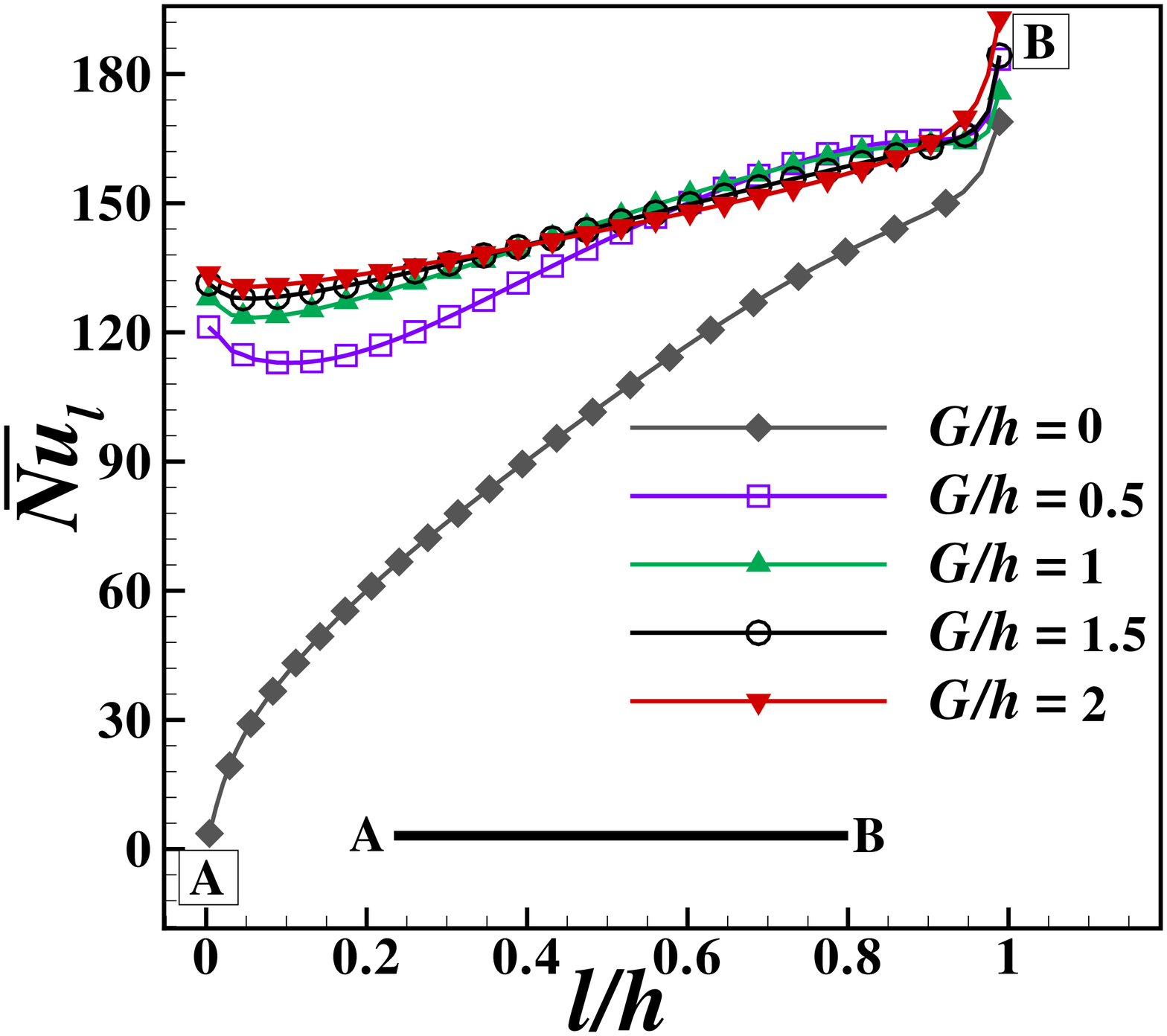} \hfill  \
\end{tabular}
$~~~~~~~~~~~~~~~~~~~~~~~$(b)$~~~~~~~~~~~~~~~~~~~~~~~~~~~~~~~~~~~~~~~~~~~~~~~~$(c)$~~~~~~~~~~~~~~~$
\\ \vspace{0.2cm}
\begin{tabular}{cc}
  \ \hfill \includegraphics[scale=0.30]{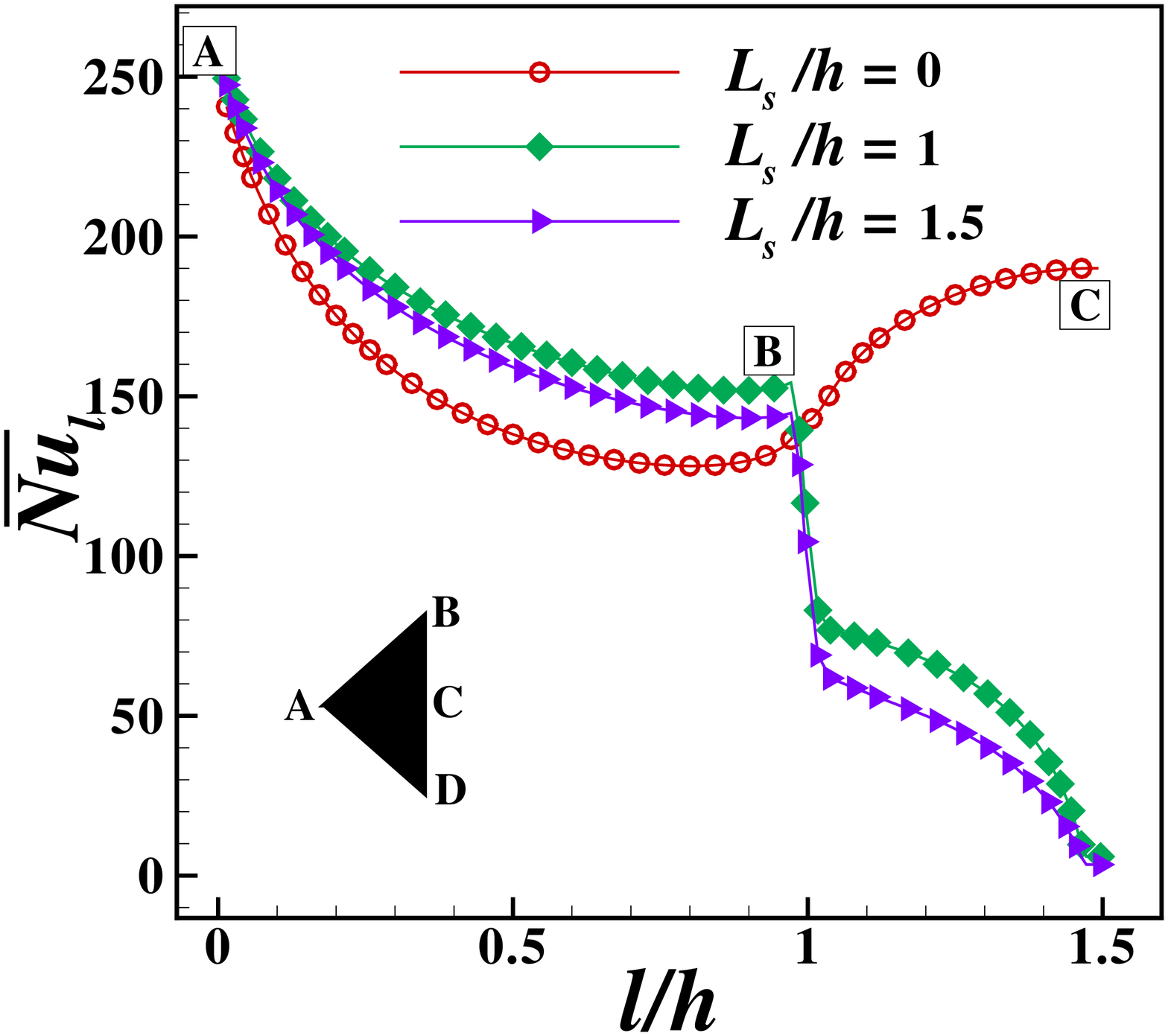} \hfill  \
  \ \hfill \includegraphics[scale=0.30]{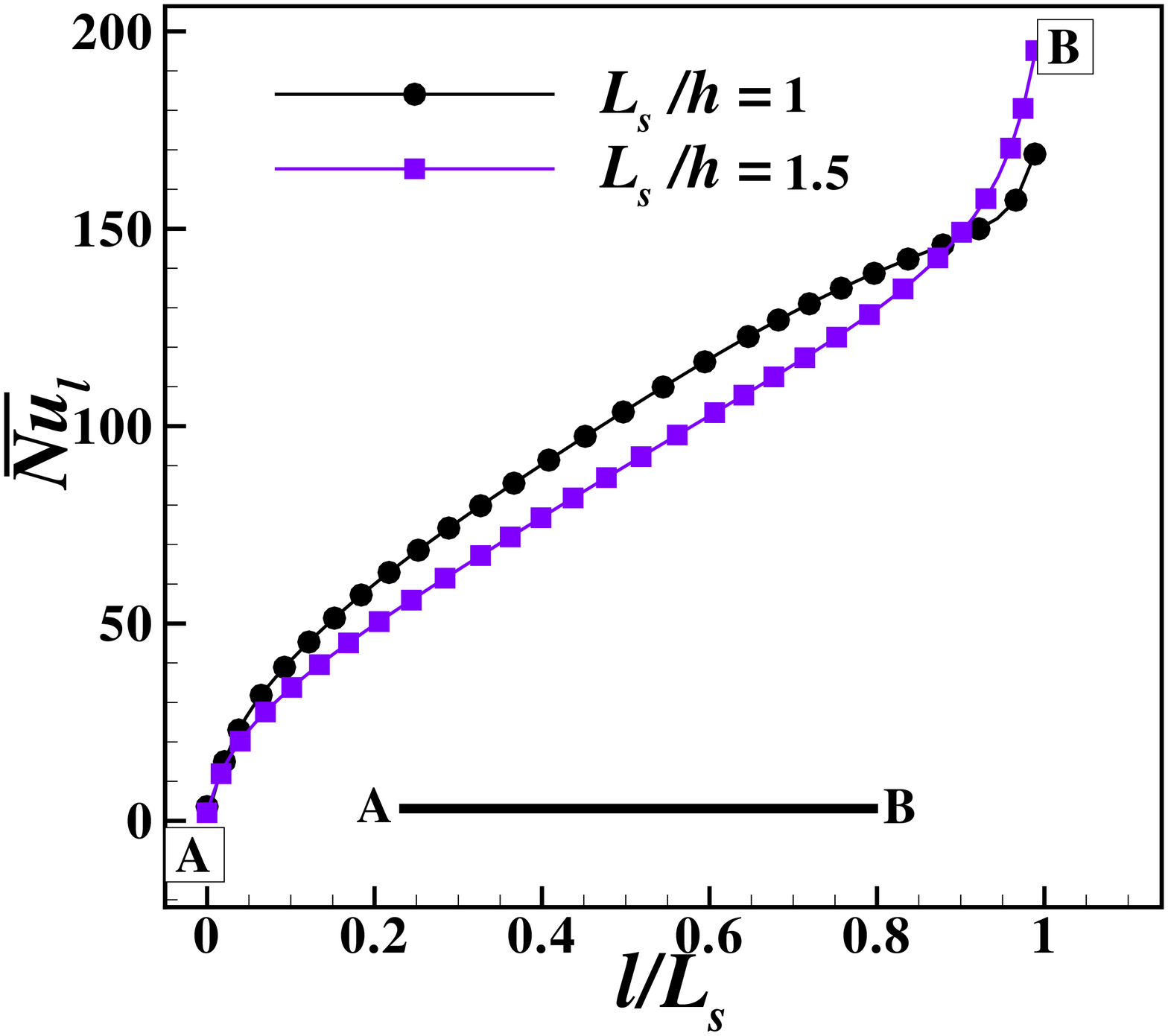} \hfill  \
\end{tabular}
$~~~~~~~~~~~~~~~~~~~~~~~$(d)$~~~~~~~~~~~~~~~~~~~~~~~~~~~~~~~~~~~~~~~~~~~~~~~~$(e)$~~~~~~~~~~~~~~~$
\caption{(a) Variation of $\overline{Nu}$ with gap ratio for 0$\leq G/h \leq$ 2, The profile of ($\overline{Nu_{l}}$) alongside the edge of the triangular cylinder (b) for $L_{s}/h=1$ and $0\leq G/h\leq 2$, (d) for $G/h=0$ and $0\leq L_{s}/h\leq 1.5$ ; and The profile of time-averaged local Nusselt number ($\overline{Nu_{l}}$) along the upper surface of the splitter plate (c) for $L_{s}/h=1$ and $0\leq G/h\leq 2$, (e) for $G/h=0$ and  $L_{s}/h$=1 and 1.5.} 
\label{fig:timeavg_local_nusselt_number}
\end{figure}

Heat transfer characteristics for various gap-ratio configurations are assessed in terms of time-averaged Nusselt number ($\overline{Nu}$), where $\overline{Nu}$ is defined over the triangular cylinder-splitter plate ($TCSP$) system, as follows in Fig. \ref{fig:timeavg_local_nusselt_number}(a). With increase in the gap-ratio from 0 to 0.5, $\overline{Nu}$ increases to a local maximum and then it decreases with further increase in gap-ratio. The aforementioned behavior suggests the presence of an optimal gap-ratio for maximum heat transfer. As depicted in the Fig. \ref{fig:timeavg_local_nusselt_number}(a), With rise in $G/h$ from 0 to 0.5, $\overline{Nu}$ shows a noticeable rise in heat transfer with a percentage increment of 30.4\%. But, the heat transfer characteristics does not show a distinguishable alteration with further increase in gap-ratio.
\\Time-averaged isotherms (as shown in Fig. \ref{fig:lambda_2_restress}) are dense near the cylinder side-wall signalling the formation of thermal boundary layer. As shown in Figure \ref{fig:timeavg_local_nusselt_number}(b), $\overline{Nu_{l}}$ does not show a significant variation along the side-wall of the cylinder, albeit variation along the rear-wall is much more prominent. For $G/h$=0, while traversing from Point-B to Point-C, $\overline{Nu_{l}}$ shows a monotonic decrease with a minimum value of of 3.4 at Point-C. Introducing a gap surprisingly improves the heat transfer (for $G/h$=0.5) with a peak value of $\overline{Nu_{l}}$=120 at Point-C. However, futhur increase in gap-ratio shows a relatively flatter local heat transfer profile though variation along trailing edge of the cylinder shows some improvement.        
\\The time-averaged local Nusselt number variation, as suggested in Fig. \ref{fig:timeavg_local_nusselt_number}(c) along the plate reveals that for $G/h$=0 case, $\overline{Nu_{l}} $ has significantly larger value at the tip of the plate than at the root (where the plate is attached) leading to a steeper slope, which indicate that owing to the presence of a sharp corner at the cylinder-plate attachment portion, the flow separates leading to a localized heat formation region. However, the inclusion of a finite gap enhances the heat transfer by a comparably larger margin.  
\subsection{Effect of $L_{s}/h$ on heat transfer performance:}
Even though exposed surface area for effective heat transfer increases with increase in length of control plate, $\overline{Nu}$ shows a dramatic reduction of $\approx 28\%$ for $L_{s}/h$ ranging from 0 to 1 (as mentioned in Table 3) and almost doesn't show much variation further.
\\Point A, being the stagnation point in Fig. \ref{fig:timeavg_local_nusselt_number} (d), experiences maximum heat transfer with the highest $\overline{Nu_{l}} $ value. Further it decreases along the side wall due to boundary layer development till it reaches the trailing edge of the cylinder (Point B). Following flow separation, two interacting shear layers lead to turbulent mixing and improvement in heat transfer (from Point B to Point C), which is reflected in Fig. \ref{fig:timeavg_local_nusselt_number} (d) for the case of $L_{s}/h$=0. However, with further increment in plate length, substantial decrement in heat transfer is observed possibly due to the presence of sharp corner at the attachment region which is evident from the $\overline{Nu}$ value presented in Table 3. An insightful observation has been made during the study that presence of splitter plate downstream of the flow affects the heat transfer upstream, as $\overline{Nu_{l}} $ improves noticeably along the side wall for cases with control plate than for a bare cylinder. 
\\Figure \ref{fig:timeavg_local_nusselt_number} (e) shows the local variation along the plate, where the length traversed is normalised with the plate length ($L_{s}$). Improvement in heat transfer can be observed towards the tip of the plate with increase in plate length which indicate that the analysis conducted here being a highly convection dominated flow, results in effective heat removal from the surface with increase in area of exposure. Instantaneous isotherms are presented in Fig. \ref{fig:plate_length_allfig}. Due to the formation of thermal boundary layer, isotherms shown are intensely clustered near to the wall. Ascribed to to a highly advecting current, heat dissipates in the fluid domain as suggested by the increasing distance between neighbouring isotherms. For $L_{s}/h=0$ isotherms show sharp bending motion at the backside of the cylinder compared to a greater stretching and narrowing of isotherms for $L_{s}/h=1$ and 1.5.      
\subsection{Instantaneous vorticity ($\omega _{z}$) contours:}
\begin{figure}[h!]
\centering
\begin{tabular}{ccc}
\ \hfill \includegraphics[scale=0.20]{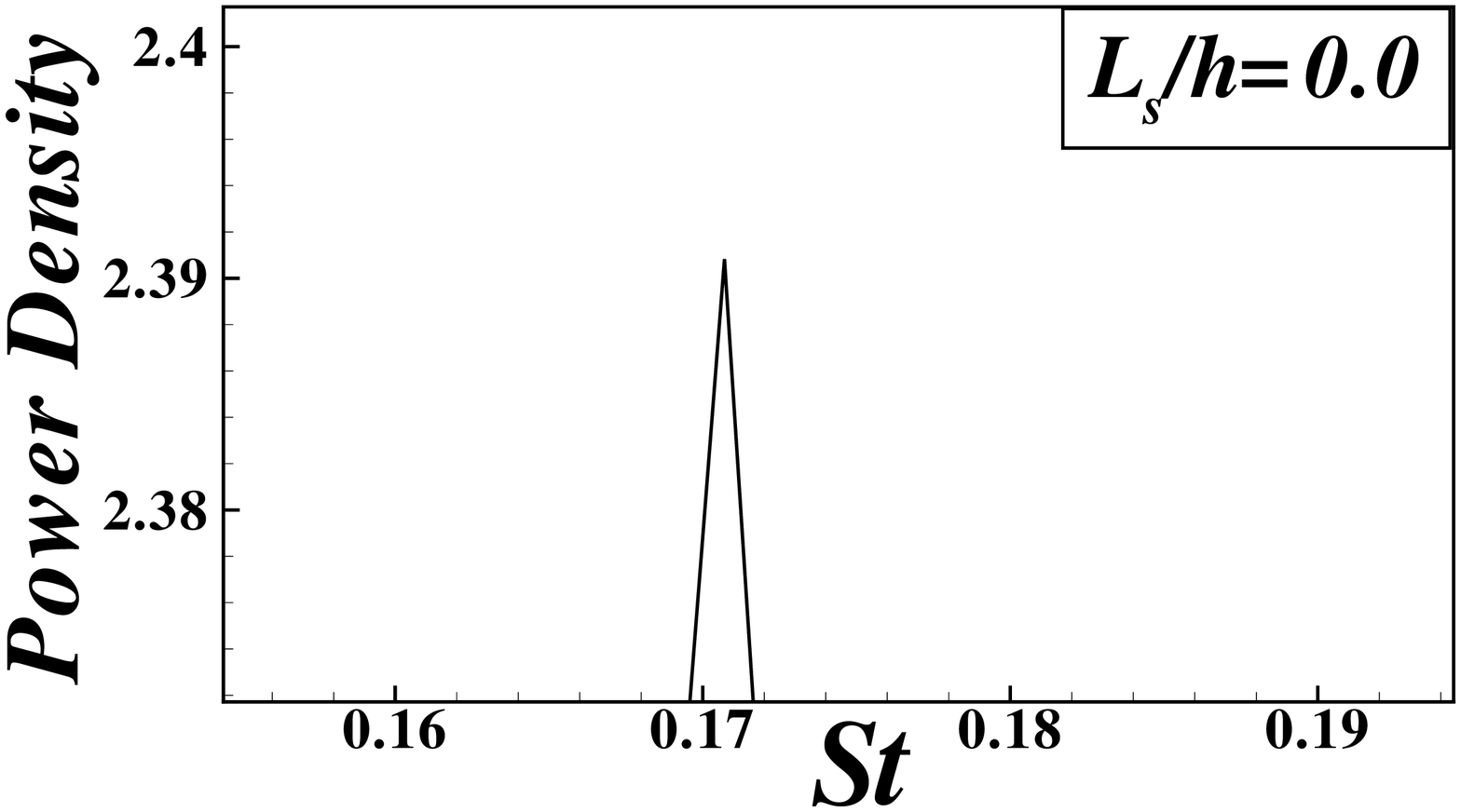} \hfill\ 
\ \hfill \includegraphics[scale=0.20]{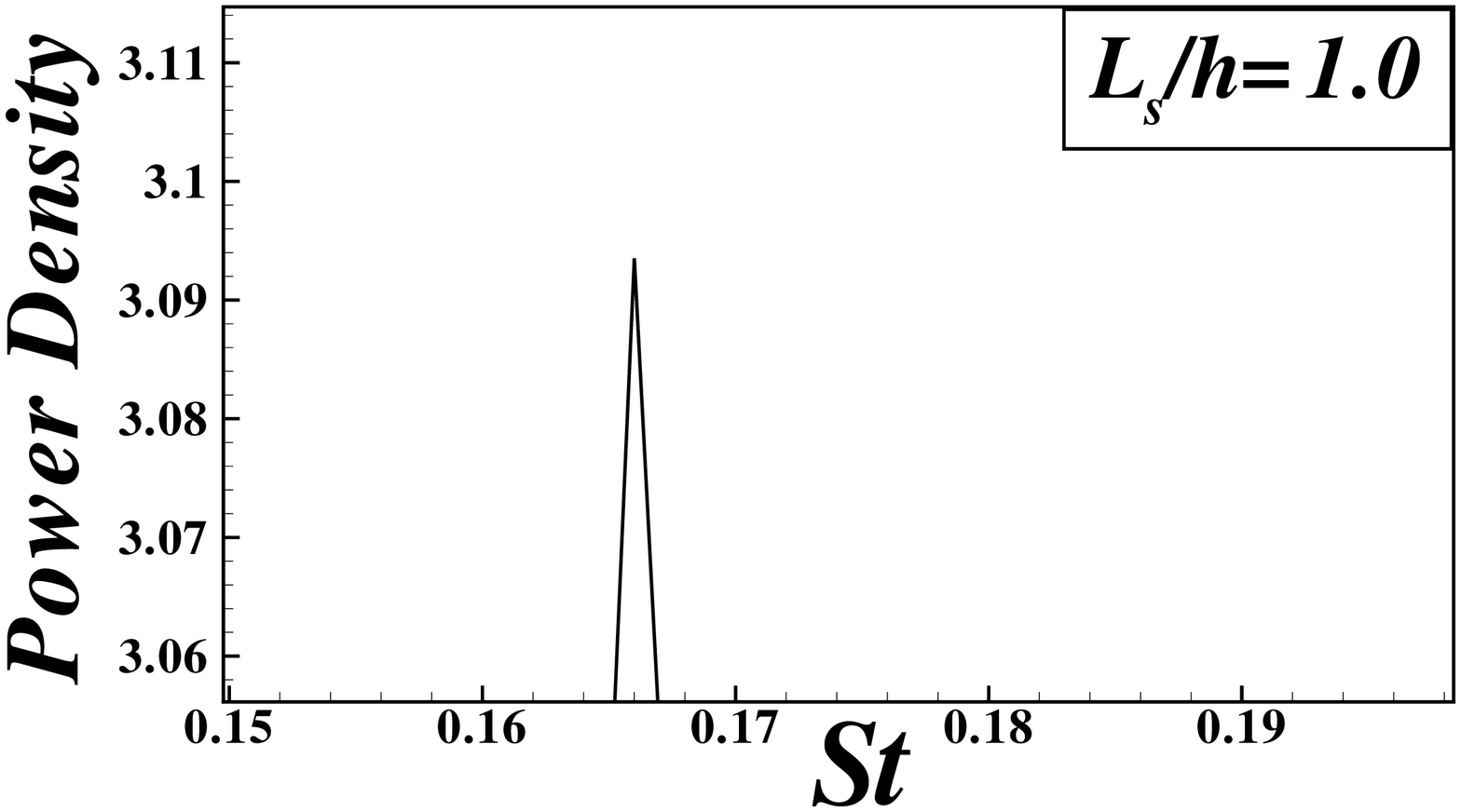} \hfill\  
\ \hfill \includegraphics[scale=0.20]{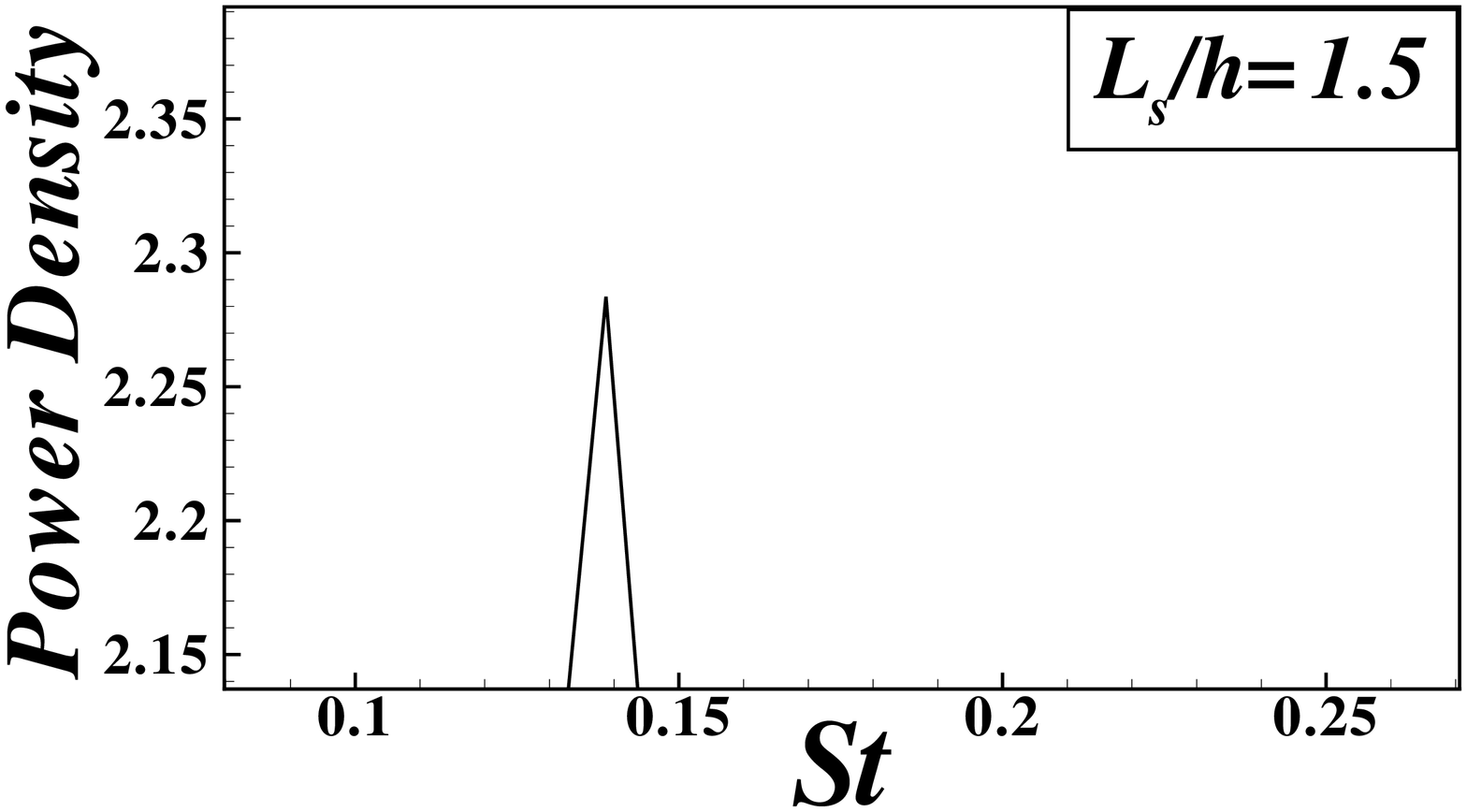} \hfill\
\end{tabular}
\begin{tabular}{ccc}
\ \hfill \includegraphics[scale=0.20]{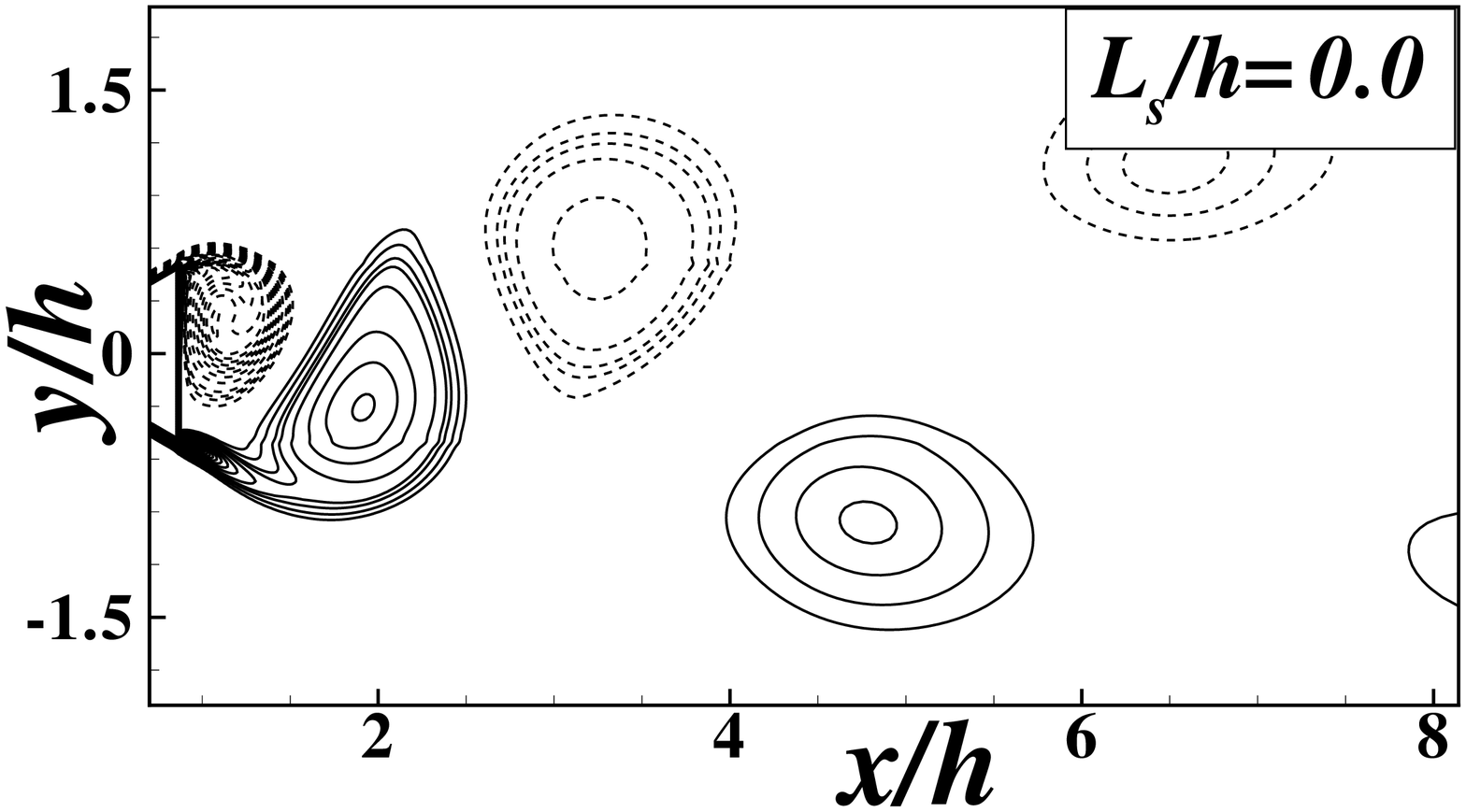} \hfill\ 
\ \hfill \includegraphics[scale=0.20]{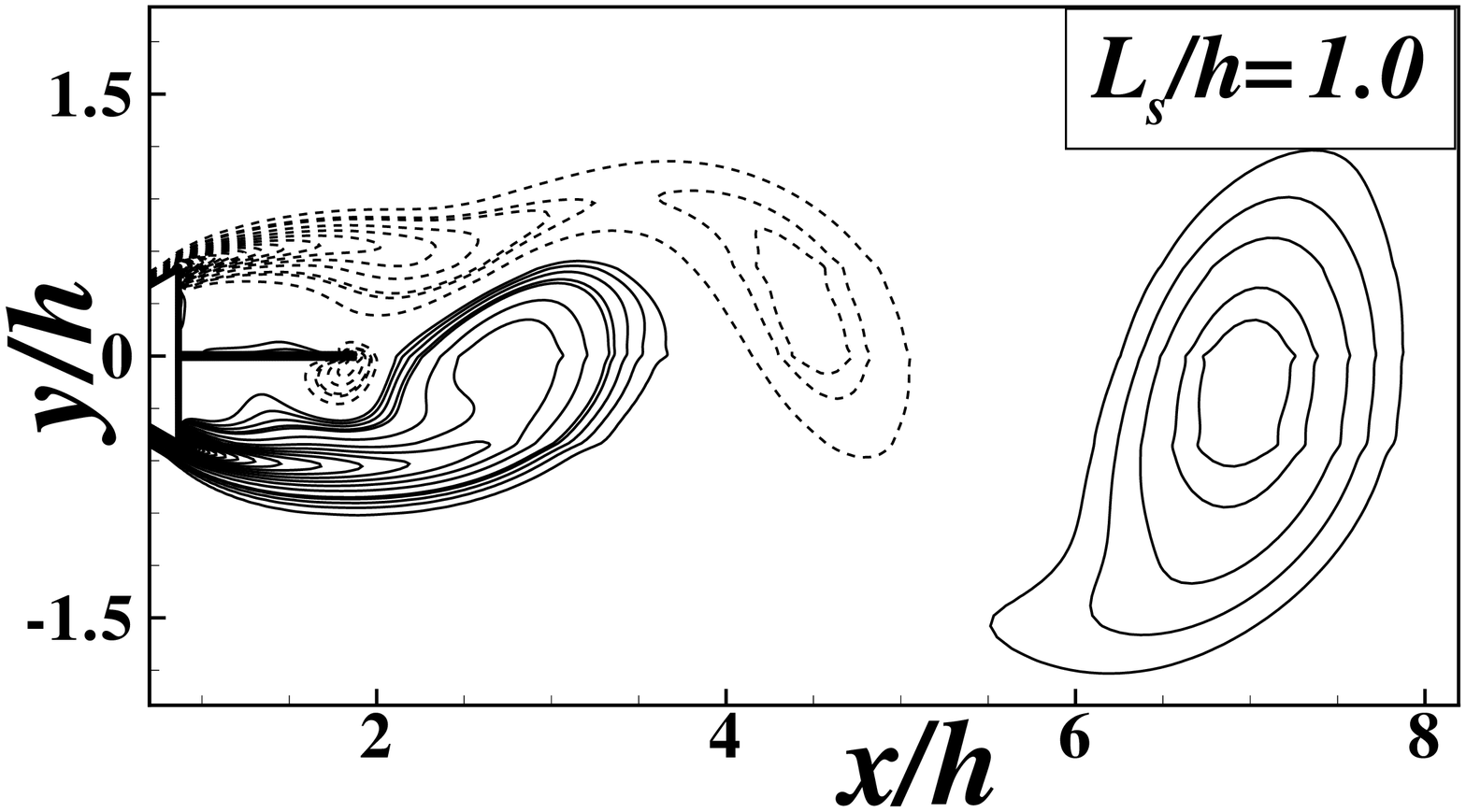} \hfill\ 
\ \hfill \includegraphics[scale=0.20]{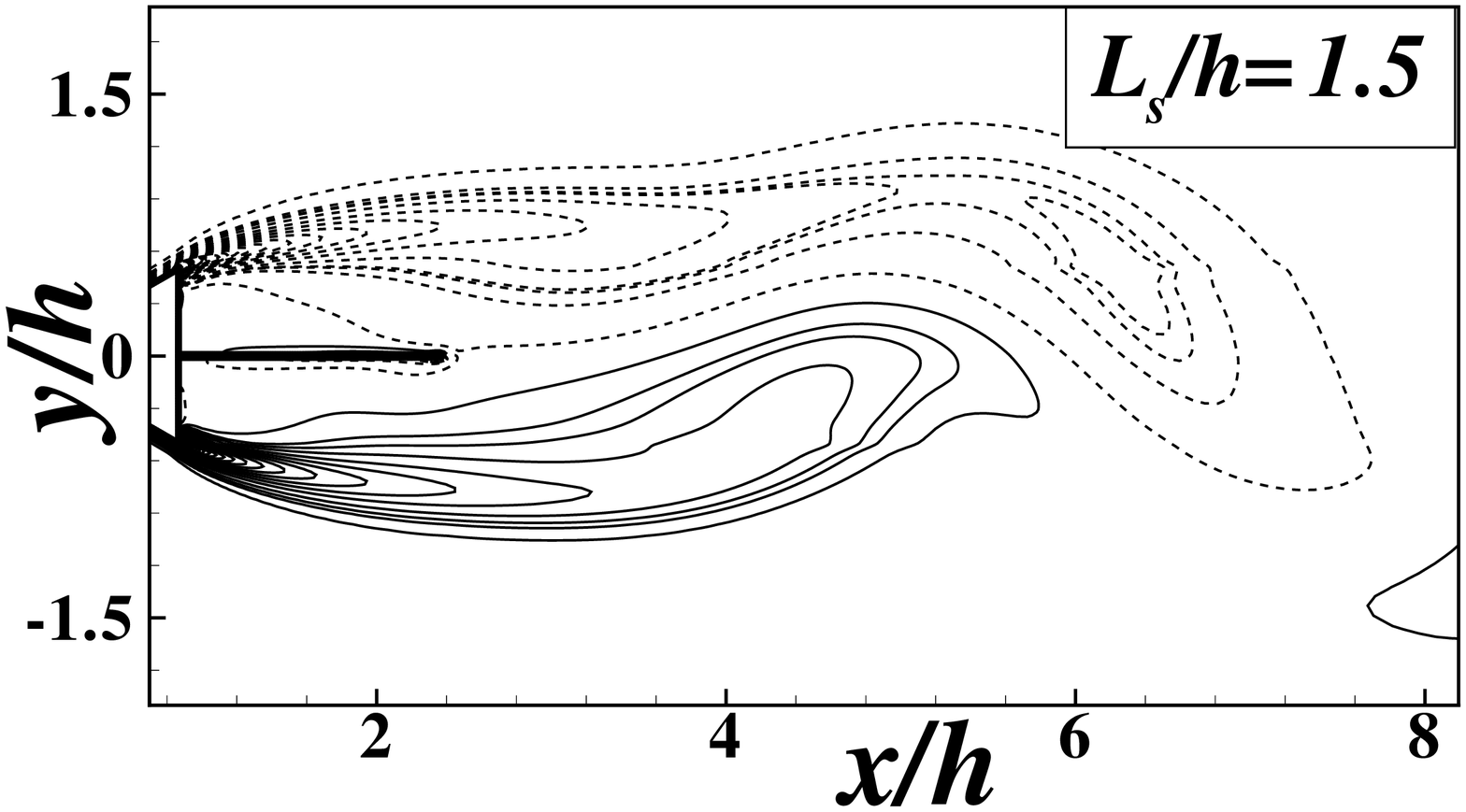} \hfill\ 
  
\end{tabular}
\begin{tabular}{ccc}
\ \hfill \includegraphics[scale=0.20]{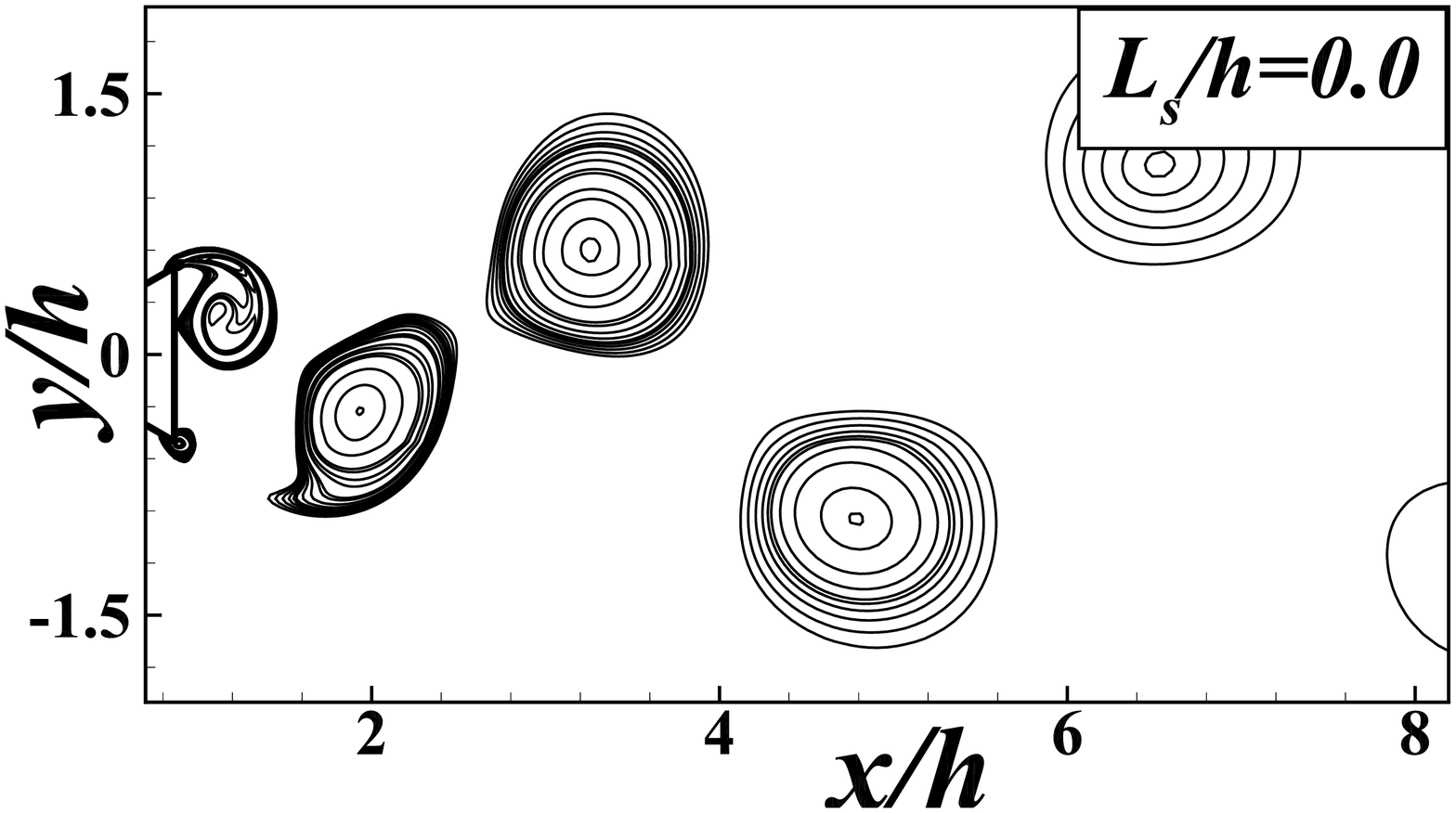} \hfill\
\ \hfill \includegraphics[scale=0.20]{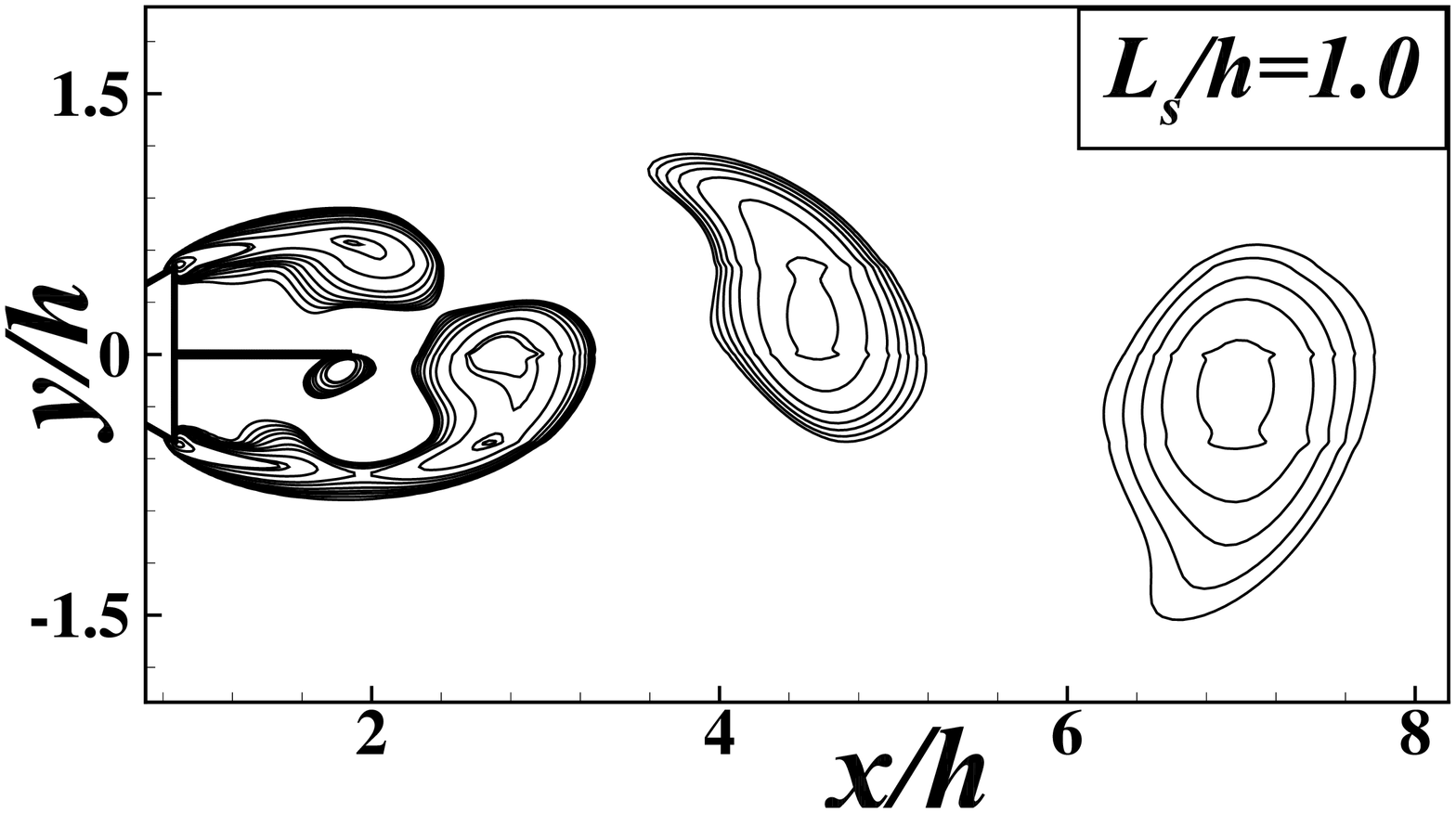} \hfill\
\ \hfill \includegraphics[scale=0.20]{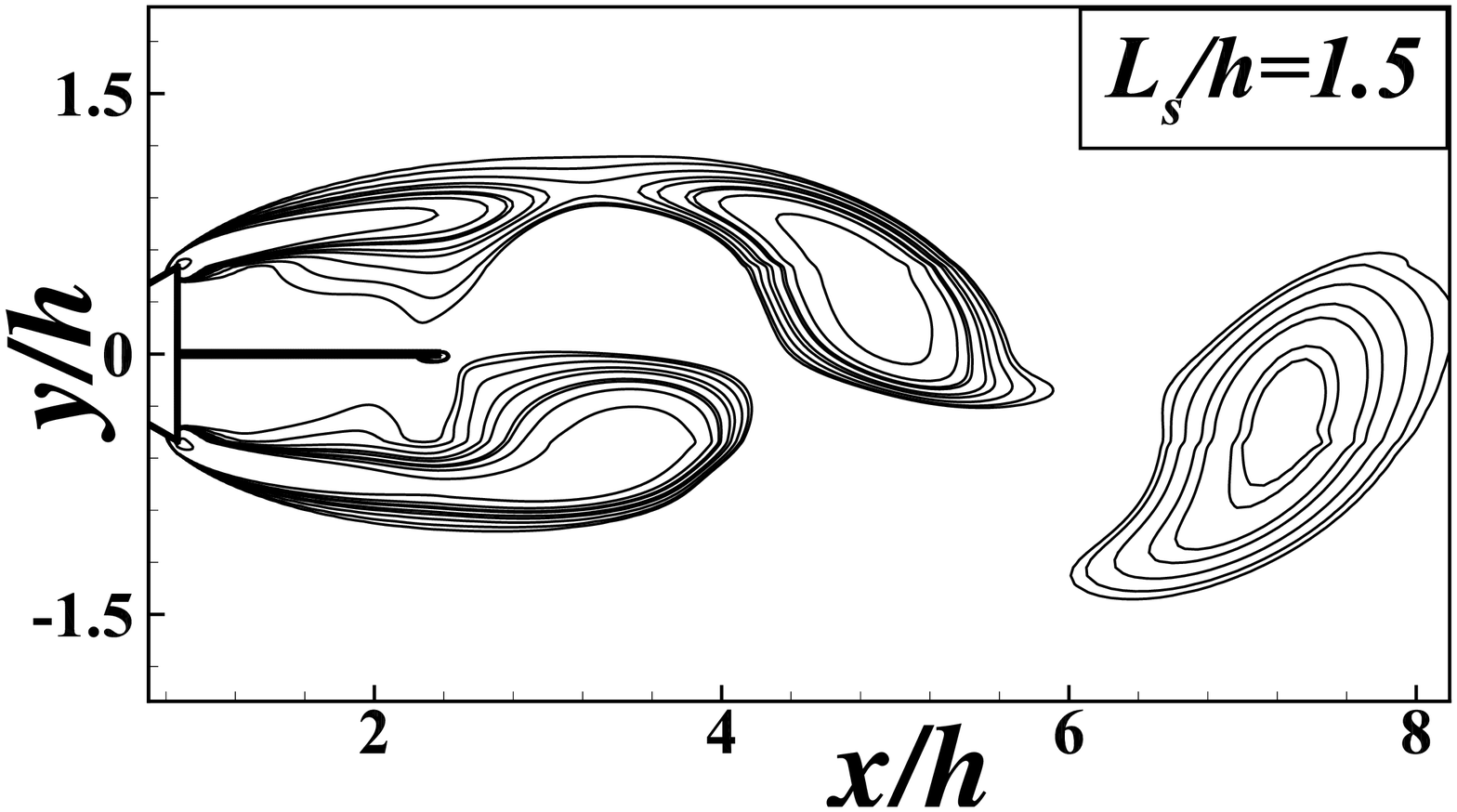} \hfill\
\end{tabular}
\begin{tabular}{ccc}
\ \hfill \includegraphics[scale=0.20]{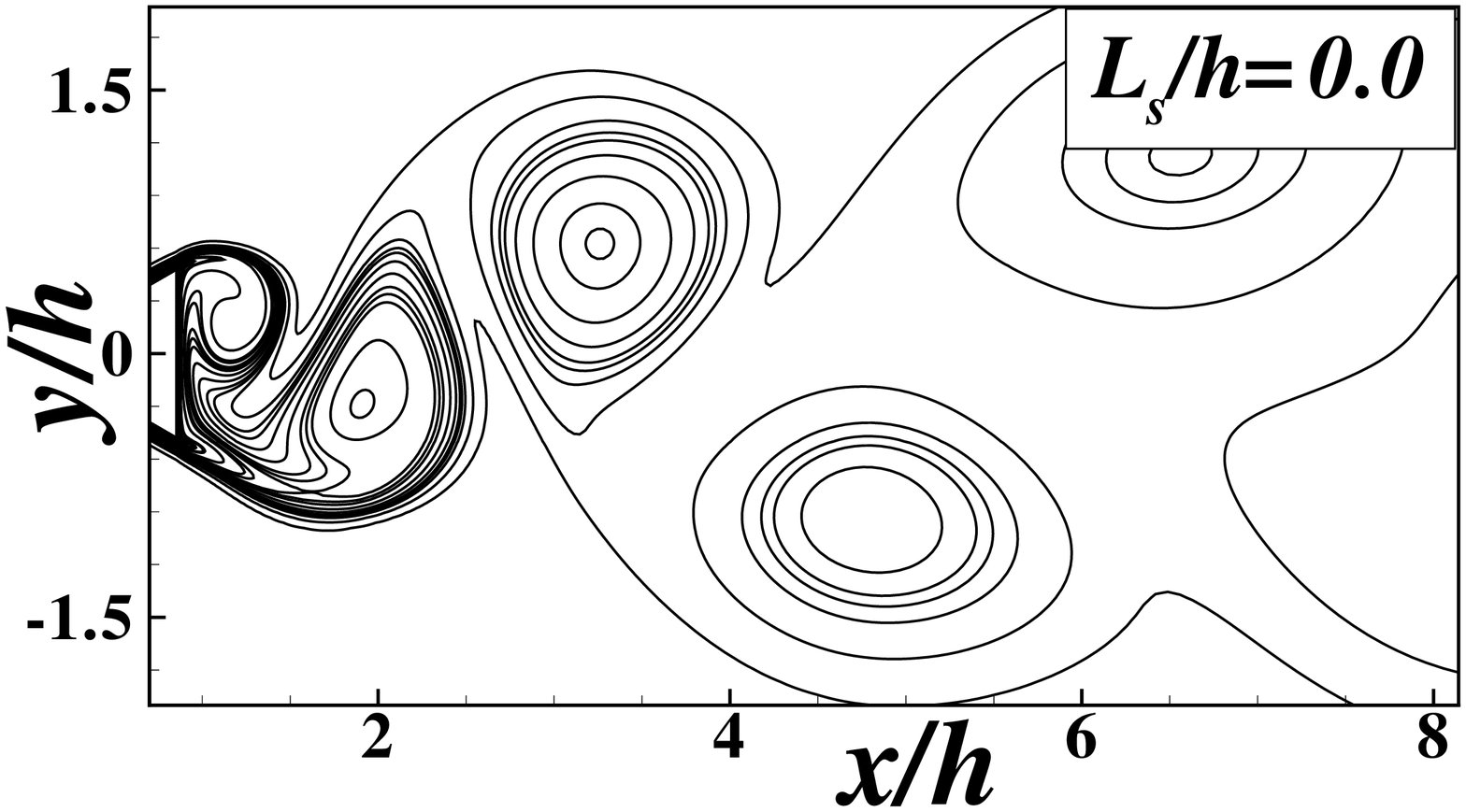} \hfill\
\ \hfill \includegraphics[scale=0.20]{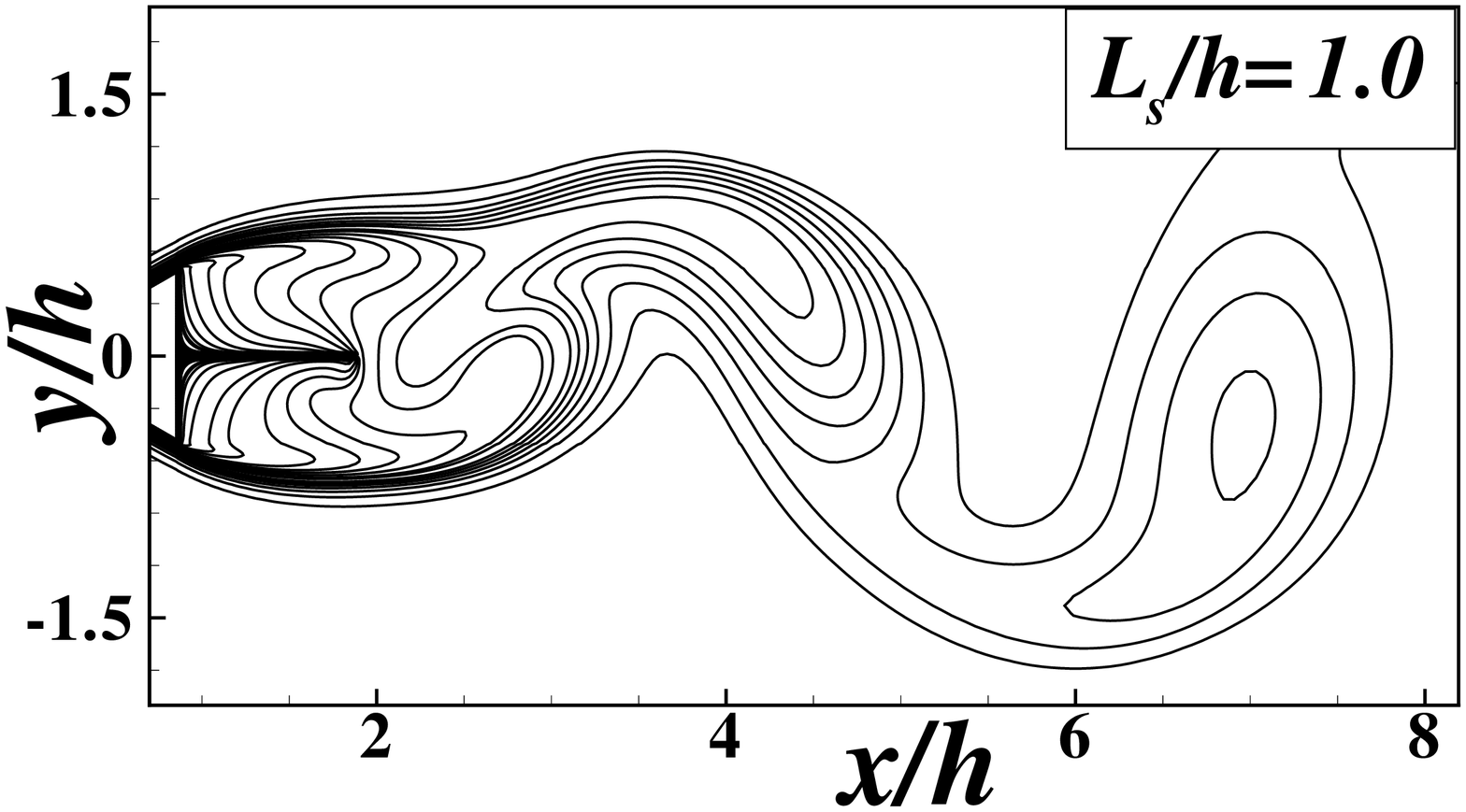} \hfill\
\ \hfill \includegraphics[scale=0.20]{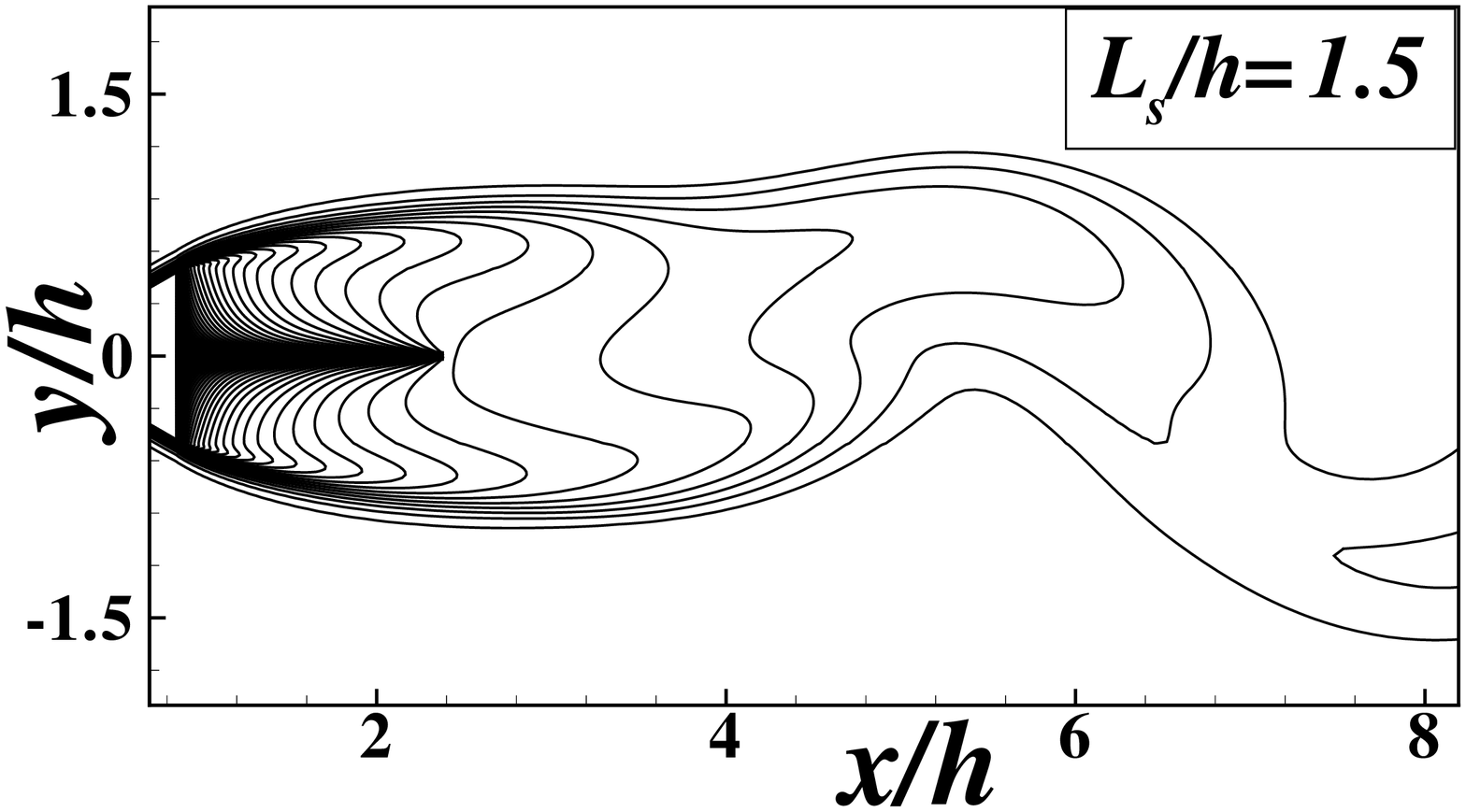} \hfill\
\end{tabular}
\caption{Zoomed view showing energy peak in a PSD plot of Lift-coefficient fluctuation against $St$ (citing the fundamental frequency of the signal); Instantaneous phase-averaged Vorticity contours ($\omega_{min}=-60s^{-1},\omega_{max}=40s^{-1}$); corresponding $\lambda_{2}$ isocontours; and corresponding instantaneous isotherms ($\theta _{min}$=0, $\Delta \theta$=0.01).}
\label{fig:plate_length_allfig}
\end{figure}

Figure \ref{fig:plate_length_allfig} shows instantaneous iso-vorticity contours for cylinder with three different plate lengths. Wake formation for flow past a bare cylinder can be observed where dashed lines show clockwise rotating vortices and solid lines show counter-clockwise rotating vortices respectively. Two rows of alternatingly rotating vortices appeared and in the absence of a splitter, the shear layers interact to form large scale $k\acute{a} rm\acute{a} n$ like structures. Results show that the extent till which the vortex elongates, while attached to one side of the cylinder, increases with increase in plate length, which indicate that with increase in plate length, restriction imposed in between two interacting shear layers increases. This leads to increment in formation length of vortex and suppression in shedding behavior. Secondary tip vortices, as been produced by Ali et al. \cite{mohamedsukri} are observed at the trailing edge of the splitter plate for different plate lengths. Simulations follow that the tip vortex reduced in size while varying the length from $L_{s}/h$=1 to $L_{s}/h$=1.5, which indicate that with increase in plate length, cross-flow is reduced at the tip. Isocontours of $\lambda_{2}$ are shown in Fig. \ref{fig:plate_length_allfig} to visualize the core of the rotating vortices. Power spectral density plot highlighting the fundamental frequency of a lift co-efficient signal is shown in Fig. \ref{fig:plate_length_allfig}.  
\section{Conclusions}
A numerical study has been performed to assess the performance of a splitter plate in modifying the flow behavior in the near wake of a triangular cylinder. The effect of gap ratio and plate geometry of a splitter plate on flow physics, local and averaged heat transfer is studied. Distinguished observations are summarized as follows: 
\begin{itemize}
  \item Present study has confirmed that the wake characteristics downstream of an equilateral triangular cylinder can be significantly altered by putting a splitter plate on the wake centreline. 
  \item Use of control plate suppressed the vortex shedding and produced a stabilizing effect. Analysis on essential flow parameters like frequency of vortex shedding, coefficient of pressure, drag coefficient, vorticity and heat transfer has been paid utmost attention.
  \item A few sets of initial case-study show that gap ratio variation has a tremendous impact on the turbulent wake regime with a minimum $\overline{C_{d}}$ and Strouhal number for $G/h$= 2. Vortex formation length increased with increase in $G/h$ as the rotating vortices has been pushed downstream, hence the wake got narrowed. 
  \item Local Nusselt number variation along cylinder and splitter plate show noteworthy improvement with increase in $G/h$. Introducing a finite gap enhanced the heat transfer manifold as the time-averaged Nusselt number showed an optimal increment of 30.4\% when $G/h$ is increased from 0 to 0.5.     
  \item Further, variation in plate length found to have substantial effect on drag reduction and shedding behavior. In comparison to a bare cylinder, $\overline{C_{d}}$ and shedding frequency reduced by 50.9\% and 18.7\% ,respectively for splitter plate having $L_{s}/h$=1.5. 
  \item Bare cylinder is found to have better heat transfer characteristics than for cases having control plate. Overall heat transfer for the system reduced, but heat transfer along the plate found to be improved with increase in plate length.
\end{itemize} 
\section{Acknowledgement}
This study is funded by a grant from the DAE-BRNS, Government of India. 

%
%
%
%
%
%

\clearpage
\noindent{\bf LIST OF TABLES}\\
\noindent Table 1:   Validation study of flow past a free-standing square cylinder at $Re=22,000$ with Bosch and Rodi \cite{Rodi}.\\ 
\noindent Table 2:   Results of grid refinement test with $Re=22,000$ for $L_{s}/h=1$ and $G/h=0$.\\
\noindent Table 3:   Variation of non-dimensional parameters for different plate lengths ($L_{s}/h$) for 0$\leq L_{s}/h \leq$1.5. 

\newpage
\begin{table}[htbp]
\centering
\label{table:bosch_rodi_validation}
\begin{tabular}{ c c c c }
\hline
\hline
\\
Methods   &  $\overline{C_{d}}$  & $St$ \\ \hline
\\
Experimental \cite{Rodi} &  2.05   & 0.135\\ 
Standard $\kappa - \epsilon$ \cite{Rodi}  & 1.555  & 0.129\\ 
Kato and Launder modified $\kappa - \epsilon$ \cite{Rodi}  & 1.789 & 0.142\\ 
Present Standard $\kappa - \epsilon$ & 1.997 & 0.112\\ \hline
\end{tabular}
\caption{Validation study of flow past a free-standing square cylinder at $Re=22,000$ with Bosch and Rodi \cite{Rodi}.}
\end{table}
\begin{table}[htbp]
\centering
\label{table:grid_refinement_table}
\begin{tabular}{ c c c c c }
\hline
\hline
\\
No. of elements   &  65,000   & 90,000 & 1,15,000 \\ \hline
\\
$\overline{C_{d}}$ &  1.2458   & 1.2366 (0.7) & 1.2376 (0.08)   \\ 
$\overline{C_{d_{p}}}$  & 1.2149  & 1.2062 (0.7) & 1.2073 (0.09)    \\ 
$C_{Lrms}$  & 0.6434 & 0.6417 (0.26) & 0.6410 (0.1)     \\ 
$St$ & 0.1630 & 0.1661 (1.8) & 0.1672 (0.6)    \\ 
$\overline{Nu}$ & 115.9877 & 115.6952 (0.25) & 115.8041 (0.09)  \\ \hline
\end{tabular}
\caption{Results of grid refinement test with $Re=22,000$ for $L_{s}/h=1$ and $G/h=0$.}
\end{table}
\begin{table}[htbp]
\centering
\begin{tabular}{ c c c c c c c c c}
\hline
\hline 
\\
$L_{s}/h$ & $\overline{C_{d}}$ & $\overline{C_{d_{p}}}$ & $C_{Lrms}$ & $St$ & $\overline{Nu }$ & $\overline{C_{p}}$ & $L_{Vf}/h$ \\ \hline
\\
0.0 & 2.021 & 1.993 & 0.290 & 0.171 & 160.813 & -4.923 & 2.105\\
1.0 & 1.237 & 1.206 & 0.642 & 0.166 & 115.695 & -0.535 & 2.777\\
1.5 & 0.991 & 0.967 & 0.231 & 0.139 & 109.413 & -0.396 & 4.454\\ \hline
\end{tabular}
\caption{Variation of non-dimensional parametrs for different plate lengths ($L_{s}/h$) for 0$\leq L_{s}/h \leq$1.5. 
\label{effect_of_change_in_plate_length_table}
}
\end{table}
\clearpage
\newpage


\clearpage
\noindent{\bf LIST OF FIGURES}\\
\noindent Figure 1:   Schematic of the computational domain showing unconfined flow past an equilateral triangular cylinder with a splitter plate.   \\
\noindent Figure 2:   Zoomed view of computational domain (a typical grid consisting of approx. 90,000 mesh elements).   \\
\noindent Figure 3:  Time-averaged stream-wise velocity ($\overline{u}$) along axial direction. \\
\noindent Figure 4:   Equispaced $\tau$/4 (where, $\tau$=time period of lift-coefficient) snapshots of phase-averaged contours showing periodicity within a shedding cycle, (a) phase-averaged vorticity contours ($\omega_{min}=-60s^{-1},\omega_{max}=50s^{-1}$), and (b) corresponding phase-averaged isotherms.\\
\noindent Figure 5:   Variation of (a) time-averaged integral parameters: $\overline{C_{d}}$, $\overline{C_{d_{p}}}$ \& $C_{Lrms}$; and (b) $St$ \& $L_{Vf}$ for 0$\leq G/h \leq$ 2.\\
\noindent Figure 6:   Zoomed view showing (a) Typical Mean Velocity ($\overline{u}/U_{in}$) isocontours and (b) Typical isocontours of turbulence intensity ($i$) for $L_{s}/h=1, G/h=0.5$. In the figure above, contours are originating from the trailing edge of the cylinder.  \\
\noindent Figure 7:  Wake envelope showing mean velocity ($\overline{u}/U_{in}=1$) profiles (a) for $L_{s}/h=1$ with different $G/h$; and (b) for $G/h=0$ with different $L_{s}/h$.  
\noindent Figure 8:   Time-averaged $\lambda_{2}$ isocontours, corresponding isocontours of normalised Reynolds Stress. Maximum and incremental values of Reynolds stress are defined as $[\overline{u'v'}/{U_{in}}^2]_{max}$=0.030 and $\Delta [ \overline{u'v'} /{U_{in}}^2]$=0.001, corresponding Isotherms (with $\theta _{min}=0$, $\Delta\theta=0.02$ and $\theta _{max}=1$ ).\\
\noindent Figure 9:   The profile of time-averaged local pressure coefficient ($\overline{C_{pl}}$) along the edge of the triangular cylinder (a) for $L_{s}/h=1$ and $0\leq G/h\leq 2$, (c) for $G/h=0$ and $0\leq L_{s}/h\leq 1.5$; (b) The profile showing variation of $\overline{C_{pl}}$ along the upper surface of the splitter plate for various gap ratio configurations; and Variation of time-averaged base pressure coefficient ($\overline{C_{bp}}$) (d) with varying $G/h$ for $0\leq G/h\leq 2$; and (e) with varying $L_{s}/h$ for $0\leq L_{s}/h\leq 1.5$.\\
\noindent Figure 10:   (a) Variation of $\overline{Nu}$ with gap ratio for 0$\leq G/h \leq$ 2, The profile of ($\overline{Nu_{l}}$) alongside the edge of the triangular cylinder (b) for $L_{s}/h=1$ and $0\leq G/h\leq 2$, (d) for $G/h=0$ and $0\leq L_{s}/h\leq 1.5$ ; and The profile of time-averaged local Nusselt number ($\overline{Nu_{l}}$) along the upper surface of the splitter plate (c) for $L_{s}/h=1$ and $0\leq G/h\leq 2$, (e) for $G/h=0$ and  $L_{s}/h$=1 and 1.5.\\
\noindent Figure 11:  Zoomed view showing energy peak in a PSD plot of Lift-coefficient fluctuation against $St$ (citing the fundamental frequency of the signal); Instantaneous phase-averaged Vorticity contours ($\omega_{min}=-60s^{-1},\omega_{max}=40s^{-1}$); corresponding $\lambda_{2}$ isocontours; and corresponding instantaneous isotherms ($\theta _{min}$=0, $\Delta \theta$=0.01).\\
\newpage
\clearpage
\clearpage
\clearpage
\clearpage

\newpage

\newpage


\newpage

\newpage
\clearpage
\newpage

\clearpage

\clearpage

\end{document}